\documentclass[twocolumn]{aastex631}

\let\bfseries\mdseries

\newcommand{\trg}[1]{{\color{red}#1}}

\newcommand{\Msolar}{\mbox{$\rm M_{\odot}\,$}}
\def\gs{\mathrel{\raise0.35ex\hbox{$\scriptstyle >$}\kern-0.6em \lower0.40ex\hbox{{$\scriptstyle \sim$}}}}
\def\ls{\mathrel{\raise0.35ex\hbox{$\scriptstyle <$}\kern-0.6em \lower0.40ex\hbox{{$\scriptstyle \sim$}}}}

\shorttitle{COSMOS2020: Identification of High-$z$ Protocluster Candidates in COSMOS}
\shortauthors{Brinch et al.}

\graphicspath{{./}{figures/}}

\begin{document}

\title{COSMOS2020: Identification of High-$z$ Protocluster Candidates in COSMOS}

\author[0000-0002-0245-6365]{Malte Brinch}
\affiliation{Cosmic Dawn Center (DAWN)}
\affiliation{DTU-Space, National Space Institute, Technical University of Denmark, Elektrovej 327, 2800 Kgs.~Lyngby, Denmark}

\author[0000-0002-2554-1837]{Thomas R.~Greve}
\affiliation{Cosmic Dawn Center (DAWN)}
\affiliation{DTU-Space, National Space Institute, Technical University of Denmark, Elektrovej 327, 2800 Kgs.~Lyngby, Denmark}

\author[0000-0003-1614-196X]{John R. Weaver}
\affiliation{Cosmic Dawn Center (DAWN)}
\affiliation{Niels Bohr Institute, University of Copenhagen, Jagtvej 128, 2200 Copenhagen, Denmark}

\author[0000-0003-2680-005X]{Gabriel Brammer}
\affiliation{Cosmic Dawn Center (DAWN)}
\affiliation{Niels Bohr Institute, University of Copenhagen, Jagtvej 128, 2200 Copenhagen, Denmark}

\author[0000-0002-7303-4397]{Olivier Ilbert}
\affiliation{Aix Marseille Univ, CNRS, LAM, Laboratoire d’Astrophysique de Marseille, Marseille, France}

\author[0000-0002-7087-0701]{Marko Shuntov}
\affiliation{Institut d’Astrophysique de Paris, UMR 7095, CNRS, and Sorbonne Universit´e, 98 bis boulevard Arago, 75014 Paris, France}

\author[0000-0002-7303-4397]{Shuowen Jin}
\affiliation{Cosmic Dawn Center (DAWN)}
\affiliation{DTU-Space, National Space Institute, Technical University of Denmark, Elektrovej 327, 2800 Kgs.~Lyngby, Denmark}

\author[0000-0001-9773-7479]{Daizhong Liu}
\affiliation{Max-Planck-Institut fur extraterrestrische Physik (MPE), Giessenbachstr. 1, 85748 Garching, Germany}

\author[0000-0001-9419-9505]{Clara Giménez-Arteaga}
\affiliation{Cosmic Dawn Center (DAWN)}
\affiliation{Niels Bohr Institute, University of Copenhagen, Jagtvej 128, 2200 Copenhagen, Denmark}

\author[0000-0002-0930-6466]{Caitlin M. Casey}
\affiliation{The University of Texas at Austin 2515 Speedway Blvd Stop C1400 Austin, TX 78712 USA}

\author[0000-0002-2951-7519]{Iary Davidson}
\affiliation{Cosmic Dawn Center (DAWN)}
\affiliation{Niels Bohr Institute, University of Copenhagen, Jagtvej 128, 2200 Copenhagen, Denmark}

\author[0000-0001-7201-5066]{Seiji Fujimoto}
\affiliation{Cosmic Dawn Center (DAWN)}
\affiliation{Niels Bohr Institute, University of Copenhagen, Jagtvej 128, 2200 Copenhagen, Denmark}

\author[0000-0002-6610-2048]{Anton M. Koekemoer}
\affiliation{Space Telescope Science Institute, 3700 San Martin Dr., Baltimore, MD 21218, USA}

\author[0000-0002-5588-9156]{Vasily Kokorev}
\affiliation{Cosmic Dawn Center (DAWN)}
\affiliation{Niels Bohr Institute, University of Copenhagen, Jagtvej 128, 2200 Copenhagen, Denmark}

\author[0000-0002-4872-2294]{Georgios Magdis}
\affiliation{Cosmic Dawn Center (DAWN)}
\affiliation{DTU-Space, National Space Institute, Technical University of Denmark, Elektrovej 327, 2800 Kgs.~Lyngby, Denmark}
\affiliation{Niels Bohr Institute, University of Copenhagen, Jagtvej 128, 2200 Copenhagen, Denmark}
\affiliation{Institute for Astronomy, Astrophysics, Space Applications and Remote Sensing, National Observatory of Athens, 15236 Athens, Greece}

\author[0000-0002-9489-7765]{H. J. McCracken}
\affiliation{Institut d’Astrophysique de Paris, UMR 7095, CNRS, and Sorbonne Universit´e, 98 bis boulevard Arago, 75014 Paris, France}

\author[0000-0003-0639-025X]{Conor J. R. McPartland}
\affiliation{Cosmic Dawn Center (DAWN)}
\affiliation{Institute for Astronomy, University of Hawaii, 2680 Woodlawn Drive, Honolulu, HI 96822, USA}
\affiliation{Department of Physics and Astronomy, University of California, Riverside, 900 University Avenue, Riverside, CA 92521, USA}
\affiliation{Niels Bohr Institute, University of Copenhagen, Jagtvej 128, 2200 Copenhagen, Denmark}

\author[0000-0001-5846-4404]{Bahram Mobasher}
\affiliation{Department of Physics and Astronomy, University of California, Riverside, 900 University Avenue, Riverside, CA 92521, USA}

\author[0000-0002-1233-9998]{David B. Sanders}
\affiliation{Institute for Astronomy, University of Hawaii, 2680 Woodlawn Drive, Honolulu, HI 96822, USA}

\author[0000-0003-3631-7176]{Sune Toft}
\affiliation{Cosmic Dawn Center (DAWN)}
\affiliation{Niels Bohr Institute, University of Copenhagen, Jagtvej 128, 2200 Copenhagen, Denmark}

\author[0000-0001-6477-4011]{Francesco Valentino}
\affiliation{Cosmic Dawn Center (DAWN)}
\affiliation{Niels Bohr Institute, University of Copenhagen, Jagtvej 128, 2200 Copenhagen, Denmark}

\author[0000-0002-2318-301X]{Giovanni Zamorani}
\affiliation{Istituto Nazionale di Astrofisica - Osservatorio di Astrofisica e Scienza dello Spazio, via Gobetti 93/3, I-40129, Bologna, Italy}

\author[0000-0002-7051-1100]{Jorge Zavala}
\affiliation{The University of Texas at Austin 2515 Speedway Blvd Stop C1400 Austin, TX 78712 USA}
\affiliation{National Astronomical Observatory of Japan, 2-21-1 Osawa, Mitaka, Tokyo 181-8588, Japan}

\author{The COSMOS Team}

\begin{abstract}
We conduct a systematic search for protocluster candidates at $z \geq 6$ in the COSMOS field
using the recently released COSMOS2020 source catalog. We select galaxies using a number of selection criteria to obtain 
a sample of galaxies that have a high probability of being inside a given redshift bin. We then apply overdensity analysis 
to the bins using two density estimators, a Weighted Adaptive Kernel Estimator and a Weighted Voronoi Tessellation Estimator. We have 
found 15 significant ($>4\sigma$) candidate galaxy overdensities across the redshift range $6\le z\le7.7$. The majority of the galaxies appear to be on the 
galaxy main sequence at their respective epochs. We use multiple stellar-mass-to-halo-mass conversion methods to obtain a 
range of dark matter halo mass estimates for the overdensities in the range of $\sim10^{11-13}\,M_{\odot}$, at the respective redshifts 
of the overdensities. The number and the masses of the halos associated with our protocluster candidates are consistent 
with what is expected from the area of a COSMOS-like survey in a standard $\Lambda$CDM cosmology. Through comparison with simulation, we 
expect that all the overdensities at $z\simeq6$ will evolve into a Virgo-/Coma-like clusters at present (i.e., with masses $\sim 10^{14}-10^{15}\,{\rm \Msolar}$). Compared to other 
overdensities identified at $z \geq 6$ via narrow-band selection techniques, the overdensities presented appear to have $\sim10\times$ 
higher stellar masses and 
star-formation rates. We compare the evolution in the total 
star-formation rate and stellar mass content of the protocluster candidates across the redshift range $6\le z\le7.7$ and find agreement with the total average star-formation rate from simulations.
\end{abstract}


\section{Introduction} \label{sec:intro}
Galaxy clusters are the most massive ($\sim 10^{13}-10^{15}\,{\rm \Msolar}$) and largest ($\sim 1-10\,{\rm Mpc}$) gravitationally-bound structures in the Universe.
In the present-day Universe, they contain up to thousands of bright ($L \geq L^{\star}$) galaxies and reside in massive dark matter halos that, according to the $\Lambda$CDM paradigm of hierarchical structure formation, mark the sites of the greatest overdensities of matter \citep{White-Rees1978}. In this paradigm, clusters are the last structures to virialize and assemble \citep[e.g.,][]{Sheth1999,Mo2010}, and they are, therefore, expected to have a complex and prolonged formation history \citep[e.g.,][]{Bower2004,Kravtsov2012,Overzier2016}. 
Observations tell us that most of the mass locked up in stars at the present time is found
in massive elliptical galaxies, which are predominantly found in clusters \citep[e.g.,][]{Dressler1980}. The same is not true for the cosmic
star-formation rate today, where clusters contribute a negligible fraction -- the bulk of their galaxy populations consist of red and quiescent ellipticals -- and most of the
star-formation takes place in spirals and late-type galaxies, which are typically found outside clusters \citep{Poggianti1999}. This alone suggests that the star-formation activity and stellar mass build-up in clusters must have peaked at earlier times ($z \geq 2$), and
%
we naturally expect the red and quiescent massive galaxies residing in
clusters today to be in star-forming overdense environments at earlier times.
In fact, young and star-forming galaxies are observed to be an increasingly dominant galaxy population in overdense regions at higher redshifts \citep[$z \gs 1.5$, e.g.,][]{Elbaz2007,Cooper2008,Scoville2013}, although red and massive quiescent galaxies continue to be present 
in some clusters up to $z\simeq 2-3$  \citep[e.g.,][]{Kodama2007, Kubo2013}.

Studies of present-day galaxy clusters are limited in their ability to probe the formation and evolution of clusters since key signatures of their formation history are erased in the final stages of assembly as the clusters, and the galaxies within them undergo transformational processes such as dynamical relaxation (virialization), mergers and tidal interactions \cite[e.g.,][]{Zabludoff1996,Kodama2001}. If we are to understand the emergence of clusters, therefore, we are best off searching for them in the high-redshift (high-$z$) Universe, where we can catch them during their formative stages as overdense regions of galaxies (often termed protoclusters; \citet{Overzier2016}). Finding and investigating distant protoclusters 
is critical to understanding the formation and evolution of present-day clusters of galaxies \cite[e.g.,][]{Toshikawa2012,Toshikawa2014, Long2020,Calvi2021}, as well as galaxy quenching caused by dense environment of clusters and formation of quiescent systems we see today \cite[e.g.,][]{Boselli2016,Foltz2018}.  
With a comoving abundance of $\sim 10^{-7}\,{\rm cMpc^{-3}}$, however, high-$z$
protoclusters are rare, and their discovery requires systematic surveys with extensive sky coverage.
At $z \geq 2$, protoclusters are typically identified using rest-frame optical color features in Lyman break 
galaxies (LBGs), narrow-band imaging and spectroscopic follow-up of Ly$\alpha$ emitters (LAEs) 
and/or H$\alpha$ emitters (HAEs) \citep[e.g.,][]{Steidel1998, Malhotra2005, Matsuda2011, Galametz2013,Capak2015}. 
However, protoclusters at $z\geq 3$ have also been found in large (sub-)millimeter surveys with single-dish telescopes 
such as the South Pole Telescope (SPT), the {\it Herschel}, and {\it Planck} space observatories, where subsequent high-resolution ALMA observations of the 
brightest sources have revealed overdensities of dust-enshrouded starburst 
galaxies \citep[e.g.,][]{Clements2014,Oteo2018,Ivison2020,Hill2020, Wang2020}.

\bigskip

According to state-of-the-art simulations of cosmic 
structure formation \citep[e.g.,][]{Chiang2017,Lovell2018, Lovell2021, Lagos2020, Springel2021}, the growth of protoclusters occurred at the peaks of the 
dark matter density distribution during the Epoch of Reionization (EoR) at $z \geq 6$, when the Universe was less than a billion years old. To probe the onset of protocluster formation, therefore, observations must be pushed back to the EoR, where galaxies, in their formative stages, are thought to have congregated around 
these density peaks to form protoclusters and eventually
developed clusters. However, only fragments of this process have been observed. One limiting factor is that to date,
there are only four spectroscopically confirmed protoclusters at $z\geq 6$ with $\geq10$ 
confirmed member galaxies\footnote{\cite{Harikane2019} gives a comprehensive list of 
high-$z$ protoclusters.}. The first one was discovered as an overdensity of $i'$-band 
dropouts, and subsequently, spectroscopically verified to reside at $z=6.01$ 
\citep{Toshikawa2012,Toshikawa2014}. The other three were discovered as 
LAE-overdensities at $z=6.54$, $z=6.6$ and $6.9$ and subsequently spectroscopically 
confirmed \citep[]{Chanchaiworawit2019,Harikane2019,Hu2021}. 
While all four protoclusters have $\geq 10$ confirmed member galaxies, none of them have 
the deep multi-band optical/near-IR data required to accurately characterize their galaxy 
populations (e.g., stellar masses, star-formation rates, and ages \citep{Conroy2013}). In addition to these 
four protoclusters, there is a tentative millimeter-selected galaxy-overdensity at 
$z\simeq 6.9$ \citep{Marrone2018}, consisting of two spectroscopically confirmed dusty 
galaxies with optical HST counterparts and three fainter sub-mm sources lacking both spectroscopic redshifts and optical/near-IR 
counterparts \citep{Wang2020}.

In this paper, we present the discovery of 15 massive protocluster candidates at $6 \leq z \leq 7.7$ 
over a 1.7\,deg$^2$ area in the Cosmic Evolution Survey (COSMOS \citep[][]{Scoville2007}) using the new release of 
the COSMOS catalog (COSMOS2020 \citep[][]{Weaver2021}). The candidates were 
identified from the source list of the near-IR ($izYJHK_{\rm S}$) selected catalog, in a manner similar to the $z=6.01$ protocluster found by 
\citet{Toshikawa2012} but, being located in the COSMOS field, they have unique multi-band wavelength coverage and depth.

The paper is organized as follows. Section \ref{section:data} briefly summarizes the COSMOS survey and our protocluster candidates selection criteria. Section \ref{section:method} explains the methods used to identify protocluster candidates through galaxy overdensity analyses. Section \ref{section:results} presents our findings and analysis of the protocluster candidates. 
We compare our candidates with galaxy overdensities found using traditional dropout selection techniques, and present estimates for the dark matter halo mass and the present-day mass of the candidates.
Section \ref{section:discussion} discusses if our dark matter estimates are in line with what we would expect for a COSMOS-like survey. We also compare the physical parameters of our candidates with other protoclusters from the literature. 
Section \ref{section:conclusions} summarizes the main findings and conclusions.
\\\\
Throughout the paper, we have adopted a standard $\Lambda$CDM cosmology
with $H_{\rm 0}=70\,{\rm km\,s^{-1}\,Mpc^{-1}}$, $\Omega_{\rm m} = 0.3$, 
and $\Omega_{\rm \Lambda} = 0.70$. All magnitudes are expressed in the 
AB system \citep{Oke1974}. A \citet{Chabrier2003} stellar Initial Mass
Function (IMF) is used to present our results. Results are reported with 68\% confidence interval uncertainties.

\section{Data}\label{section:data}
\subsection{The COSMOS Survey}\label{subsection:cosmos-survey}
The Cosmic Evolution Survey (COSMOS) covers 2\,deg$^2$ and boasts a plethora of 
deep multi-wavelength data in over 40 bands from the world's major facilities
\citep{Scoville2007}. The new version of the COSMOS source catalog, COSMOS2020 
\citep{Weaver2021}\footnote{The COSMOS2020 catalog can be downloaded from \url{https://cosmos2020.calet.org}}, constitutes 
a major improvement over the previous catalog \citep{Laigle2016}, in that it includes 
detections and new ultra-deep optical/NIR imaging and multi-waveband photometry of 
1.7 million sources over the entire COSMOS field, with $\sim 89000$ measured in all the available 
broadband filters. Key additions for COSMOS2020 include new ultra-deep optical data from the 
Hyper Suprime-Cam Subaru Strategic Program (HSC-SSP) public data release 2 (PDR2; \citet{Hiroaki2019}), new data from the Visible Infrared Survey Telescope for Astronomy (VISTA) data release 4 (DR4; \citet{McCracken2012}), reaching up to one magnitude deeper over the full area than previous data, and the inclusion of all {\it Spitzer} IRAC data in the COSMOS \citep{Moneti2021}. Legacy data sets 
(such as the Suprime-Cam imaging) have also been reprocessed (see \cite{Weaver2021} for details). 

COSMOS2020 consists of two independent photometric catalogs. 
The first is the {\tt Classic} 
catalog, where standard aperture photometry is performed \citep{Bertin1996} on PSF-homogenized optical/NIR images, 
except for the IRAC images where the software {\tt IRACLEAN} 
\citep{Hsieh2012} is used to do PSF photometry. The second catalog is created using a new 
profile-fitting photometric software developed specifically for COSMOS2020 called {\sc The Farmer}. Using {\tt The Tractor} software \citep{Lang2016} for the source modelling, {\sc The Farmer} generates reproducible source detection and photometry to generate a full multi-wavelength catalog. In this paper, we have used 
{\sc The Farmer} catalog throughout since it has more accurate photometry in different bands for fainter sources than the {\tt Classic} catalog and appears to have a higher density of sources with photometric redshifts $> 6$ by almost a factor of two in the faintest NIR magnitude bins \citep{Weaver2021}. This also translates to more sources for {\sc The Farmer} catalog, as seen in Fig.~\ref{fig:zdistribution}. This difference in high redshift sources could be explained by two factors, one is {\sc The Farmer} is able to deblend more sources than {\tt Classic} and the other is that for the {\tt Classic} catalog, the apertures can be contaminated by stray blue (optical) light, while the same is not the case in  {\sc The Farmer} catalog, leading to more robust high photometric redshift solutions.  
\subsection{Photometric redshifts: LePhare and EAZY}\label{subsection:physical-parameters}
Photometric redshifts and physical parameters, including stellar masses and star-formation rates, for the COSMOS2020 sources, were derived using two independent photometric redshift codes, LePhare \citep{Arnouts2002,Ilbert2006} and EAZY \citep{Brammer2008}. Both codes were used to fit photometric redshifts (photo-$z$) and UV/optical to near-IR spectral energy distributions (SEDs) to the objects in COSMOS2020. The codes fit galaxy SED templates to the photometric data, corrected for Galactic extinction using the \citet{Schlafly2011} dust map. 

The codes have similar approaches to determining the photo-$z$ as they iteratively derive multiplicative corrections to both the individual photometric bands and the SED templates. The main differences consist of the stellar population synthesis templates used and the method by which they are fit to the observed photometry. A list of all the magnitude offsets used to optimize the absolute calibration of the photometry for each band for LePhare, and EAZY and both {\sc The Farmer} and {\tt Classic} COSMOS2020 catalogs are given in \cite{Weaver2021}. While results between the two codes are in good overall agreement, LePhare has a lower outlier percentage ($\eta$)\footnote{The outlier percentage $\eta$ is defined as the percentage of galaxies that have $|\Delta z|>0.15(1+z_{\rm spec})$, where $|\Delta z|$ is the absolute difference between $z_{\rm phot}$ and $z_{\rm spec}$.} as a function of the 
Normalized Median Absolute Deviation ($\sigma_{\rm NMAD}$) and $i$-band magnitude bin \citet{Weaver2021}. 
This is particularly the case for the faintest magnitude bin ($25-27$ magnitude in the $i$-band, see Fig.~15 in \citep[][]{Weaver2021}), where both $\sigma_{\rm NMAD}$ and $\eta$ are lower when applying LePhare on  {\sc The Farmer} catalog. 
For this reason we mainly use the LePhare photo-$z$, which, unless stated otherwise, we will henceforth refer to as the photo-$z$ (for more details on the calculation of the outlier percentage and the differences between the COSMOS2020 catalog versions, see \S5.3 in \citet{Weaver2021}). To test the consistency of the results we have also done our overdensity analysis with the EAZY photo-$z$ and found general agreement between the maps at the $1\sigma$ level. The relevant overdensity maps are available upon request

\subsection{Physical parameters: LePhare and EAZY}
The physical properties of the COSMOS2020 galaxies, such as stellar masses and star-formation rates, are also available from the two relevant codes (see \cite{Weaver2021} for details on how these quantities are calculated). 
In this paper, however, we did not use the SFR estimates provided by the two codes. Tests showed
significant run-to-run variation in the SFR-estimates,  particularly for LePhare, and between the two codes. 
In the case of LePhare, the code is run twice, once to obtain the photo-$z$ and then again with the photo-$z$ fixed to obtain physical parameters such as stellar mass and SFR. For more details, see \S6 in \cite{Weaver2021}. 
LePhare imposes a declining SFH for both star-forming and quiescent galaxies (though at the redshifts we are considering, the delayed SFH is likely to be in the rising SFR epoch), whereas EAZY makes no assumption about the SFH. This has no effect on the stellar masses, where the two codes show good agreement, but it can strongly affect the SFR estimates. 
Neither of the codes includes far-IR photometry in the SED fitting, which may also lead to significant 
uncertainties in SFR estimates (see \cite{Laige2019}).  
Since we use the LePhare photo-$z$ for our galaxy selection and to ensure our analysis is done as 
self-consistent as possible, we will use the stellar masses from LePhare throughout the paper. 
We will use the Ultra Violet (UV) luminosity to SFR conversion from \citet{Barro2019} for the SFR estimates, corrected for dust attenuation: 
\begin{equation}
    {\rm SFR}_{\rm UV}^{\rm corr} = (1.09\times10^{-10})(10^{0.4A_{\rm 2800}})(3.3 L_{\rm 2800}/{\rm L_{\odot}}),
\label{equation:sfr-uv}
\end{equation}
where $L_{\rm 2800}$ and $A_{\rm 2800}$ are the UV luminosity and dust attenuation at rest-frame $\lambda = 2800$Å, respectively. Assuming a \cite{Calzetti2000} attenuation law, the UV attenuation at $2800$Å can be inferred directly from the best-fit model to the overall SED so that $A_{\rm 2800} = 1.8A_{\rm V}$, where $A_{\rm V}$ is the extinction in the $V$-band. The COSMOS2020 catalog provides $L_{\rm 2800}$ and $A_{\rm 2800}$ from the best fit SED made with EAZY. The same values are not provided for LePhare in the catalog. However, it is possible to obtain the flux (and thereby the luminosity, assuming a luminosity distance from the photo-$z$) at rest-frame $\lambda=2800$Å from the best fit SED before the dust attenuation using the \cite{Calzetti2000} law is applied, effectively using eq.\,\ref{equation:sfr-uv} with $A_{\rm 2800}=0$. Since they are already available in the catalog and we require the photo-$z$ from the two codes to be similar (thereby requiring the SED fits to be similar by proxy), we will use the two EAZY properties to estimate the star-formation rates. The star-formation rates we get from this conversion and the LePhare fit both scatter around the expected main sequence from \citet{Speagle2014}, with the conversion having a greater scatter overall.
The \cite{Barro2019} relation has a scatter of $0.32$ dex when compared with the \cite{Wuyts2011} relation that includes both the IR and UV. We have chosen not to include this scatter in our estimate and use the uncertainty for $L_{\rm 2800}$ and $A_{\rm 2800}$ given by EAZY.
\subsection{Selection criteria and redshift  binning}\label{subsection:selection-criteria}
Our search for high-$z$ protoclusters in the COSMOS2020 catalog focuses on galaxies at  $z\geq 6$, as it goes from the End of Reionization (EoR) to the highest photo-$z$ provided by the LePhare. The photo-$z$ assigned to a given galaxy and used in its selection is the median of its redshift probability distribution, $p(z)$, from LePhare. We use the flags in  {\sc The Farmer} catalog, so as to only include galaxies and not sources such as stars, sources that coincide with an X-ray source (based on Chandra COSMOS Legacy; \citet{Civano2016}), sources in masked-out region as defined in \citet{Weaver2021} (combining Hyper Suprime-Cam, Suprime-Cam and UltraVISTA masks) or sources where LePhare fails to fit the SED. We checked the catalog for any AGN sources with X-ray detection's and a LePhare AGN fit and found seven sources between $6.3<z<9.7$. None of the sources are in an overdense environment and would not be inside any of our overdensities given our selection criteria, so we have chosen not to incorporate them in our analysis further.

\begin{figure}
  \includegraphics[width=0.47\textwidth]{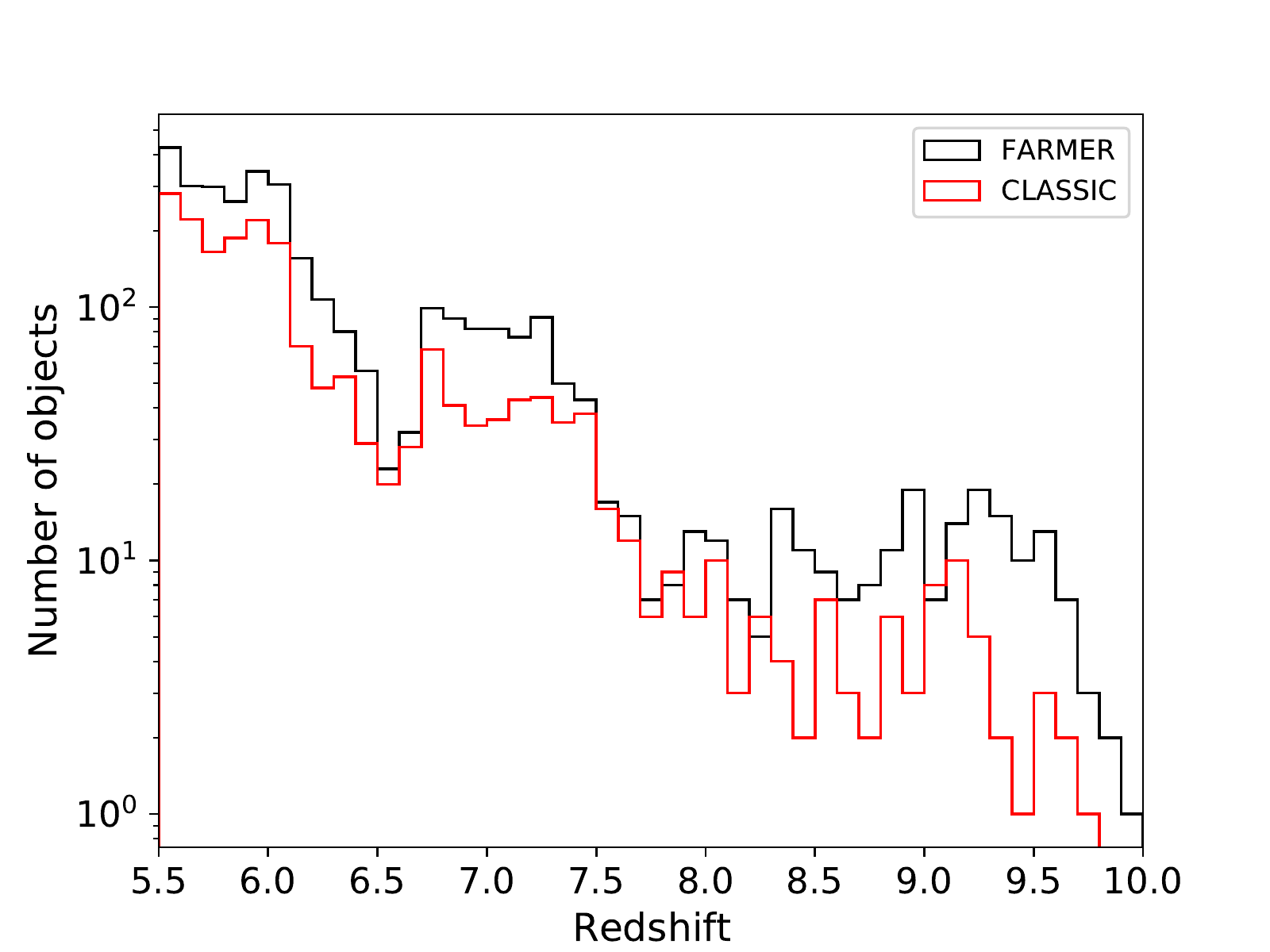}
  \caption{The redshift distribution of galaxies between $z=5.5-10$ in the COSMOS2020 catalog, based on the LePhare median photo-$z$ and the selection criteria specified in {\S}\ref{subsection:selection-criteria}. The histogram bin width is $\Delta z=0.1$. The black curve is for the {\tt FARMER} catalog and the red curve is for the {\tt Classic} catalog (see \S\ref{subsection:cosmos-survey}).}
  \label{fig:zdistribution}
\end{figure}

To ensure that the LePhare photo-$z$ we work with comes from a secure sample with good SED fits, we require the reduced $\chi^2<5$ for all sources. The $\chi^2$ criteria accounts for a small number of outliers with very high $\chi^2$ values. 
We also investigated the relationship between the reduced $\chi^2$ from LePhare and the difference between the median photo-$z$ from LePhare and EAZY, to explore if there was a positive correlation between the two. We found no such correlation, with the majority of the sources having an absolute difference in photo-$z$ of $<0.5$ between the two codes. There appears to be smaller groups of galaxies where the absolute difference in photo-$z$ between the two codes is $>5$. This is because EAZY places many ill-constrained SEDs at $z>11.5$ or between $z=0-1$, whereas LePhare places the same galaxies at $z\sim6$, as its $p(z)$ is constrained between $z=0-10$. To account for the possible bias that comes with using a single SED fitting code, we require that the absolute difference between the median LePhare and EAZY photo-$z$ is $<0.5$. This accounts for the majority of the galaxies with median LePhare photo-$z$ between $z=6-10$. The total number of galaxies in the sample is 3256 across $z=5.5-10$. The redshift distribution of these galaxies is shown in Fig.~\ref{fig:zdistribution} using the {\tt FARMER} and {\tt CLASSIC} catalogs.

Next, we will describe the redshift bins we are using to search for protocluster candidates, we initially describe our lowest redshift bin and then generalise to higher redshifts.
We initially target a redshift bin centered on $z=6.05$ with a bin width of $\Delta z=0.2$. At this redshift, the bin width is equivalent to a distance of $\approx80\,{\rm cMpc}$, which we will use for all the other bins (i.e., the redshift range will increase with redshift to a maximum of $z=9.81\pm0.19$ for the highest bin). 
This bin width is the same as the one used by \cite{Harikane2019} to search for LAE overdensities. Semi-analytical models in \cite{Chiang2017} suggest that the average protocluster size at $z\simeq 6$ is $\approx10\,{\rm cMpc}$, so we should be able to determine overdensities 
within our adopted redshift bin width. This bin size also encompasses the redshift range of other protoclusters found at $z\approx 6$ \citep[]{Toshikawa2014,Chanchaiworawit2019,Harikane2019,Hu2021}.    

To select galaxies with a high probability of being inside our redshift bin, 
we require that all galaxies inside the bin have their maximum
$p(z)$ value within 5\% of their median value. This excludes four groups of $p(z)$
that could otherwise affect our overdensity analysis: 1) $p(z)$, which have narrow low-$z$ 
solutions with peak values higher than the broader peak found inside the $z$-bin, 2)
$p(z)$ with broad distributions and median redshifts inside our $z$-bin, but with a peak on top
of the broad trend that is more than 5\% away from the median value, 3) double-peaked
$p(z)$ with median values that fall within the $z$-bin and 4) $p(z)$ with a plateau with equivalent
values on either side of the bin that span multiple redshifts, but has a peak more
than 5\% away from the median value. This selection does not take $p(z)$ into account
that have a plateau-like probability distribution similar to 4), but with their maximum $p(z)$ value within 5\% of the median
value. We describe in {\S}\ref{subsection:WAK} how to account for these objects. These selection criteria aim to obtain a sample with the best trade-off between having many galaxies, so as to have a representative estimate for the mean density field, and to have secure galaxies that have a high probability of being in our redshift bins.

In LePhare, the $p(z)$ is only defined between $z=0-10$, and we find a group of galaxies that are placed right at the $z=10$ boundary with a narrow $p(z)$. Furthermore, galaxies that fall within redshift bins very close to $z=10$ will have a significant part of their $p(z)$ cut off at $z > 10$. This can lead to overestimating the galaxy overdensity near these galaxies since their weights will be biased high because less of the total $p(z)$ will be outside the redshift bin due to the cut-off at $z=10$. We, therefore, chose to discard redshift bins higher than $z=9.23$ since beyond this cut-off, it is not possible to say anything meaningful about the overdensities. This is because a large number of the galaxies in the higher bins ($z=9.59\pm0.18$ and $z=9.81\pm0.19$) either have a large fraction of its $p(z)$ outside the $z=10$ boundary for LePhare or in the case of the highest bin, the $p(z)$ peaks right at the $z=10$ boundary showing the code is not adequately able to model the probability distribution for these sources.  
\subsection{Possible false detections using {\sc The Farmer}}
{\sc The Farmer} and {\tt CLASSIC} use a combined $izYJHK_{\rm s}$ (denoted CHI\_MEAN) to detect the sources. This makes it possible to detect sources in the combined image, which would not be detected at the same confidence in the individual images. There are also sources where there are no individual detection in the images, but there is a detection in the combined image. A small group of galaxies in our sample have no individual detections in the grizYJHKs bands {\sc The Farmer} uses or the bands LePhare uses for SED fitting (see \cite{Weaver2021} for a full list), These galaxies all have low weights (see \S\ref{subsection:WAK} for definition of weights) with none having a weight above 0.5. We also see an increase in the weight of galaxies as a function of the number of bands with detections. While {\sc The Farmer} uses multiple bands to locate sources in the field, it is still possible for it to falsely detect spurious sources, especially around very bright sources. While bright sources have been masked out, there is always a trade-off regarding how much of the region around a bright object should be masked. Within the UltraVista area of the COSMOS field, 
23079 sources detected with {\sc The Farmer} catalog are not in the {\tt Classic} catalog (for our sample, compare the number distributions at high redshifts in Fig.\,\ref{fig:zdistribution}). These sources are mainly faint sources or sources that {\sc The Farmer} de-blended.
Most of the 23079 sources are fainter than the detection limit in the UltraVISTA bands while being brighter than the detection limit in the IRAC\_CH1 and HSC $i$ bands. 
We produce $izY$\footnote{The $Y$-band is from UltraVISTA.} and $JHK$ true color images for all significant overdensities we find in a given redshift bin. We searched for any sources located inside overdensities with no individual detection in the six bands that {\sc The Farmer} uses, but did not find any. Upon inspection, we found that sources in the overdense regions that were only detected with {\sc The Farmer} are real sources, clearly visible in the true-color images in at least one of the izYJHKs bands. We highlight the $KsHJ$-band (red, green, blue) true color images in appendix \ref{sec:allOD}.
\subsection{Accounting for lower redshift interlopers}
An indicator of possible low-$z$ interlopers is strong FIR or mm detections. We cross-matched all the protocluster candidate galaxies we found with the {\it Spitzer}/MIPS $24\,{\rm \mu m}$/radio catalogs in COSMOS and found no robust detections in either. 

We have also cross-matched all protocluster candidate galaxies with the A3COSMOS photometry and found no single dust continuum detection above a Signal-to-Noise ratio ($S/N$) of $4.3$. This $S/N$ threshold is the criterion for a $50\%$ false detection rate. There are in total 15 sources with dust continuum upper limits from the A3COSMOS photometry. There is only one $S/N>3$ source, COSMOS2020-ID871193 with $S/N\sim4$, so considering it as an upper limit is still reasonable.

\section{Method}\label{section:method}
\subsection{Weighted Adaptive Kernel}\label{subsection:WAK}
To search for galaxy overdensities, which indicate the presence of protoclusters, we use a Weighted Adaptive Kernel (WAK) Estimator \citep{Darvish2015}. This estimator uses an iterative process that computes the galaxy surface density field and takes into account the uncertainties related to the photometric redshifts of the galaxies by assigning a weight, $w_{i}$, to each galaxy.  
The weight of a galaxy is assigned according to the percentage of its $p(z)$ that falls within the chosen redshift bin. By only selecting galaxies whose median and maximum $p(z)$ values fall within a given bin (see \S\ref{subsection:selection-criteria}) and further weighing them in the aforementioned manner, we are conservative with our selection, preferentially selecting galaxies that have a high probability of being at their assigned redshift. A more lenient approach would be to assign a weight to all galaxies whose $p(z)$ overlaps with the chosen redshift bin \citep[e.g.,][]{Darvish2015, Darvish2020}. In Appendix \ref{sec:weights} we show the distribution of the weights for all the redshift bins we consider in our analysis (Fig. \ref{fig:weights}).

\begin{figure*}
\centering
  \includegraphics[width=\textwidth]{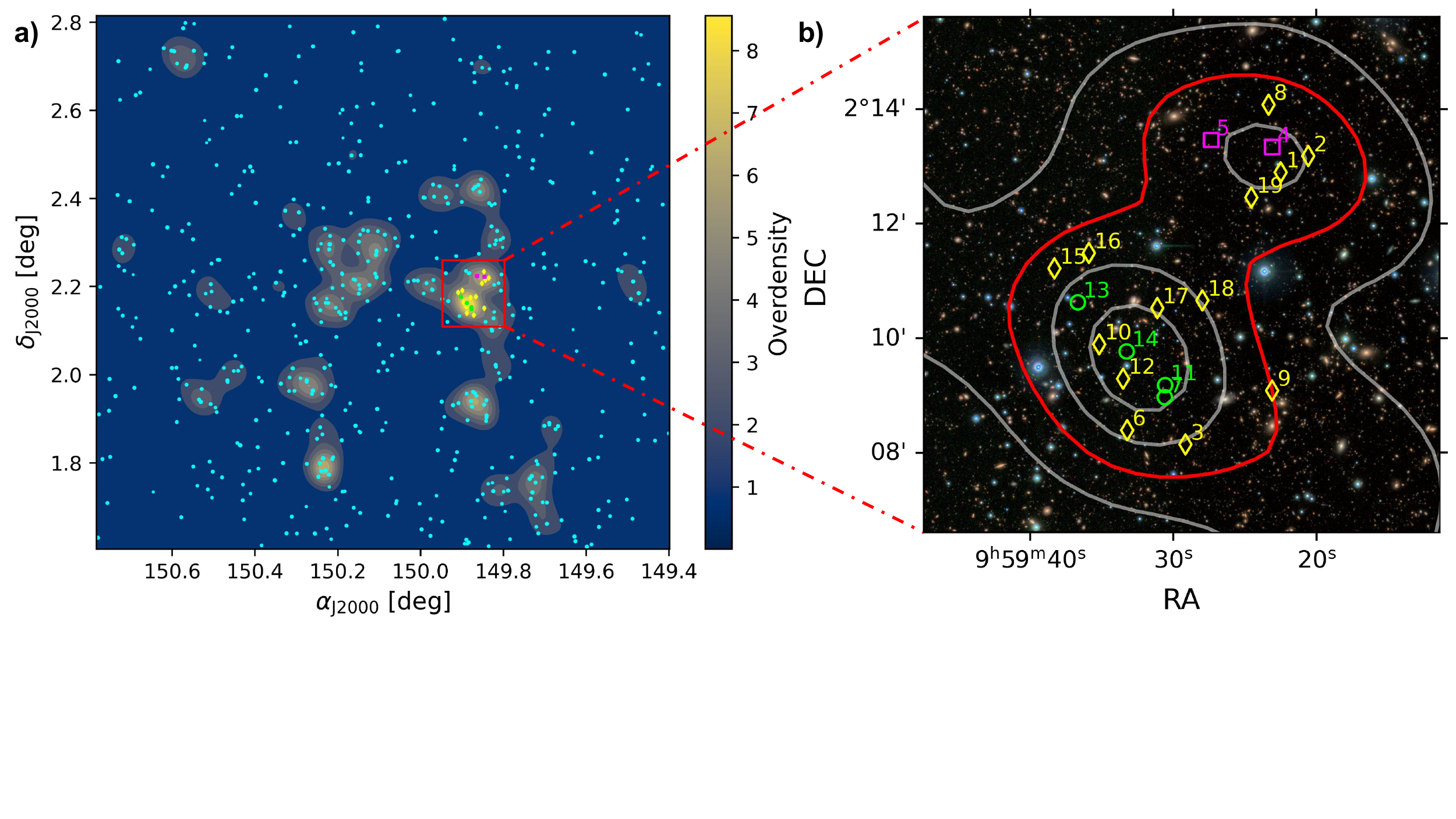}
  \vspace{-2.5cm}
  \caption{{\bf a)} The galaxy overdensity field at $z=6.05\pm 0.10$ in COSMOS, derived from the Weighted Adaptive Kernel method described in \S\ref{subsection:WAK}. contours are in steps of $1\sigma$ and all grid points below $1\sigma$ have the same color. Individual galaxies are indicated as black dots, except for the colored dots which indicate the 19 galaxies, which make up the highly significant ($\delta_{\rm OD, max} \simeq 9$, $\sim 8\sigma$) galaxy overdensity, COSMOS2020-PCz6.05-01 (see Tables \ref{tab:show19data} and \ref{tab:fulldata}).  
  {\bf b)} RGB ($KsHJ$-band) true color image of the $13'\times 18'$ cluster region, with galaxy overdensity ($\delta_{\rm OD}$) contours (in steps of $2\sigma$) shown. The red contour indicates the $4\sigma$ level. Protocluster members are indicated according to their stellar mass: $M_{\rm \star}=10^{8.0-9.0}\,{\rm M_{\rm \odot}}$ (magenta squares), $M_{\rm \star}=10^{9.0-10.0}\,{\rm M_{\rm \odot}}$ (yellow diamonds), and $M_{\rm \star} = 10^{10.0-11.0}\,{\rm M_{\rm \odot}}$ (green circles). We estimate the dimensions of the protocluster be roughly $10.7\,{\rm cMpc}\times 14.3\,{\rm cMpc}\times 66.3\,{\rm cMpc}$ (RA, Dec, $z$).}
  \label{fig:adaptive kernel}
\end{figure*}

The galaxy number surface density, $\hat{\Sigma}_i$, at some grid-position, $i$, in a given redshift bin is calculated by summing over the weighted fixed kernels placed at the positions of the galaxies, $j$, within the bin, where $i \neq j$, i.e.:
\begin{equation}
    \hat{\Sigma}_i=\frac{1}{\sum^{N}_{j=1,\ i \neq j}w_j}\sum^{N}_{j=1,\ i \neq j} w_j K(r_i,r_j,h),
\end{equation}
where $N$ is the number of galaxies in the bin, $K(r_i,r_j,h)$ is
the fixed kernel, $r_i$ is the position of the galaxy for which the estimate of surface density is measured, $r_j$ is the
position of all other galaxies in the bin and $w_j$ the weights of all other galaxies in the redshift bin. The kernel function used is a 2D symmetric Gaussian, with a kernel width, $h$, that controls the smoothing. The kernel function is defined as follows:
\begin{equation}
    K(r_i,r_j,h)=\frac{1}{2\pi h^2}\exp\left(\frac{-|r_i-r_j|^2}{2h^2}\right).
\end{equation}
To select an optimal global kernel width, $h$, we maximize the so-called Likelihood Cross-Validation (LCV) quantity \citep{Chartab2020}, which is defined as:
\begin{equation}
{\rm LCV}(h)=\frac{1}{N}\sum^{N}_{k=1}\log(\sigma_{-k}(r)),    
\end{equation}
where $N$ is the total number of galaxies in the given redshift bin
and $\sigma_{-k}(r)$ is the kernel estimator computed at position $r$
excluding the $k$'th galaxy\footnote{This is equivalent to the surface density $\hat{\Sigma}_i$}. 
Determining $h$ in this way provides an optimal trade-off between a high-variance estimate (under-smoothing) 
and a high-bias estimate (over-smoothing) \citep{Chartab2020}. 
We perform a grid search from $0.0001\,{\rm deg}$ to $0.1\,{\rm deg}$ with 1000 steps to determine the optimal $h$-value that maximizes ${\rm LCV}(h)$. An adaptive kernel width ($h_i$), which is a measure of the local surface density associated with each galaxy, is then used to account for the fact that a fixed kernel width would underestimate the
surface density in crowded regions while overestimating in sparsely populated areas. The adaptive kernel width is defined as $h_i = h \times \lambda_i$, where $\lambda_i$ is calculated as:
\begin{equation}
    \lambda_i=\sqrt{G/\hat{\Sigma}(r_i)},
\end{equation}
where G is the geometric mean of all the $\hat{\Sigma}(r_i)$ values. By using the adaptive kernel, the surface density field can now be calculated on a 2D grid, $r=(x,y)$ as:
\begin{equation}
 \Sigma(r)=\frac{1}{\sum^{N}_{i=1}w_i}\sum^{N}_{i=1}w_i K(r_i,r,h_i).
\end{equation}
The overdensity is then calculated as $\delta_{\rm OD}=\frac{\Sigma-\langle \Sigma\rangle}{\langle \Sigma \rangle}$ for all 
the grid points, as a measure of how high the surface density is over the background.\footnote{The overdensity 
can also be defined by using the median instead of the mean. Both definitions were tried and they gave virtually identical results.} The resulting overdensity field corresponding to the 
$z=6.05\pm 0.1$ redshift-bin is shown in Fig.~\ref{fig:adaptive kernel}a. Contours are in steps of $1\sigma$ and all grid points below $1\sigma$ have the same color. All galaxies inside a given 4$\sigma$ overdensity level contour are selected and investigated as potential protocluster member candidates. Here, $\sigma$ is defined as the standard deviation of the overdensity value for the entire field. 

To meaningfully investigate the properties of the protocluster candidates, we require that 
there are at least five or more galaxies inside a given 4$\sigma$ contour. If there are fewer galaxies than that, we do not include the overdensity as a protocluster candidate. This approach of identifying multiple (in our case $\geq 5$) galaxies with a high probability of being close together is supported
by simulations. Simulations show that the best indicator of whether a protocluster at $z > 4$ will end up as a massive cluster with a present-day virial mass of $M_{\rm vir}>10^{15}M_{\odot}$ is not an extreme star-formation rate or stellar mass \textbf{of the galaxies at $z>4$}, nor the presence of a massive bright central galaxy (BCG). Instead, the best indicator is \textbf{the galaxy number overdensity}
(\textbf{\cite{Chiang2013,Muldrew2015,Remus2022}}). This is because the number of galaxies and satellites associated with the protocluster traces the range of accretion of material from cosmic filaments.
\subsection{Weighted Voronoi Tessellation Estimator}\label{subsection:voronoi2D}
\begin{figure}
    \centering
    \includegraphics[width=0.5\textwidth]{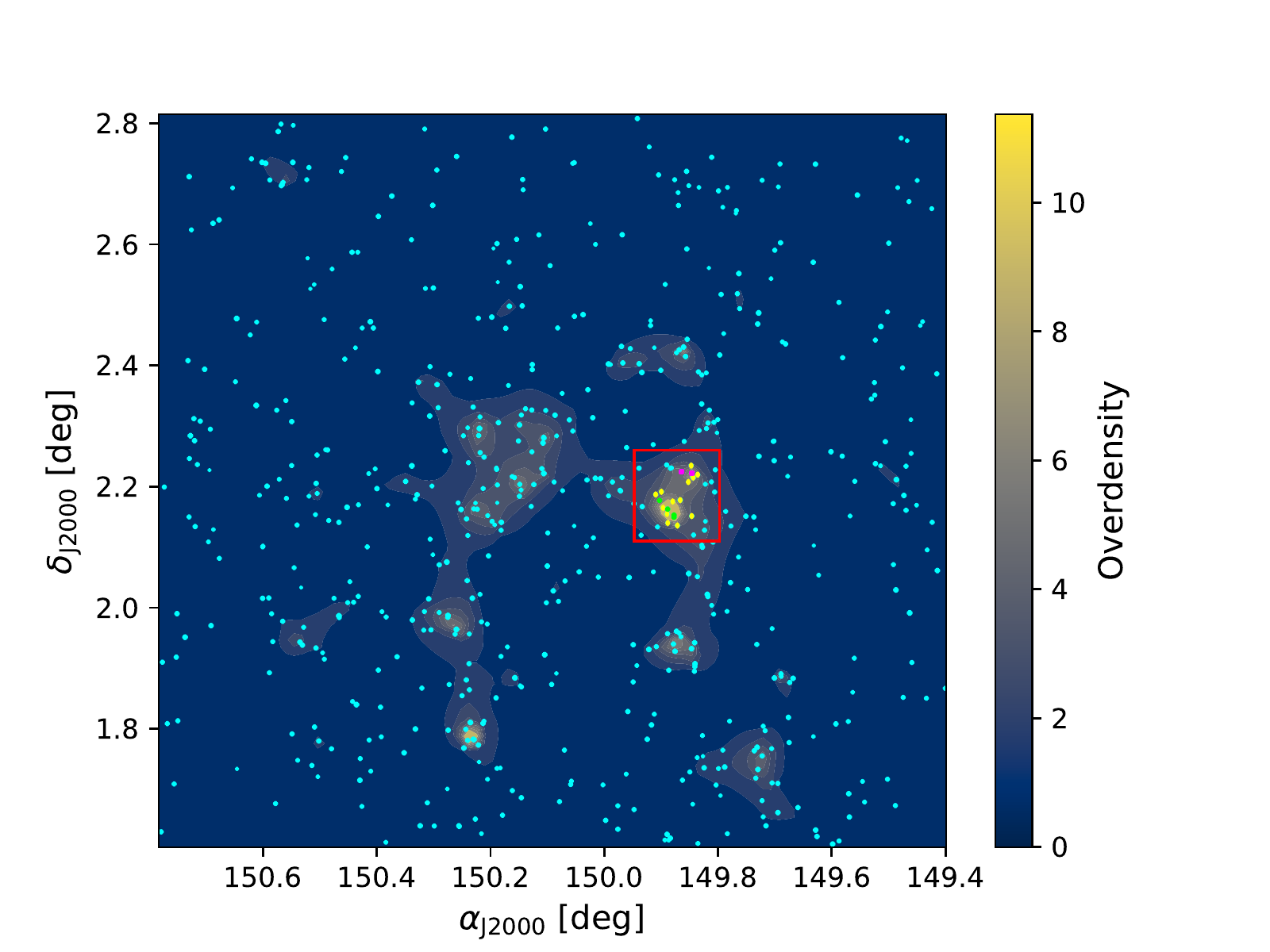}
    \caption{Voronoi tessellation of the galaxy overdensity field in the $z = 6.05\pm0.1$ redshift 
    bin in COSMOS. As in Fig.~\ref{fig:adaptive kernel}a, individual galaxies are shown as black dots, except for the 19 galaxies shown as colored dots, which make up COSMOS2020-PCz6.05-01 (highlighted by the same red rectangle as in Fig.~\ref{fig:adaptive kernel}a).}
    \label{fig:2DVoronoi}
\end{figure}

It is important when using kernel estimation to locate overdensities to check if the resulting galaxy overdensity map gives reasonable results.
The most straightforward approach would be to inspect the distribution of galaxies in and around the overdensities, as we expect a relatively high number of galaxies associated with an overdensity compared to the rest of the field. This approach is complicated by the fact that galaxies 
can have low weights \textbf{(i.e., a low probability of being in the redshift bin)} and therefore contribute little to the overdensity. Another way to gauge the robustness of our overdensity maps and to test for potential biases is to use an independent estimator and check if it recovers overdensities at the same locations.
We adopt the Weighted Voronoi Tessellation (WVT) Estimator as a \textbf{verification} of overdensities identified by the Weighted Adaptive Kernel method. Voronoi tessellation sections the field into regions, so-called Voronoi cells, so that each galaxy has an area associated with it \citep{Darvish2015,Shi2021}. The Voronoi cell of a galaxy is defined as all points in the redshift bin plane closer to that galaxy than any other galaxy. Consequently, in dense regions of the field, the Voronoi cells will be smaller relative to more sparsely populated regions. This means that the inverse of the area of a galaxy's Voronoi cell informs us about the local surface density at its location, which can be written as: 
\begin{equation}
    \Sigma(r_i)=\frac{1}{A_i},
\end{equation}
where $A_i$ is the Voronoi cell area associated with a given galaxy, $i$. The weights of each galaxy are taken into account by using a Monte-Carlo acceptance–rejection process that works as follows:
\begin{enumerate}
    \item We start by generating a random number, $R_i$, from a uniform distribution between the minimum and maximum weight values in a given redshift bin.
    \item We then check if $w_i>R_i$, and if that is the case, we use the galaxy in the density estimation.
    \item Using  Voronoi tessellation, we calculate the surface density only for those selected galaxies.
    \item We use Sibson natural neighbor interpolation to estimate the surface density for the grid points of a regularly spaced 2D grid. The result is a Monte Carlo estimate for the surface density field $\tilde{\Sigma}(r)$.
    \item We repeat this procedure $N$ times and take the mean of all the Monte-Carlo density fields as the actual density field $\Sigma(r)$:
    \begin{equation}
        \Sigma(r)=\frac{1}{N}\sum^{N}_{m=1}\tilde{\Sigma}_m(r).
    \end{equation}
\end{enumerate}
We choose to set $N=20$ to save on computational time. A trial run with $N=100$ performed on the $z=6.05$ bin yielded no significant differences in the overdensity map. The contours are chosen the same way as the Weighted Adaptive Kernel Estimator. The resulting overdensity field for the $z=6.05\pm 0.1$ redshift bin is shown in Fig. \ref{fig:2DVoronoi}. An advantage of Voronoi tessellation is its scale independence and its ability to span a wide range of physical lengths. In addition, it does not make any assumptions about the geometry and morphology of the structures in the density field. This means that it is not susceptible to the same bias as the Weighted Adaptive Kernel since we do not need to worry about the choice of global kernel width. We investigate the drawbacks of WVT and the differences between the two estimators in {\S}\ref{section:results}. \textbf{It is also possible to account for substructure within a redshift by constructing 3D (RA, Dec, $z$) overdensity maps in comoving space, though this requires sub-bin-width photo-$z$ uncertainties, which with our $1\sigma$ uncertainties on the order of the bin size is currently infeasible.}
\begin{figure}
    \centering
    \includegraphics[width=\linewidth]{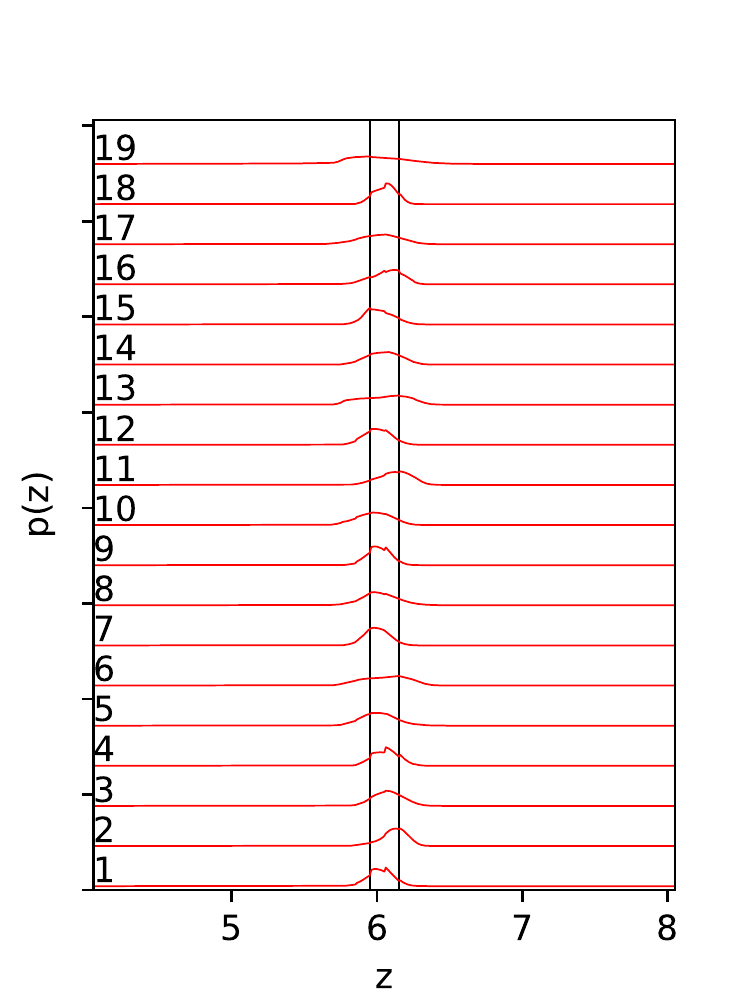}
    \caption{LePhare photometric redshift probability distributions ($p(z)$) for the 19 protocluster candidates of PCz6.05-01. The bottom $p(z)$ corresponds to the first entry in table \ref{tab:show19data}, the second lowest one to the second entry and so on. The vertical black bars correspond to the $z=6.05\pm0.1$ redshift bin.}
    \label{pcz6.05-01_pz}
\end{figure}
\subsection{Accounting for ultra-deep coverage at high-$z$}
Observing the distribution of galaxies and their resulting density field in bins at $z=6.4$ and above, we noticed that galaxies were predominantly located within vertical stripes in the field, with fewer galaxies found between them. The reason for this is that at $z \geq 6.4$ galaxies in COSMOS are mainly detected due to the UltraVISTA ultra-deep survey, which covers (part of) the COSMOS field in four separate vertical stripes \citep{McCracken2012, Weaver2021}. This will affect our overdensity estimates since the mean galaxy surface density is calculated from all the grid points in the field. Including grid points where we lack coverage would overestimate the overdensity since the mean surface density would be lower. It would also affect our $4\sigma$ selection criterion, since the standard deviation of the overdensity across the grid would be lower. To account for this and obtain a more representative overdensity estimate of the field, we only calculate the mean surface density and overdensity standard deviation within the stripes covered by Ultra-Deep for all $z \geq 6.4$ bins. \textbf{We tested if the ultra-deep coverage would affect our overdensity estimates at $z<6.4$ by using the method mention above and found the difference to be minimal, with 1-2 galaxies no longer part of the overdensities at the $4\sigma$ level and a slightly lower peak overdensity in the $z=6.05\pm0.1$ bin ($\delta_{\rm max}=8.0$ vs. $\delta_{\rm max}=9.2$ as seen with the first entry in table \ref{tab:fulldata}). We still argue that the overdensity at $z<6.4$ should be calculated using the area covered by UltraVISTA, since we observe clear overdensities that are not within the ultra-deep stripes (half of PCz6.05-01 is outside the stripes), as opposed to $z \geq 6.4$ where all the overdensities are concentrated on the stripes.} 

\section{Results \& Analysis}\label{section:results}
\subsection{Overdensity maps}
To present our findings, we initially highlight one of the overdensities we have located in the COSMOS field 
and then, subsequently, present our results for all the redshift bins considered.  In the galaxy overdensity map for the $z=6.05\pm 0.1$ 
bin derived from the WAK estimator, we 
identify a significant ($> 4\sigma$) overdensity ($\delta_{\rm OD}=9.2$) consisting 
of 19 galaxies at $z\simeq 6.0$ with 14 of them being $i$-band dropouts (see Fig.~\ref{fig:adaptive kernel}b). 
This overdensity spans a volume of roughly $10.7\,{\rm cMpc}\times 14.3\,{\rm cMpc}\times 66.3\,{\rm cMpc}$, corresponding to the maximal extent in the RA, Dec, and $z$ directions, respectively. We verify the
overdensity using the WVT estimator, see Fig.~\ref{fig:2DVoronoi}. Both estimators show a clear overdensity at the same location in the field, 
lending additional credence to the presence of a protocluster. We note that the Voronoi estimator of the overdensity is somewhat higher ($\delta_{\rm OD} \simeq 11$ at the peak) than the WAK estimator, which is more smoothed out at the center. 
This galaxy overdensity -- which we hereafter refer to as COSMOS2020-PCz6.05-01 -- was not identified before, as 
the majority of the sources are too faint in the NIR to have been included in the previous COSMOS catalog \citep{Laigle2016}. It has also not been identified in any other surveys (e.g., narrow-band 
searches for Lyman-$\alpha$ emitters at $z\sim 6$). The properties of the 19 galaxies are given in Table \ref{tab:show19data}.

Having shown our approach to locating and analyzing overdensities in the COSMOS field at $z=6$, we now extend our search of protocluster candidates to higher redshifts. To search for protocluster candidates between $z=6-10$, we repeat our overdensity analysis for redshift bins corresponding to $80\,{\rm cMpc}$ in size. The number of galaxies in the higher redshift bins is of course lower. In the $z=6.05\pm 0.1$ bin, there is $\approx 600$ objects after applying all our selection criteria, but applying the same criteria to the higher redshift bins leaves us with $\approx 20-230$ galaxies.
We show the overdensity maps for all bins using both the WAK and WVT estimators in Fig.~\ref{fig:WAVT} in Appendix \ref{sec:maps}, highlighting the $4\sigma$ contours that fulfil our $\geq5$ galaxies selection criterion. In total, we find 15 overdensities at $\geq 4\sigma$ across the redshift range $z=6.0-7.7$ using the WAK estimator. At $z >7.7$, we do not identify any galaxy overdensities at a significance of $\geq 4\sigma$.
We note that, in some cases (Fig.~\ref{fig:WAVT}k-m), overdensities are found at the same location in the field across adjacent redshift bins, suggestive of either extended structures \textbf{or possible overdensities near the edge of a bin that are smeared out due to the photo-$z$ uncertainty.}

From Fig.~\ref{fig:WAVT}a-z, it is seen that the two overdensity estimators are in excellent agreement and capture the same galaxy overdensities in the COSMOS field in nearly all of the redshift bins investigated. 
One aspect of using WVT that is clear from the maps is that it struggles to capture overdensities close to the edge of the field 
(see Fig.~\ref{fig:WAVT}m-n and o-p). 
This is due to the way the areas of the Voronoi cells are calculated. The cells closest to the edge of the field will have one side facing the edge, which means these cells will be open facing polygons, and we cannot calculate their area. \textbf{The same problem does not exist with the WAK, since it calculates the surface density by essentially using Gaussians, which can be arbitrarily close to the edge.}
We could alleviate this by defining boundary points around the edge of the field\textbf{, which would allow us to then calculate the area of the cells spanned between the points in the field and the boundary points}, but it would not capture the true distribution of galaxies outside the field and give a false impression of the overdensity near the edge. \textbf{Not knowing the true distribution of galaxies outside the field affects both the WAK and WVT kernel estimators, but the effect inside the field close to the edge is different.}
In one instance, an overdensity is located at the same position using both estimators, but the $4\sigma$ criterion is not met by one of the estimators. This is seen in Fig.~\ref{fig:WAVT}o-p,
where a $4\sigma$ is present using WAK but not present using WVT.
We have not discarded this overdensity from our selection but merely caution that since no independent ($4\sigma$ level) verification has been made, it could be the effect of using a biased kernel. \textbf{It is also worth noting that the $z=7.69\pm0.14$ bin only has 26 galaxies in total, meaning the ability to discern what is an galaxy in an overdensity and what is a field galaxy becomes difficult.}

\begin{table*}
\caption{Properties of the 19 galaxies in COSMOS2020-PCz6.05-01 (Fig.\,\ref{fig:adaptive kernel}) from LePhare and EAZY.
The galaxies with a * indicate they are $i$-dropouts according to the criteria in \cite{Ono2018}.}
\label{tab:show19data}
\centering
\begin{tabular}{l l l l l l l l l}
\hline
\hline
\# & ID & RA      & Dec     & $z_{\rm Lephare}$ & $z_{\rm EAZY}$ & $\log(M_{\rm \star, LePhare}/M_{\odot})$ & ${\rm SFR}_{\rm UV}$ & $K_{\rm S}$ \\ 
   &    & {\it hh:mm:ss.ss} & {\it dd:mm:ss.ss} &                   &                &                                                ${\rm dex}$        &                        ${\rm [M_{\odot}\,yr^{-1}]}$                          &     ${\rm [AB\,mag]}$                         \\ 
\hline
1* & 127337 & 09:59:22.49 & 02:12:53.54 & 6.02$^{+0.09}_{-0.08}$ & 6.06$^{+0.05}_{-0.05}$ & 9.4$^{+0.1}_{-0.2}$ & 48$^{+16}_{-15}$ & 26.57\\
2* & 142959 & 09:59:20.56 & 02:13:10.62 & 6.12$^{+0.08}_{-0.10}$ & 6.12$^{+0.06}_{-0.05}$ & 9.1$^{+0.1}_{-0.2}$ & 9$^{+1}_{-1}$ & 26.17\\
3 & 156780 & 09:59:29.14 & 02:08:08.38 & 6.08$^{+0.10}_{-0.10}$ & 6.08$^{+0.05}_{-0.04}$ & 9.3$^{+0.1}_{-0.2}$ & 13$^{+1}_{-1}$ & 26.68\\
4 & 187778 & 09:59:23.08 & 02:13:20.10 & 6.06$^{+0.09}_{-0.09}$ & 6.09$^{+0.04}_{-0.04}$ & 8.7$^{+0.2}_{-0.2}$ & 5$^{+1}_{-1}$ & 26.28\\
5 & 220530 & 09:59:27.36 & 02:13:27.56 & 6.01$^{+0.12}_{-0.11}$ & 6.03$^{+0.05}_{-0.06}$ & 8.2$^{+0.3}_{-0.2}$ & 2$^{+1}_{-1}$ & 28.08\\
6* & 225263 & 09:59:33.21 & 02:08:23.24 & 6.07$^{+0.14}_{-0.18}$ & 5.91$^{+0.05}_{-0.06}$ & 9.4$^{+0.2}_{-0.2}$ & 32$^{+6}_{-7}$ & 25.47\\
7* & 361608 & 09:59:30.58 & 02:08:57.78 & 6.00$^{+0.09}_{-0.08}$ & 5.99$^{+0.05}_{-0.04}$ & 10.4$^{+0.1}_{-0.1}$ & 15$^{+1}_{-1}$ & 25.46\\
8 & 369661 & 09:59:23.33 & 02:14:04.42 & 6.01$^{+0.13}_{-0.11}$ & 6.05$^{+0.06}_{-0.06}$ & 9.4$^{+0.2}_{-0.2}$ & 41$^{+4}_{-6}$ & 27.05\\
9* & 413243 & 09:59:23.09 & 02:09:04.73 & 6.01$^{+0.08}_{-0.08}$ & 5.99$^{+0.05}_{-0.05}$ & 9.8$^{+0.1}_{-0.2}$ & 44$^{+44}_{-24}$ & 25.80\\
10* & 441761 & 09:59:35.16 & 02:09:53.31 & 5.98$^{+0.12}_{-0.12}$ & 5.90$^{+0.05}_{-0.06}$ & 10.0$^{+0.1}_{-0.2}$ & 125$^{+41}_{-26}$ & 25.43\\
11* & 444487 & 09:59:30.56 & 02:09:10.48 & 6.13$^{+0.11}_{-0.12}$ & 6.08$^{+0.05}_{-0.05}$ & 10.0$^{+0.1}_{-0.1}$ & 16$^{+1}_{-1}$ & 25.89\\
12* & 482804 & 09:59:33.50 & 02:09:17.17 & 6.00$^{+0.09}_{-0.09}$ & 6.00$^{+0.06}_{-0.06}$ & 9.3$^{+0.2}_{-0.1}$ & 10$^{+1}_{-1}$ & 25.32\\
13* & 573604 & 09:59:36.65 & 02:10:37.28 & 6.06$^{+0.15}_{-0.20}$ & 5.94$^{+0.20}_{-0.20}$ & 10.1$^{+0.1}_{-0.1}$ & 9$^{+1}_{-1}$ & 26.57\\
14* & 582186 & 09:59:33.25 & 02:09:45.81 & 6.05$^{+0.11}_{-0.12}$ & 6.10$^{+0.07}_{-0.07}$ & 10.2$^{+0.1}_{-0.1}$ & 117$^{+27}_{-21}$ & 25.88\\
15* & 694706 & 09:59:38.29 & 02:11:12.90 & 6.00$^{+0.11}_{-0.09}$ & 5.98$^{+0.05}_{-0.04}$ & 9.8$^{+0.2}_{-0.2}$ & 73$^{+61}_{-37}$ & 25.83\\
16* & 742465 & 09:59:35.88 & 02:11:29.06 & 6.08$^{+0.10}_{-0.13}$ & 6.05$^{+0.05}_{-0.05}$ & 9.6$^{+0.1}_{-0.2}$ & 24$^{+1}_{-1}$ & 25.08\\
17* & 759747 & 09:59:31.11 & 02:10:31.32 & 5.99$^{+0.15}_{-0.22}$ & 6.04$^{+0.06}_{-0.07}$ & 9.2$^{+0.2}_{-0.2}$ & 11$^{+1}_{-1}$ & 25.97\\
18* & 783817 & 09:59:27.96 & 02:10:39.35 & 6.06$^{+0.07}_{-0.09}$ & 6.03$^{+0.07}_{-0.06}$ & 9.8$^{+0.1}_{-0.1}$ & 59$^{+25}_{-15}$ & 25.48\\
19 & 958367 & 09:59:24.54 & 02:12:27.15 & 5.97$^{+0.23}_{-0.19}$ & 5.96$^{+0.16}_{-0.09}$ & 9.6$^{+0.2}_{-0.2}$ & 39$^{+4}_{-4}$ & 27.17\\
\hline
\end{tabular}
\end{table*}

\subsection{Galaxy overdensity properties}
In this section, we examine the properties of the galaxies
associated with the overdensities identified in the previous section. We use the physical properties (e.g., stellar masses and star-formation rate) as described in Section \ref{subsection:physical-parameters}. As in the previous section, we first present our results for COSMOS2020-PCz6.05-01, followed by a summary of our findings for the remaining overdensities at higher redshifts. 

\begin{figure}
    \centering
    \includegraphics[width=\linewidth]{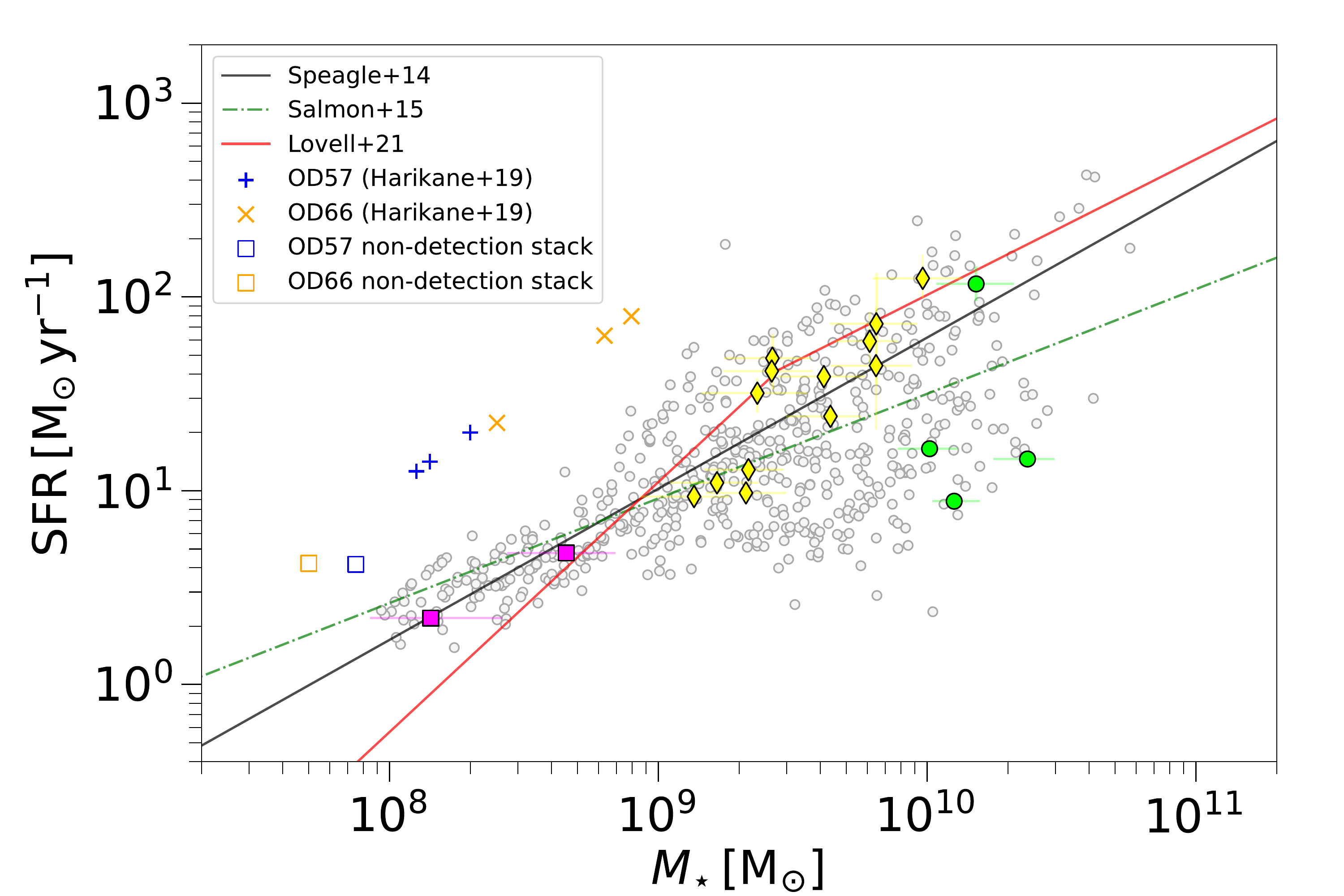}
    \caption{SFR vs $M_{\rm \star}$ for all $z_{\rm phot} = 6.05\pm 0.10$ galaxies in COSMOS2020 (grey circles) and for the 19 galaxies in COSMOS2020-PCz6.05-01 (shown as filled squares, diamonds, and circles -- color coded according to their mass as in Fig.~\ref{fig:adaptive kernel}). The galaxies follow best estimates of the galaxy main sequence at $z\simeq 6$ \citep{Speagle2014, Salmon2015, Lovell2021}. For comparison, the $z=5.7$ and $z=6.6$ spectroscopically confirmed LAE-overdensities reported by \citep{Harikane2019} are also shown (shown as blue $+$ and yellow $\times$ symbols, respectively), along with the stacked LAE non-detection subsamples of OD57 and OD66 (blue and yellow open squares, respectively). The LAEs have low stellar masses and lie $\sim 5\times$ above the main sequence, suggesting that they probe a different protocluster galaxy population than those in COSMOS2020-PCz6.05-01.}
    \label{fig:mainsequence}
\end{figure}

A $JHK_{\rm S}$ true-color image of a $13\arcmin \times 18\arcmin$ region centered on COSMOS2020-PCz6.05-01 is shown in Fig.~\ref{fig:adaptive kernel}b, 
with the 19 galaxies that make up the overdensity highlighted. 
The galaxies span a stellar-mass range of 
$M_{\rm \star} \simeq 10^8 -2.4\times 10^{10}\,{\rm M_{\rm \odot}}$ and a range in star-formation 
rate of ${\rm SFR} \simeq 2-125\,{\rm M_{\rm \odot}\,yr^{-1}}$.
The photo-$z$ probability distribution, $p(z)$, of the 19 protocluster galaxies is shown in Fig.~\ref{pcz6.05-01_pz}. 
Most of the galaxies in COSMOS202-PCz6.05-01 are consistent with being normal star-forming galaxies and agree with estimates of the galaxy main sequence at $z\simeq 6$ as seen in Fig.~\ref{fig:mainsequence} (note also the different main sequences, e.g., \citet{Speagle2014, Salmon2015, Lovell2021}). 

\textbf{For comparison,}
the galaxies residing in one of the most distant spectroscopically confirmed protocluster at $z=6.6$ (hereafter denoted OD66 \citep{Harikane2019}), lie $\sim 5\times$ above the galaxy main sequence at this epoch (Fig.~\ref{fig:mainsequence}). This may be explained by the fact that OD66 was selected as an LAE overdensity, which is biased towards dust-poor young, actively star-forming galaxies. 
We note that the $i'$-band dropouts in the $z=6.01$ protocluster discovered by \citet{Toshikawa2012,Toshikawa2014}, the protocluster at $z=6.54$ from \citet{Chanchaiworawit2019} and the one at $z=6.9$ from \citet{Hu2021} do not have reliable stellar mass estimates. Narrow-band Ly$\alpha$ selected sources are biased towards unobscured star-formation activity, and (sub)-mm selected sources are biased towards obscured star-formation activity (Fig.~\ref{fig:SFRoverz}). 
COSMOS2020-PCz6.05-01 is NIR-selected (specifically the $izYJHK_{\rm S}$ bands) and thus biased towards blue stars and unobscured star-formation in a similar manner to the LAEs. For a comparison with a different selection method, most galaxies in COSMOS2020-PCz6.05-01 are $\gtrsim\,10\times$ more massive than those in OD66 (Fig.~\ref{fig:mainsequence}). 
The total stellar mass and star-formation rate of the 19 galaxies in COSMOS2020-PCz6.05-01 is $\sim 1\times 10^{11}\,{\rm M_{\odot}}$ and $\sim 700\,{\rm M_{\odot}\,yr^{-1}}$, respectively.  
The total star-formation rate is, therefore, $\sim 6\times$ higher than that of OD66 (Fig.~\ref{fig:SFRoverz}). 
\begin{table*}
\caption{Properties of the overdensities found at $z=6-10$. This includes (1) overdensity name, (2) number of galaxies inside the $4\sigma$ overdensity contour, \textbf{with the parenthesis showing the number classified as LBGs} (3) mean redshift of overdensity members, (4,5) RA and Dec.~of the peak overdensity value inside the $4\sigma$ contour, (6) peak overdensity value inside the 4$\sigma$ contour with the upper lower $1\sigma$(Corresponding to Gaussian statistics $1\sigma=0.8413$) Poisson uncertainties \textbf{\citep{Gehrels1986}}, (7) total stellar mass of all overdensity members, (8) total SFR of all overdensity members}
\label{tab:fulldata}
\centering
\hspace{-2.0 cm}
\begin{tabular}{l l l l l l l l}
\hline
\hline
\multicolumn{1}{l}{Name} & \#$_{\rm gal}$ & $z_{\rm mean}$ & \multicolumn{1}{l}{RA$_{\rm max}$} & \multicolumn{1}{l}{Dec$_{\rm max}$}  & $\delta_{\rm max}$ & $\log(M_{\rm \star, tot}/M_{\odot})$ & ${\rm SFR}_{\rm UV,tot}$\\ 
     &                &                & \multicolumn{1}{l}{{\it hh:mm:ss.ss}}        & \multicolumn{1}{l}{{\it dd:mm:ss.ss}}           &                    &     \multicolumn{1}{c}{${\rm dex}$}                                              &   \multicolumn{1}{l}{${\rm [M_{\odot}\,yr^{-1}]}$}\\ 
\multicolumn{1}{l}{(1)} & \multicolumn{1}{l}{(2)} & \multicolumn{1}{l}{(3)} & \multicolumn{1}{l}{(4)} & \multicolumn{1}{l}{(5)} & \multicolumn{1}{l}{(6)} & \multicolumn{1}{l}{(7)} & \multicolumn{1}{l}{(8)}\\\hline
PCz6.05-01 & 19(13) & 6.04 & 09:59:32.61 & 02:09:28.29 & 9.2$^{+3.9}_{-3.1}$ & 11.06$^{+0.12}_{-0.14}$ & 692$^{+234}_{-162}$\\
PCz6.05-02 & 8(7) & 6.03 & 09:59:27.60 & 01:56:19.92 & 6.1$^{+2.2}_{-3.3}$ & 10.62$^{+0.12}_{-0.13}$ & 240$^{+78}_{-55}$\\
PCz6.05-03 & 6(3) & 6.02 & 09:59:25.93 & 02:25:09.96 & 4.7$^{+3.6}_{-1.9}$ & 10.83$^{+0.08}_{-0.09}$ & 298$^{+124}_{-84}$\\
PCz6.05-05 & 5(2) & 6.03 & 10:00:36.06 & 02:12:23.49 & 4.7$^{+3.6}_{-1.9}$ & 10.61$^{+0.13}_{-0.16}$ & 301$^{+24}_{-23}$\\
PCz6.05-06 & 8(5) & 6.02 & 10:00:56.10 & 01:47:12.44 & 6.8$^{+4.0}_{-2.4}$ & 10.39$^{+0.13}_{-0.14}$ & 82$^{+21}_{-15}$\\
PCz6.05-08 & 5(3) & 6.05 & 10:01:04.45 & 01:58:31.32 & 5.0$^{+3.3}_{-2.2}$ & 10.62$^{+0.12}_{-0.13}$ & 208$^{+66}_{-47}$\\
PCz6.69-02 & 5(2) & 6.76 & 10:00:42.74 & 02:02:54.11 & 8.5$^{+4.6}_{-2.4}$ & 10.69$^{+0.13}_{-0.12}$ & 570$^{+127}_{-113}$\\
PCz6.92-01 & 6(0) & 6.93 & 09:57:47.41 & 01:49:45.74 & 11.1$^{+4.3}_{-3.4}$ & 9.66$^{+0.20}_{-0.20}$ & 120$^{+50}_{-36}$\\
PCz6.92-04 & 5(2) & 6.89 & 10:00:32.72 & 02:10:33.99 & 6.6$^{+4.2}_{-2.2}$ & 9.94$^{+0.18}_{-0.18}$ & 93$^{+11}_{-12}$\\
PCz6.92-05 & 7(3) & 6.96 & 10:02:06.23 & 02:34:39.34 & 13.0$^{+4.7}_{-3.6}$ & 10.74$^{+0.12}_{-0.13}$ & 405$^{+187}_{-123}$\\
PCz7.17-01 & 5(0) & 7.20 & 09:57:44.07 & 02:03:16.01 & 5.4$^{+2.9}_{-2.6}$ & 10.62$^{+0.18}_{-0.19}$ & 306$^{+30}_{-29}$\\
PCz7.17-03 & 10(0) & 7.16 & 09:58:52.53 & 02:42:19.22 & 8.5$^{+4.6}_{-2.4}$ & 9.97$^{+0.12}_{-0.12}$ & 327$^{+37}_{-31}$\\
PCz7.17-04 & 5(0) & 7.25 & 10:00:51.09 & 02:24:26.16 & 5.3$^{+3.0}_{-2.5}$ & 9.66$^{+0.20}_{-0.22}$ & 57$^{+23}_{-16}$\\
PCz7.42-01 & 8(0) & 7.39 & 09:57:54.09 & 02:38:18.33 & 5.8$^{+2.5}_{-3.0}$ & 9.72$^{+0.30}_{-0.32}$ & 37$^{+5}_{-5}$\\
PCz7.69-01 & 5(0) & 7.69 & 09:57:49.08 & 02:26:59.45 & 6.6$^{+4.2}_{-2.2}$ & 10.66$^{+0.18}_{-0.17}$ & 71$^{+24}_{-6}$\\
\hline
\end{tabular}
\end{table*}
Table \ref{tab:fulldata} shows the global properties of all the overdensities identified in the COSMOS field that meet our selection criteria.
In Appendix \ref{sec:allOD}, we highlight the individual galaxies within each overdensity $4\sigma$-contour overlaid on a $JHK_{\rm S}$-band true-color image of the region they occupy. We also show the location of the galaxies in the ${\rm SFR}-M_{\rm \star}$ plane,  their stellar mass Cumulative Distribution Function, and the $p(z)$ of the galaxies.  
Of the 15 overdensities identified, 9 have either 5 or 6 galaxies inside their $4\sigma$ contour, meaning they are just above our selection criterion of $\geq5$ galaxies. The maximum overdensity values vary significantly from 4.7 for COSMOS2020-PCz6.05-05 to 13.0 for COSMOS2020-PCz6.92-05. The total stellar masses and star-formation rates of the overdensties also vary significantly, spanning a range of $M_{\rm \star, tot} = 10^{9.7-11.1}\,{\rm M_{\odot}}$ and ${\rm SFR_{\rm UV, tot}} = 37-692\,{\rm M_{\odot}yr^{-1}}$, respectively.

\begin{figure*}
\centering
  \includegraphics[width=0.329\textwidth]{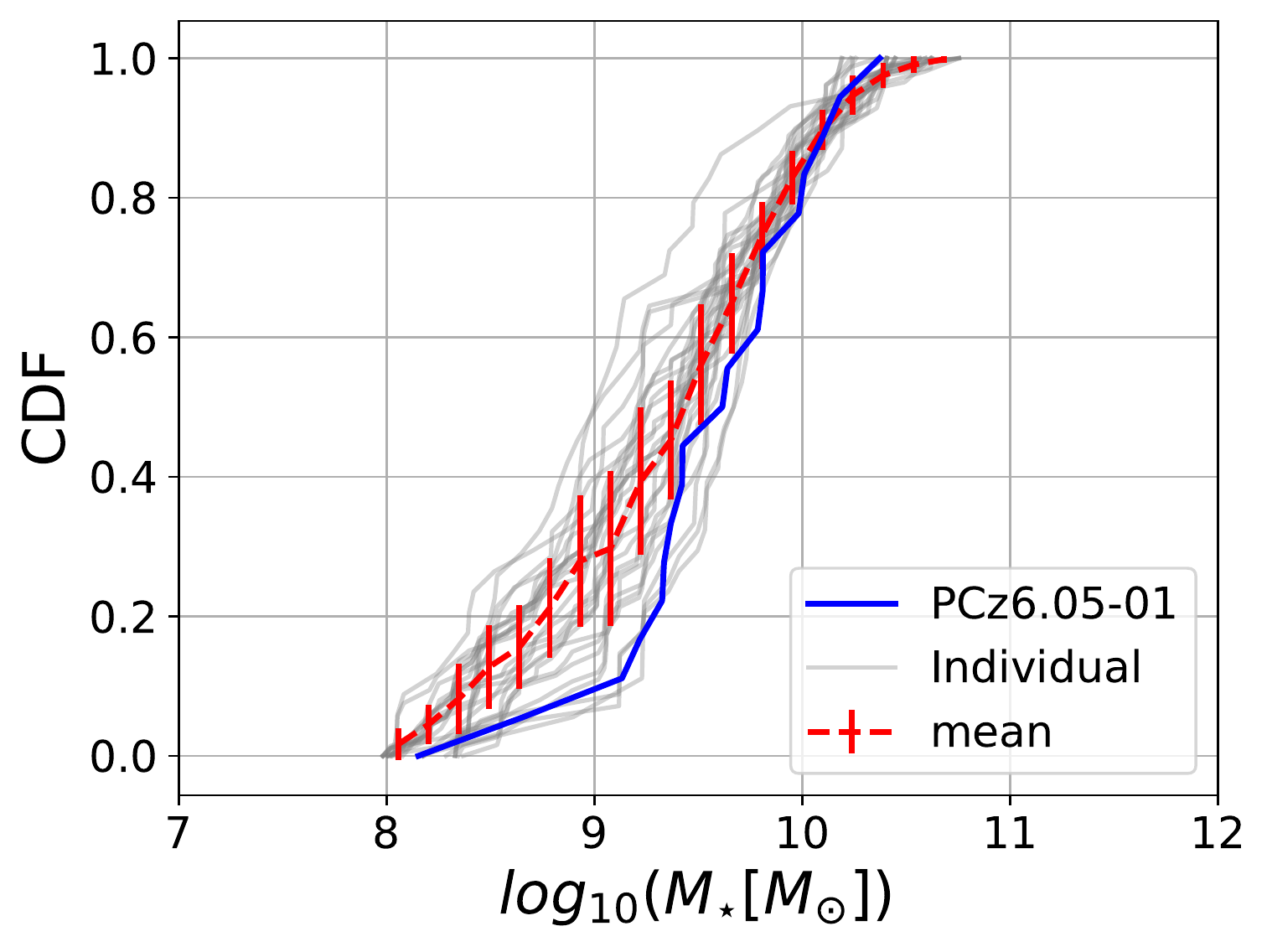}
  \includegraphics[width=0.329\textwidth]{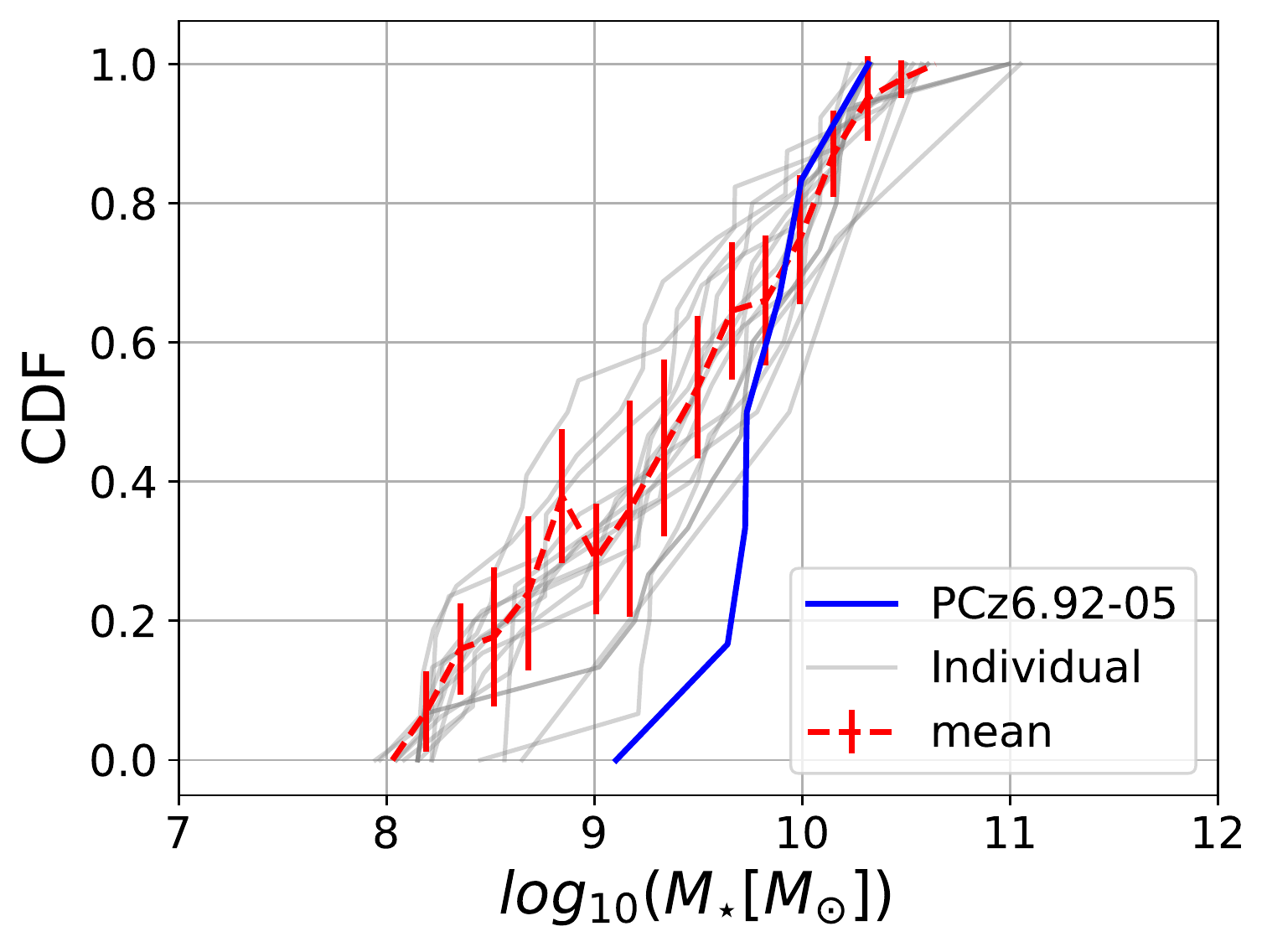}
  \includegraphics[width=0.329\textwidth]{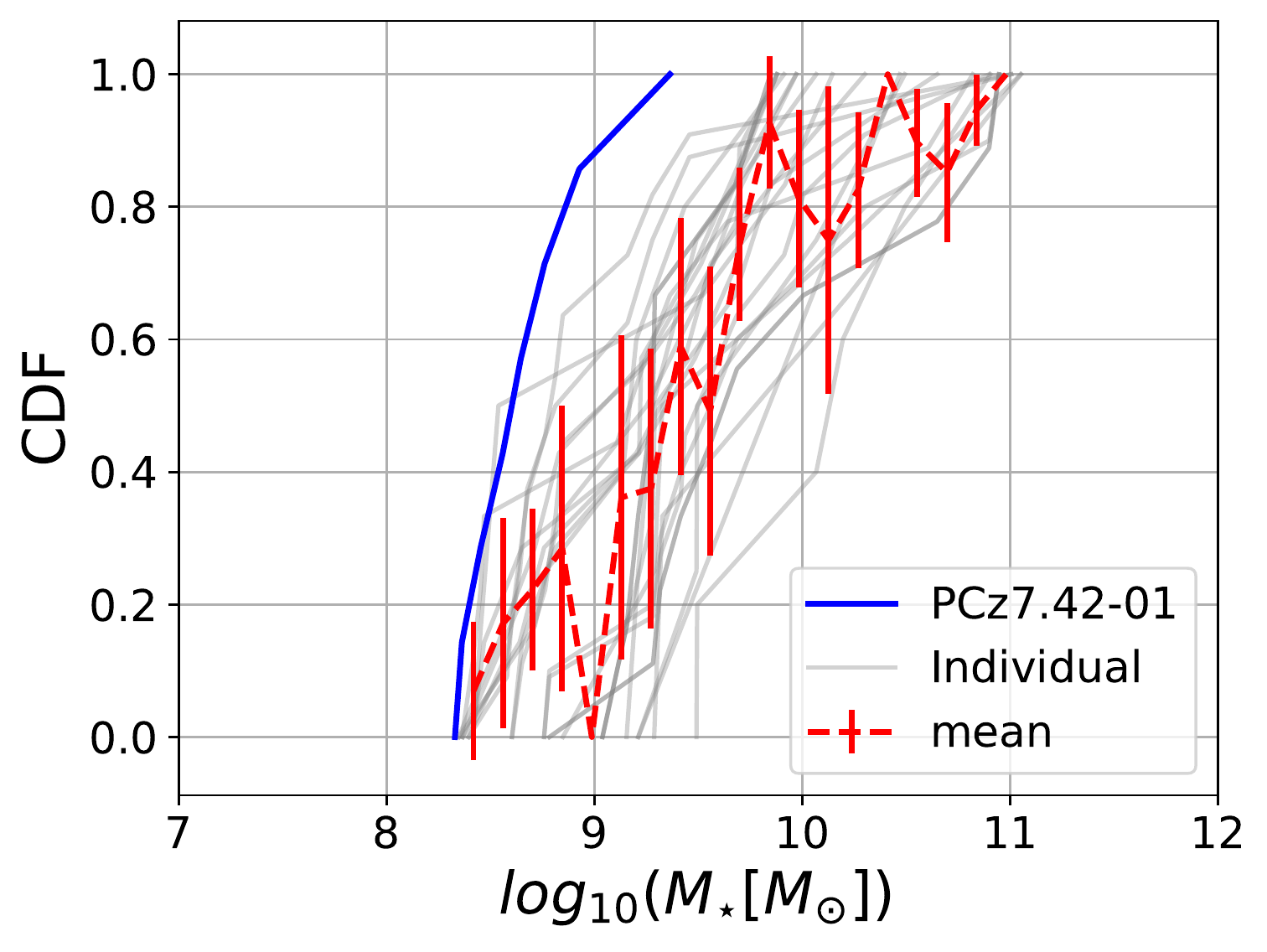}
  \includegraphics[width=0.329\textwidth]{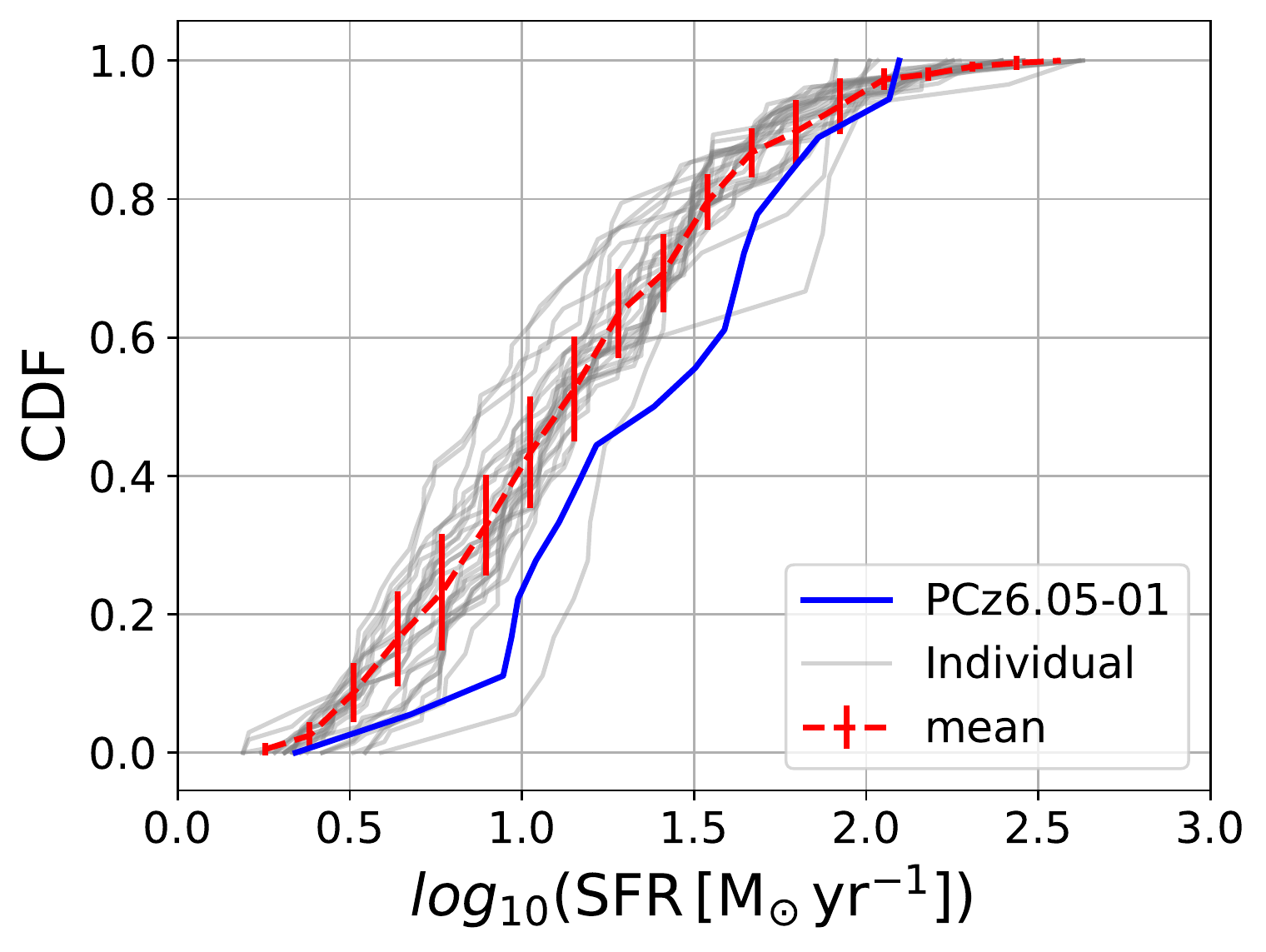}
  \includegraphics[width=0.329\textwidth]{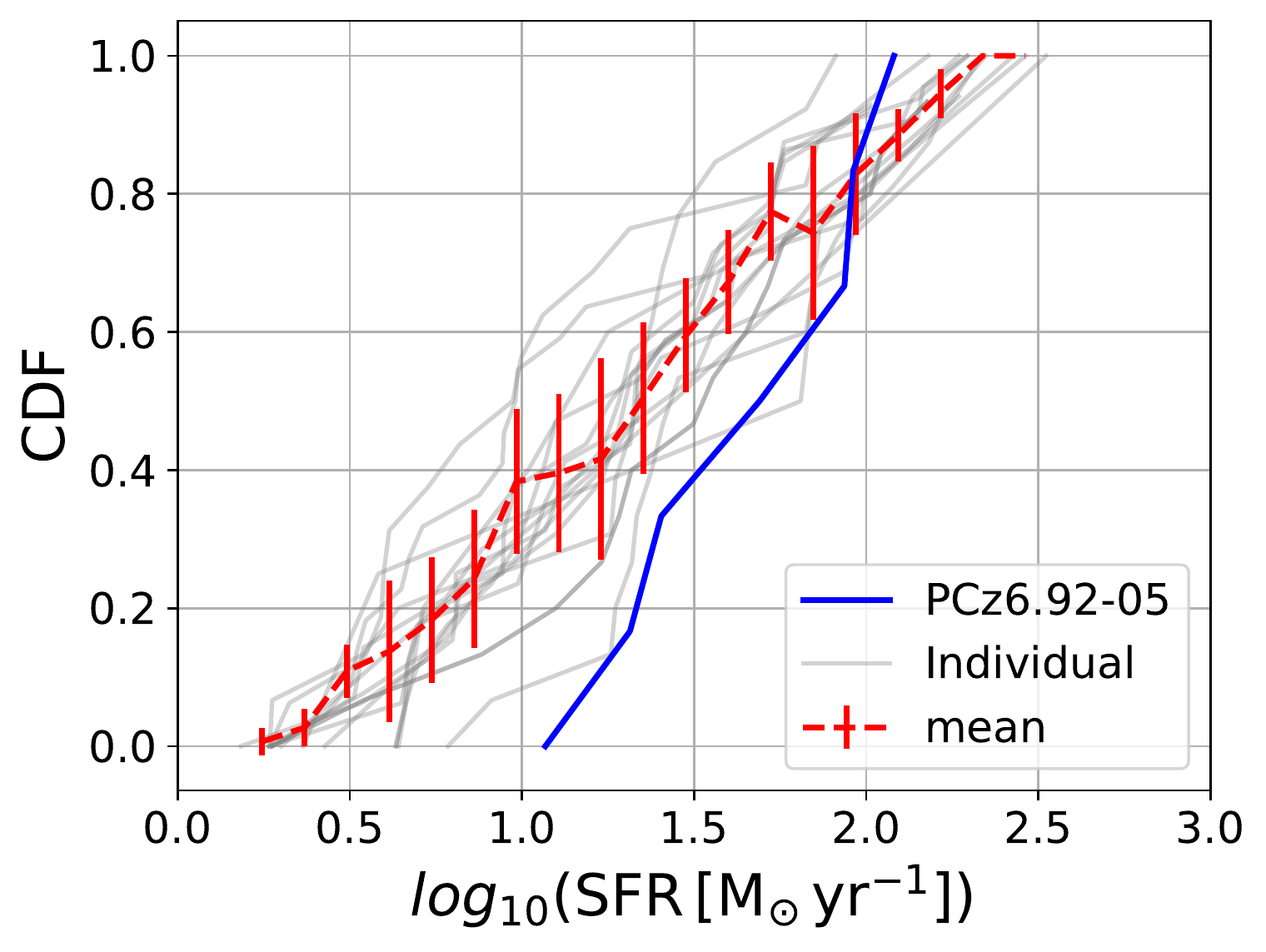}
  \includegraphics[width=0.329\textwidth]{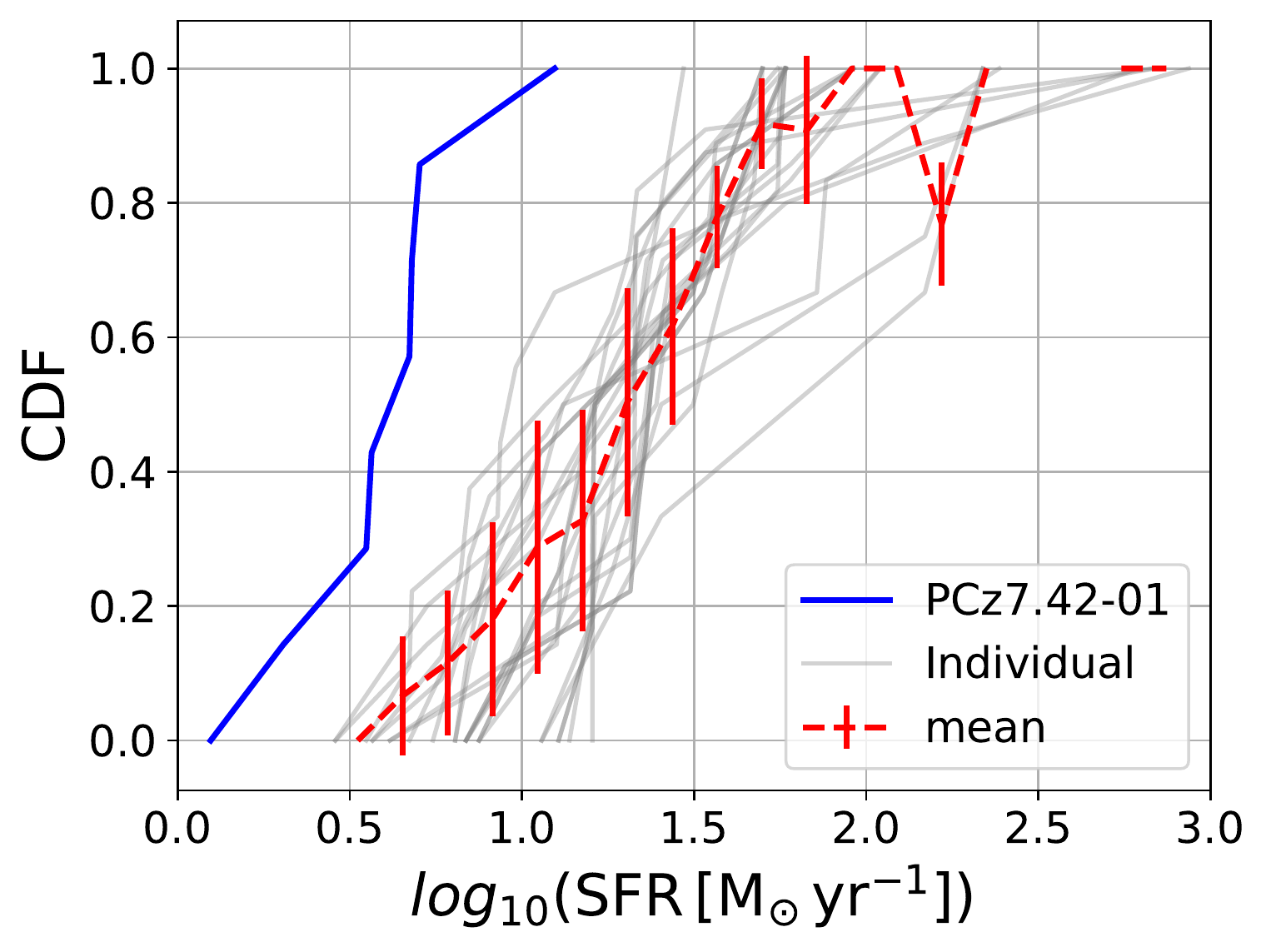}
  \caption{Stellar mass and SFR cumulative distribution functions for the three overdensities COSMOS2020-PCz6.05-01, COSMOS2020-PCz6.92-05, and COSMOS2020-PCz7.42-01, compared with field galaxies of similar mass in the same $z$-bin. The individual CDFs for the field galaxies are shown in gray, whereas the mean CDF is shown in red with associated errorbars.}
  \label{fig:MstarSFRCDF}
\end{figure*}

Multiple protocluster candidates appear to have one massive galaxy surrounded by less massive ones, most obviously seen in COSMOS2020-PCz6.05-01, COSMOS2020-PCz6.92-05, and COSMOS2020-PCz7.42-01. To test that this is not simply an effect of the fact that the number density of less massive galaxies is higher than that of massive galaxies, we examine whether the galaxies in the 
overdensities are different from the galaxies in the field. Specifically, we compute the Cumulative Distribution Functions (CDF) for the stellar mass and SFR of these three overdensities and compare them to the environments of field galaxies with a central galaxy of similar stellar mass as the most massive galaxy in each overdensity (within $\pm0.2\,{\rm dex}$). For each overdensity, we do this by defining a circular region that has the same area as the 
$4\sigma$ contour of the overdensity in a given $z$-bin and selecting all galaxies within that region to determine the CDFs. We only target field galaxies, that is we mask out all galaxies inside $4\sigma$ overdensity contours. Fig.\,\ref{fig:MstarSFRCDF} shows the CDFs for the three overdensities, as well as the individual and mean CDFs of the field galaxies. We see that both COSMOS2020-PCz6.05-01 and COSMOS2020-PCz6.92-05 skew towards more massive and more starforming galaxies than the field. The CDFs for COSMOS2020-PCz7.42-01, however, are noticeably shifted towards lower masses and lower SFR than the field. While there is difference between how the individual overdensities skew, it is clear that they appear different than the majority of environments with central galaxies of similar mass in the field. 

The most massive galaxies in our overdensities could be the \textbf{progenitor} Bright Cluster Galaxies (BCG) of the protoclusters. There is evidence to suggest that there is a downsizing effect for galaxy clusters where, on average, the BCGs (cores) for the most massive (proto-)clusters assemble at earlier times than the ones at lower masses \citep{Rennehan2020}. BCGs are formed in high galaxy density protocluster cores and among our candidates, COSMOS2020-PCz6.05-01 seem to fit that description best, {\bf with its most massive galaxy appearing to lie below be the main sequence and in a central position in the cluster (see Figs.\,\ref{fig:adaptive kernel} and  \ref{fig:mainsequence}, and Table \ref{tab:show19data})}. This suggests a highly evolved (proto-)BCG has formed already at $z\simeq6$, pushing back the evolution of BCGs further than previously expected \citep{Ito2019,Rennehan2020}. \textbf{Keep in mind that that there may exist multiple possible pathways for the formation of a BCG, as seen by protoclusters dominated by a single massive object like many (sub-)millimeter-selected sources \citep{Wang2020}, or a closely clustered association of objects that will rapidly merge to form a BCG (e.g., what is suggested in \cite{Kubo2021}). There are also protoclusters without any evidence for a massive (proto-)BCG in them (e.g., \cite{Toshikawa2016,Harikane2019}).}
\subsubsection{The 3D distributions of protocluster galaxies} 
Looking at the 3D distribution (RA,Dec,$z$) of the protocluster candidates (tables \ref{tab:PCz6.05-01data}-\ref{tab:PCz7.69-01data}), almost all of them appear to be elongated structures, with their extend in RA and Dec.~being about an order of magnitude smaller than their extend in the $z$-direction. The extend in the $z$-direction is, in most cases (the exceptions being COSMOS2020-PCz6.05-03 and COSMOS2020-PCz7.17-03), not near $80\,{\rm cMpc}$, so this is not a problem related to the choice of bin size for the overdensity maps. This could be expected, since we are not working with a 3D overdensity map and can therefore not distinguish the position of galaxies inside the bin, only by proxy through their $p(z)$ weighting. The explanation is that this is due to a combination of two factors. One is that we have chosen to only look at galaxies inside a $4\sigma$ contour and therefore only get the central galaxies of the overdensity. Suppose we relaxed the restriction to $3\sigma$ or $2\sigma$, the extent in RA and Dec would increase. The other factor is the uncertainty associated with the $p(z)$ of the galaxies can make their distribution in space to be more elongated that they actually are. Keeping these two factors in mind, we may be seeing the infall of galaxies from the cosmic web, which would appear as an elongated structure inside our bin. This should be more prevalent as we go to higher redshift since the galaxies have had less time to move in from the cosmic web and coalesce into a protocluster structure. We do not expect this elongation to only be present in the $z$-direction, but since our $4\sigma$ contour selection limits the extent of the structure in RA and Dec, and there is an uncertainty associated with the $p(z)$ of the galaxies, we will only see the elongation in the $z$-direction.     
\subsubsection{Stellar mass cumulative distribution functions}
\begin{figure}
    \centering
    \includegraphics[width=0.5\textwidth]{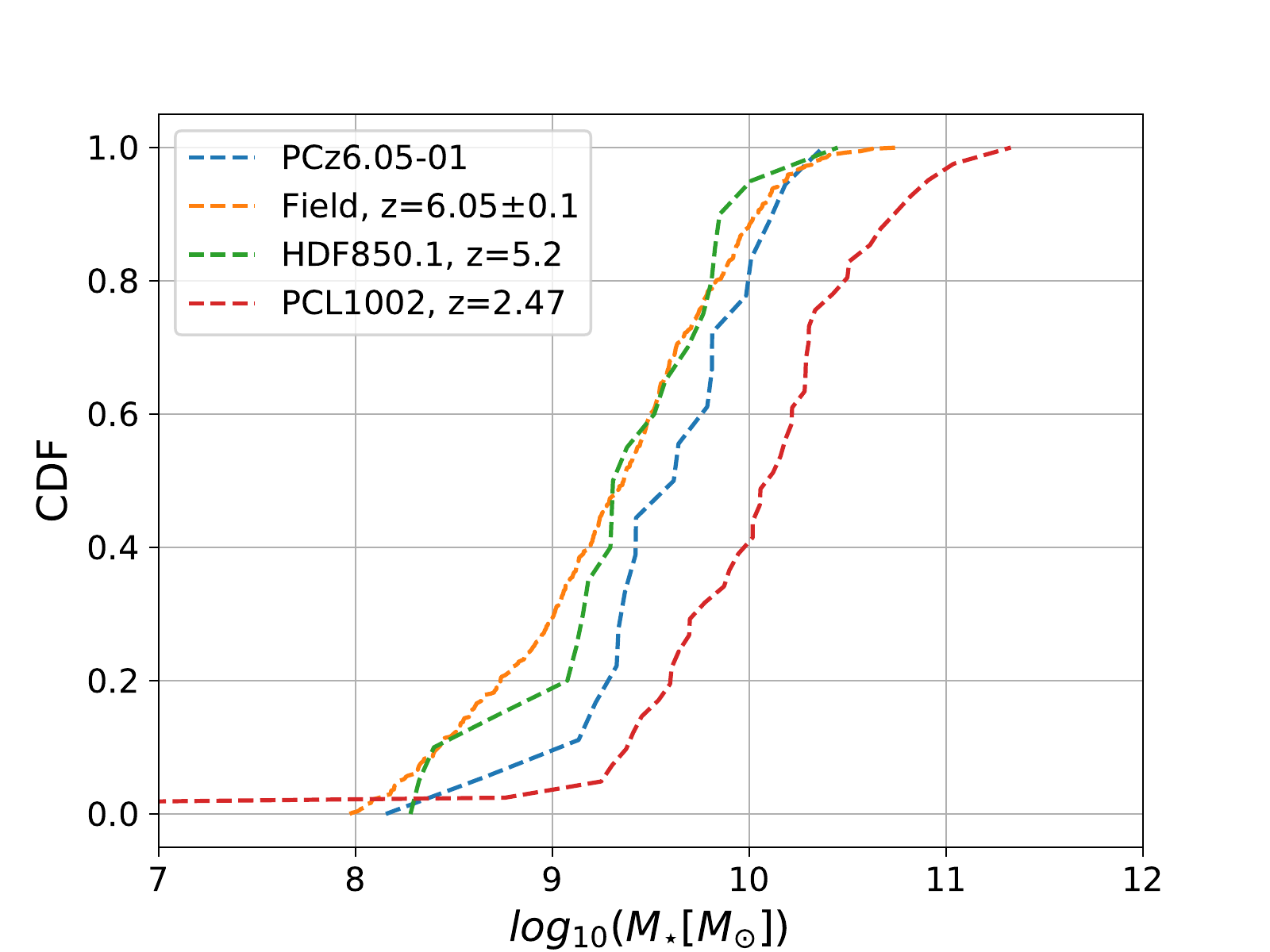}
    \caption{The stellar mass cumulative distribution function (CDF) of the 19 galaxies in COSMOS2020-PCz6.05-01
    (dashed blue line) and of the field (orange dashed line) consisting of 544 galaxies at $z_{\rm phot} = 6.05\pm 0.10$. For comparison, we also show the stellar mass CDF from the spectroscopically verified 
    protoclusters HDF850.1 at $z=5.2$ \citep{Calvi2021} and PCL1002 at $z=2.47$ \citep{Casey2015}.}
    \label{fig:CDF}
\end{figure}

Fig.~\ref{fig:CDF} compares the stellar mass Cumulative Distribution Function (CDF) for the 19 galaxies in COSMOS2020-PCz6.05-01 with that of the field galaxies (i.e., the ones not in other protocluster candidates) 
at the same redshift bin. It is seen that the latter has a broader distribution with tails at low and high masses. $\sim90\%$ of the field galaxies are in the same mass range as COSMOS2020-PCz6.05-01, with $\sim5\%$ having a lower mass and $\sim5\%$ having a higher mass. Generally, we see that COSMOS2020-PCz6.05-01 is skewed towards higher masses than the field CDF. To check if the overdensity and field are drawn from different distributions, we can perform a two-sample Kolmogorov-Smirnov (KS) test \citep{Darling1957} using the stellar masses with the null hypothesis that the two independent samples are drawn from the same continuous distribution. We obtain a KS statistic of $0.28$ and a $p$-value of $0.10$ and therefore cannot reject the null hypothesis at $10\%$ level. This means that we cannot state that protocluster candidate galaxies are from a separate population with a different stellar mass distribution from the field population.  We also compare with the protocluster HDF850.1 at $z=5.2$ \citep{Calvi2021} and a lower redshift comparison with PCL1002 at $z=2.47$ \citep{Casey2015}. Despite being at $z=5.2$, HDF850.1 is skewed towards lower masses than COSMOS2020-PCz6.05-01, possibly indicating that our candidate will evolve into a more massive system. We also see the difference in mass with a more evolved structure in PCL1002, which is skewed towards masses higher than both our field, COSMOS2020-PCz6.05-01 and HDF850.1, as expected of a structure at half the redshift.\\
Comparing the CDFs for the other protocluster candidates (see appendix \ref{sec:allOD}), we note that for all the candidates in the $z=6.05$ bin, there are more massive galaxies ($>10^{10}M_{\odot}$) than in the higher redshift bins. The higher redshift bins have steeper CDFs with masses typically lower than those in the $z=6.05$ bin. The trend is also visible in the main sequence plots for $z>6.9$, where there are fewer massive galaxies compared to the lower redshift bins. This cannot be an issue related to the sensitivity of the survey since we should target the most massive (and brightest) galaxies, which means this could hint at an evolution in the most massive galaxies between the lowest redshift bin and the higher ones. This evolution is investigated further in ${\S}$\ref{section:discussion}.
\subsubsection{Star-formation rate evolution}
\begin{figure*}
\centering
  \includegraphics[width=\textwidth]{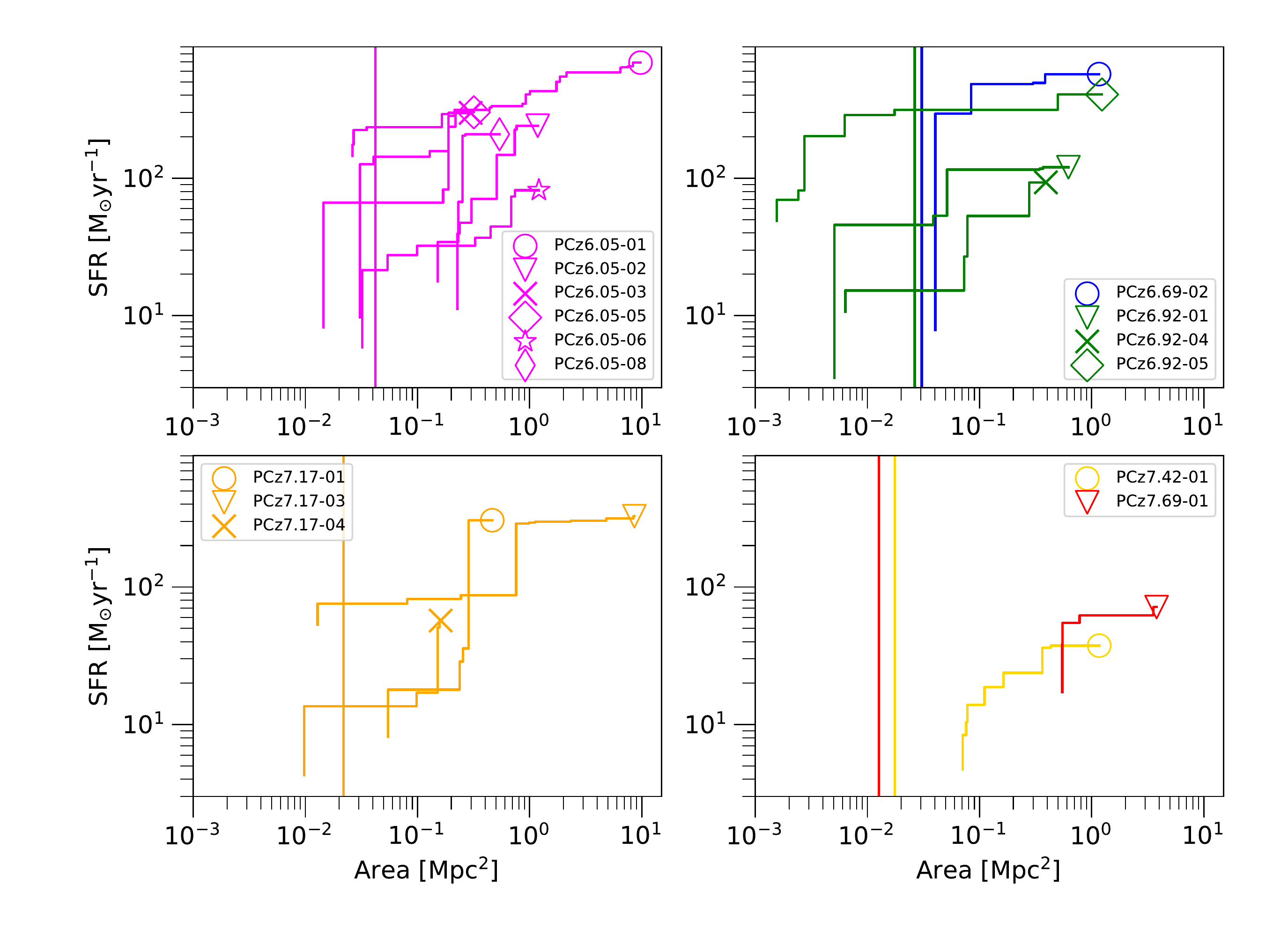}
  \caption{Cumulative SFR enclosed within circular apertures centered on the overdensity peaks of the protocluster candidates. For each protocluster candidate, the cumulative SFR is plotted as a function of the area of the aperture. Candidates are separated into four redshift bins from: $z=6.0-6.5$ (upper left), $z=6.5-7.0$ (upper right), $z=7.0-7.5$ (lower left), and $z=7.5-8.0$ (lower right). \textbf{The average protocluster core size (R$_{200}$) from \cite{Chiang2017} at each redshift bin is plotted as vertical lines, with the color corresponding to the same bin as the protocluster candidates. When multiple lines are present in a plot, the higher redshift bin line always has the lower area.}}
  \label{fig:SFRvsAREA}
\end{figure*}

To compare the evolution of the SFR for our protocluster candidates across the various redshift bins, we calculate the cumulative SFR as a function of the area from the peak overdensity value of each candidate. The resulting curves displayed in Fig. \ref{fig:SFRvsAREA} show a general trend of decreasing SFR with redshift and a tendency for the lower redshift bins to have a sharper increase at the center of the overdensity, whereas the higher $z$-bins either have larger increases further out from the center of the overdensity or a more gradual increase as seen in the highest $z$-bins. This could indicate the central cores of the protoclusters being more evolved at lower redshift.   
\subsection{Dark matter halo masses}\label{section:DMHM}
\begin{table*}
\caption{Estimates of the total dark matter halo mass associated with the protocluster candidate at the redshift of the overdensity and the most massive individual halo at the redshift of the overdensity using \cite{Behroozi2013,Behroozi2018} and \cite{Shuntov2022}.}
\centering
\begin{tabular}{l l l l l l l}
\hline
\hline
\multicolumn{1}{l}{Paper} & 
\multicolumn{1}{l}{Behroozi} & 
\multicolumn{1}{l}{} & 
\multicolumn{1}{l}{} & 
\multicolumn{1}{l}{} & 
\multicolumn{1}{l}{Shuntov}\\ 
\multicolumn{1}{l}{Name} & 
\multicolumn{1}{l}{$\log(M_{\rm DM,tot}(z))$} & 
\multicolumn{1}{l}{} &
\multicolumn{1}{l}{$\log(M_{\rm DM,cent}(z))$} & 
\multicolumn{1}{l}{} &
\multicolumn{1}{l}{$\log(M_{\rm DM,tot}(z))$} & 
\multicolumn{1}{l}{$\log(M_{\rm DM,cent}(z))$}\\
\multicolumn{1}{l}{Method} &
\multicolumn{1}{l}{$5\%M_{\rm Bar}/M_{\rm DM}$} &
\multicolumn{1}{l}{Abundance Matching} & 
\multicolumn{1}{l}{$5\%M_{\rm Bar}/M_{\rm DM}$} & 
\multicolumn{1}{l}{Abundance Matching} & 
\multicolumn{1}{l}{HOD model} & 
\multicolumn{1}{l}{HOD model}\\
\multicolumn{1}{l}{Unit} &
\multicolumn{1}{l}{${\rm dex}$} & 
\multicolumn{1}{l}{${\rm dex}$} & 
\multicolumn{1}{l}{${\rm dex}$} & 
\multicolumn{1}{l}{${\rm dex}$} & 
\multicolumn{1}{l}{${\rm dex}$} &
\multicolumn{1}{l}{${\rm dex}$}\\ 
\hline
PCz6.05-01 & $>12.36^{+0.12}_{-0.14}$ & $12.86^{+0.28}_{-0.19}$ & $>11.67^{+0.10}_{-0.13}$ & $12.38^{+0.44}_{-0.38}$ & $12.53^{+0.09}_{-0.09}$ & $12.18^{+0.05}_{-0.06}$\\
PCz6.05-02 & $>11.92^{+0.12}_{-0.13}$ & $12.41^{+0.23}_{-0.15}$ & $>11.55^{+0.11}_{-0.12}$ & $12.05^{+0.35}_{-0.23}$ & $12.29^{+0.06}_{-0.07}$ & $12.12^{+0.05}_{-0.05}$\\
PCz6.05-03 & $>12.13^{+0.08}_{-0.09}$ & & $>11.92^{+0.06}_{-0.05}$ & & $12.40^{+0.05}_{-0.05}$ & $12.30^{+0.03}_{-0.02}$\\
PCz6.05-05 & $>11.91^{+0.13}_{-0.16}$ & $12.63^{+0.59}_{-0.38}$ & $>11.71^{+0.14}_{-0.15}$ & $12.53^{+0.66}_{-0.50}$ & $12.28^{+0.07}_{-0.08}$ & $12.20^{+0.07}_{-0.07}$\\
PCz6.05-06 & $>11.69^{+0.13}_{-0.14}$ & & $>11.05^{+0.12}_{-0.14}$ & & $12.18^{+0.06}_{-0.07}$ & $11.92^{+0.04}_{-0.04}$\\
PCz6.05-08 & $>11.92^{+0.12}_{-0.13}$ & & $>11.66^{+0.11}_{-0.10}$ & & $12.29^{+0.06}_{-0.06}$ & $12.18^{+0.05}_{-0.05}$\\
PCz6.69-02 & $>11.99^{+0.13}_{-0.12}$ & & $>11.69^{+0.11}_{-0.12}$ & & $12.32^{+0.07}_{-0.06}$ & $12.19^{+0.05}_{-0.06}$\\
PCz6.92-01 & $>10.96^{+0.20}_{-0.20}$ & $11.59^{+0.12}_{-0.11}$ & $>10.60^{+0.15}_{-0.17}$ & $11.09^{+0.09}_{-0.10}$ & $11.89^{+0.06}_{-0.05}$ & $11.80^{+0.30}_{-0.30}$\\
PCz6.92-04 & $>11.24^{+0.18}_{-0.18}$ & $11.73^{+0.13}_{-0.10}$ & $>10.98^{+0.14}_{-0.16}$ & $11.35^{+0.13}_{-0.12}$ & $11.98^{+0.07}_{-0.06}$ &$11.90^{+0.05}_{-0.04}$\\
PCz6.92-05 & $>12.04^{+0.12}_{-0.13}$ & & $>11.62^{+0.10}_{-0.09}$ & & $12.35^{+0.07}_{-0.06}$ & $12.16^{+0.05}_{-0.04}$\\
PCz7.17-01 & $>11.92^{+0.18}_{-0.19}$ & & $>11.89^{+0.19}_{-0.19}$ & & $12.29^{+0.09}_{-0.09}$ & $12.28^{+0.10}_{-0.09}$\\
PCz7.17-03 & $>11.27^{+0.12}_{-0.12}$ & $11.84^{+0.07}_{-0.07}$ & $>10.82^{+0.09}_{-0.13}$ & $11.21^{+0.07}_{-0.09}$ & $11.99^{+0.05}_{-0.05}$ & $11.85^{+0.02}_{-0.03}$\\
PCz7.17-04 & $>10.96^{+0.20}_{-0.22}$ & $11.54^{+0.13}_{-0.13}$ & $>10.75^{+0.19}_{-0.21}$ & $11.15^{+0.15}_{-0.13}$ & $11.89^{+0.06}_{-0.05}$ & $11.83^{+0.05}_{-0.04}$\\
PCz7.42-01 & $>11.02^{+0.30}_{-0.32}$ & $11.65^{+0.18}_{-0.16}$ & $>10.66^{+0.32}_{-0.35}$ & $11.09^{+0.25}_{-0.21}$ & $11.91^{+0.11}_{-0.08}$ & $11.82^{+0.08}_{-0.05}$\\
PCz7.69-01 & $>11.96^{+0.18}_{-0.17}$ & & $>11.76^{+0.15}_{-0.15}$ & & $12.31^{+0.09}_{-0.08}$ & $12.22^{+0.07}_{-0.07}$\\
\hline
\end{tabular}
\label{tab:MDM}
\end{table*}

\textbf{There exist multiple approaches in the literature to derive estimates of the dark matter halo masses $M_{\rm DM}$ \citep{Behroozi2013,Long2020,Calvi2021}. We first consider three different methods to derive the halo masses laid out in \cite{Long2020,Calvi2021} as well as the method laid out in \cite{Shuntov2022} and discuss which ones are appropriate in our case. After introducing the methods we discuss how we handle uncertainties on our dark matter halo mass estimates and compare the different methods. The resulting estimates are shown in Table \ref{tab:MDM}.}

\begin{figure}
    \centering
    \includegraphics[width=0.49\textwidth]{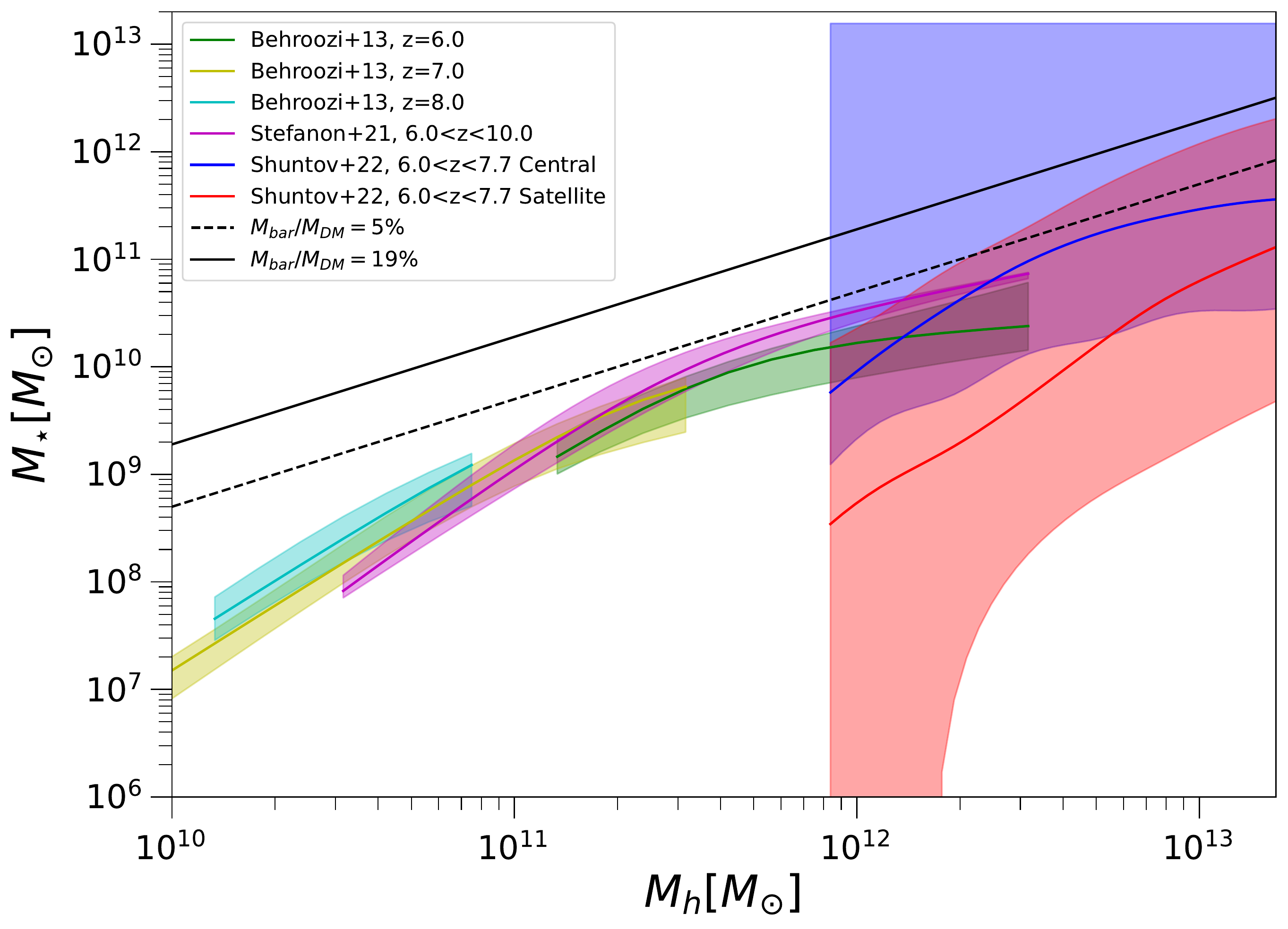}
    \caption{Different stellar mass to halo mass relationships from the literature. The green, yellow and cyan curves are from \cite{Behroozi2013}, the magenta curve is from \cite{Stefanon2021} and the blue and red curves are from \cite{Shuntov2022}. \cite{Shuntov2022} curves are split into central and satellite halos. \textbf{For comparison a fixed $M_{\rm bar}/M_{\rm DM}$ of $5\%$ \citep{Behroozi2018} and $19\%$ \citep{Planck2016} are shown.}}
    \label{fig:MDMall}
\end{figure}

\bigskip
First, we derive a modest estimate of \textbf{the total mass in dark matter halos associated with the protocluster using method:\\}
\noindent {\it i) Individual Abundance Matching}\\ The first method estimates the \textbf{total dark matter halo mass associated with the protocluster candidate} by summing the halo masses of the constituent galaxies. This estimate assumes that individual galaxies are self-bound objects with halos closer to virialization than the protocluster itself and that each galaxy formed its halo prior to coalescing in this overdensity. The halo masses of the galaxies are derived by applying the \textbf{stellar mass to halo mass relationship (SHMR) from abundance matching as detailed in \citet{Behroozi2013} (Be13, see Fig.\,\ref{fig:MDMall})}, which is developed assuming that the bulk of the baryonic mass in dark matter halos is locked up in stars and that massive galaxies trace massive halos. 

\bigskip

\noindent {\it ii) Summed Abundance Matching}\\
The second method assumes that the galaxies in overdensity are all residing in and evolving as part of the same overall dark matter halo.
In this scenario, the stellar masses of the galaxies are summed into a single total stellar mass for the halo and then a stellar-to-halo abundance matching relationship from \citet{Behroozi2013} is used to convert the stellar mass into a dark matter halo mass. A problem with this method is that the relationship in \citet{Behroozi2013} generally does not extend to the large total stellar masses found in protoclusters, nor for the overdensities we identify at $z \geq 6$. To circumvent this problem, \cite{Long2020,Calvi2021} uses the stellar-to-halo mass ratios in \citet{Behroozi2013} for the most massive individual galaxy halo mass estimate from method {\it i)}. \textbf{It has been argued that it is possible to use this method if the galaxies occupy the same single massive halo, which would require spectroscopic redshift confirmation and radial velocity measurements for the galaxies, as seen in \cite{Long2020}.}
\textbf{A potential problem with
this method is that simulations of protoclusters at $z\geq 6$ do not consist of a single dark matter halo but large agglomerations of individual dark matter halos \citep{Chiang2017}. Note also that the virial radius of any of these dark matter halos (few tens to $100\,{\rm ckpc}$
at most) will be an insignificant fraction of the structure size of the protocluster (many ${\rm cMpc}$). For these reasons we have chosen not to use this method.}

\bigskip

\noindent {\it iii) 5\% $M_{\rm Bar}/M_{\rm DM}$}\\
The third and final method assumes a 5\% baryonic-to-dark-matter fraction \citep[BS18]{Behroozi2018} to convert the total stellar mass of the protocluster candidate galaxies to an estimate of the halo mass.  
Note, however, that this method does not consider the gaseous component of the galaxies, and the dark matter halo mass should therefore be considered a lower limit. While the conversion ratio has been applied to lower redshift results \citep{Long2020,Calvi2021}, it does not take the evolution of the ratio between baryonic and dark matter at higher redshifts into account. \textbf{It is also worth nothing the difference between the observationally motivated $5\%$ conversion ratio and the universal baryon fraction obtained from cosmological results \citep[see Fig. \ref{fig:MDMall}]{Planck2016,Behroozi2018}}

\bigskip

A potential caveat associated with methods {\it i)} (and {\it ii)}) is that the redshift-dependent $M_{\rm \star} - M_{\rm DM}$ parametrization given by \cite{Behroozi2013}
are only supported over a specific mass range. For example, at $z=6$, a valid parametrization is only provided over the stellar mass range  $\approx10^{9}-2\times10^{10} M_{\rm \odot}$ (see Fig.~7 in \cite{Behroozi2013}). When our stellar masses exceed this range, we obtain estimates of orders of magnitude higher than expected.
In such cases, where one or more of the galaxy stellar mass estimates are outside the supported range for \cite{Behroozi2013}, we use method {\it iii)} to estimate a lower limit for the dark matter halo mass.

\bigskip
\noindent {\it iv) HOD-based modelling}\\
\textbf{
Finally, we use the SHMR provided by \cite{Shuntov2022} (Sh22). This relationship is determined using the stellar mass function and clustering of galaxies at $6.0<z<7.7$ in the COSMOS2020 catalog and fitting them with a halo occupation distribution-based (HOD) model where a parameterized form of the SHMR is used. Whereas the \cite{Behroozi2013} relationships are each for one redshift, the method used here is similar to the "universal" relationship between $6.0<z<10.0$ of \cite{Stefanon2021} because they both consider a range of redshifts. There are two curves which describe the SHMR for the centrals and satellites, that is, the total stellar mass of centrals/satellites in a halo of a given mass. To estimate the total halo mass of an overdensity we take the total stellar mass of the galaxies and use the sum of the central and satellite curves to convert to a dark matter halo mass. We consider this approach reasonable, because the total stellar mass includes both the central(s) and its(their) satellites.}
\\

\textbf{The SHMRs that we have used are shown in Fig.~\ref{fig:MDMall}. We see that the \cite{Shuntov2022} SHMR generally gives higher halo mass estimates than the \cite{Behroozi2013} and \cite{Stefanon2021} curves, the main exception being the \cite{Behroozi2013} curve at $z=6.0$. We also note that the uncertainty on the stellar mass for the \cite{Shuntov2022} SHMR is substantially larger than the other SHMRs shown and that the SHMR does not go below dark matter halo masses of $~10^{11.8}\,{\rm M_{\rm \odot}} $.}

\textbf{To estimate the uncertainties on our dark matter halo masses, we adopt the mean SMHRs from \cite{Behroozi2013} and \cite{Shuntov2022} and propagate the uncertainties on our stellar mass estimates (see Table \ref{tab:MDM}). This is an underestimate of the true uncertainties since both SHMRs have intrinsic uncertainties (shaded regions in Fig.~\ref{fig:MDMall}.}

\textbf{The dark matter halo mass estimates using the different methods are shown in Table \ref{tab:MDM}.}
\textbf{Using the Be13 method, eight galaxy overdensities presented in this paper have abundance matching results, with total dark matter halo mass estimates in the range $M_{\rm DM} \simeq 3.5\times 10^{11} - 7.2\times 10^{12}\,{\rm \Msolar}$ (see Table \ref{tab:MDM}). For the remaining seven overdensities, we only have an estimate using the BS18 method, corresponding to lower limits for the total dark matter halo masses.
Using the Sh22 method, the range of total halo mass estimates is $M_{\rm DM} \simeq 7.8\times 10^{11} - 3.4\times 10^{12}\,{\rm \Msolar}$ for all of the 15 galaxy overdensities.}

\textbf{Comparing the Be13 and Sh22 estimates in Table \ref{tab:MDM}, we see that for the lowest $z$-bin where is good agreement between the estimates, with the Be13 estimates being more massive, but also more uncertain (not accounting for the uncertainty in the $M_{\rm \star}$ of the \cite{Shuntov2022} curves), while for the higher $z$-bins the relationship is flipped and the Sh22 estimates are higher, though the uncertainty on the Be13 estimates is still larger. This is in line with what we would expect from the different SHMR in Fig.~\ref{fig:MDMall}. In both cases the estimates are higher than the lower limits estimated using the 5\%$M_{\rm bar}$/$M_{\rm DM}$ ratio.}

The estimates presented in Table \ref{tab:MDM} correspond to large {total dark matter halo masses associated with the protoclusters} at redshifts $\geq 6$. COSMOS2020-PCz6.05-01, in particular, may represent a very massive structure early in the Universe's history. For comparison, the spectroscopically confirmed galaxy overdensities associated with the (sub-)millimeter-selected sources SPT\,0311$-$58 at $z=6.9$ and HDF850.1 at $z=5.2$ were found to have dark matter halo masses of $\simeq 1.4-7.0\times 10^{12}\,{\rm \Msolar}$ and $\simeq 1.9-8.0\times 10^{12}\,{\rm \Msolar}$, respectively \citep{Walter2012,Marrone2018, ArrabalHaro2018,Calvi2021}. At later times, when the halos have had time to grow, more massive (sub-)millimeter-selected protoclusters have been found, such as SPT\,2349$-$56 at $z=4.3$ and the Distant Red Core at $z=4.0$, which have dark matter halo masses of $\simeq 10^{13}\,{\rm \Msolar}$ and $\simeq 10^{14}\,{\rm \Msolar}$, respectively \citep{Miller2018, Long2020}.

We caution that our halo estimates must be viewed in light of the inherent model uncertainty when using the stellar masses from COSMOS2020 and its effect on the halo mass estimates.  
Furthermore, the above dark matter halo mass estimates could be underestimated for the following reasons. First, the estimates do not take the cluster members that are not detected in COSMOS2020 into account. Galaxies that are likely to have been missed are low-mass (and thus faint) galaxies and massive quiescent (and thus faint in the rest-frame UV) galaxies \citep{Muldrew2015}. The latter could have a more significant effect on observations of the central regions of protoclusters where downsizing, i.e., the most massive galaxies form earlier than less massive ones, might be an important factor \citep{Rennehan2020}.
The second reason our dark matter halo mass estimates might be biased low is that the abundance matching method of Be13 is developed on the basis that the dominating mass component in all halos is the stellar one. It is important to keep in mind that the galaxies in high-$z$ 
overdensities may contain large gas reservoirs \citep{Strandet2017,Marrone2018}, which can be significant (or even the main baryonic) contributors to the overall mass budget \citep{Long2020}. While none of the galaxies in  COSMOS2020-PCz6.05-01, or indeed any of the galaxies associated with the other overdensities presented in this paper, have robust detections at (sub-)millimeter wavelengths, this does not rule out the possibility that they could harbour significant amounts of gas (and dust). 
\subsection{Estimating the $z=0$ cluster masses}
\subsubsection{Descendant halo masses}
Using Millenium Simulations of 3000 galaxy clusters and their evolution across cosmic time, \citet{Chiang2013} presented evolutionary tracks of  protocluster progenitor masses with redshift (see also \citet{Muldrew2015}). 
The \citet{Chiang2013} models predict that overdensities residing in dark matter halos with masses $\gs 10^{12}\,{\rm \Msolar}$ at $z \geq 6$ are expected to evolve into present-day descendent galaxy clusters similar to a Virgo or Coma-like cluster with a total mass $\sim 10^{15}M_{\odot}$. 
Since \textbf{\cite{Chiang2013} considers the main progenitor halo in a cluster merger tree, we need to identify the most massive individual dark matter halo in our protocluster candidates. We have done this by using the most massive galaxy in each overdensity to estimate a central dark matter halo mass ($M_{\rm DM,cent}$) using the Be13 and Sh22  methods (see Fig.~\ref{fig:MDMall} and Table \ref{tab:MDM})} 

\textbf{
Note, for the Sh22 method, we use the central SHMR to estimate the (sub)halo mass of the single most massive galaxy in each overdensity.
The dark matter halo mass estimates of the Be13 and Sh22 can also be seen as a subhalo mass, in the case of Be13 if the horizontal axis in Fig.~\,\ref{fig:MDMall} is interpreted as the halo mass at the time of accretion and in the case of Sh22 because the SHMR functional form used in \cite{Shuntov2022} has been used to also link the satellite galaxy mass to its subhalo mass \citep{Behroozi2010,Behroozi2019}.}
In essence, the central SHMR can be considered the same for satellites if the halo mass is taken as the subhalo mass at the time of accretion. We therefore argue the central SHMR is applicable in this case, knowing that the estimated mass would be the halo mass at the time of accretion due to the fact that once a subhalo is accreted onto a more massive one, mass is tidally stripped as time goes on.

\textbf{
Considering the abundance matching results of Be13, of the eight overdensities where we were able to use the curves of Fig. \ref{fig:MDMall}, three have halo masses in excess of $10^{12}\,{\rm \Msolar}$ and would therefore be expected to evolve into $\sim 10^{15}\,{\rm \Msolar}$ clusters at $z=0$. These three overdensities are in the lowest $z$-bin.
Using the Sh22 method, out of the 15 $z \geq 6$ galaxy overdensities we have identified, 9 have halo masses in excess of $10^{12}\,{\rm \Msolar}$. The smaller scatter and difference in redshift we see with the Sh22 method (all galaxies have $M_{\rm DM}\approx10^{12}\,{\rm \Msolar}$ within $0.3\,{\rm dex}$) would suggest that our higher redshift overdensities will evolve to be even more massive than a Virgo- or Coma-like cluster at present.}

We caution that the above descendant halo mass estimates require several assumptions that are not currently well constrained by observations. For example, the \citet{Chiang2013} halo growth curves depend heavily on the presumed volumes of the observed galaxy overdensities. 
\subsubsection{Galaxy richness and growth}
\textbf{It is important to keep in mind that it is not enough for a dark matter halo to be in the excess of $10^{12}\,{\rm \Msolar}$ at $z \geq 6$ to grow into a Virgo or Coma-like cluster at present, they also need to be in an environment that allows them to grow over time \citep{Overzier2009,Angulo2012}. From simulations like Magneticum \citep{Remus2022} it has been argued that galaxy richness, which can be characterised by determining the number overdensity, is the best tracer of this growth. This is because it traces the feeding of galaxies into the main dark matter halo from the cosmic web. What this means for us is we should consider not only which overdensity has the most massive halo estimate, but also which ones are in the most overdense environments.
COSMOS2020-Pz6.05-1 stands out in this regard not only by having the second most prominent dark matter halo mass estimates (see Table \ref{tab:MDM}) next to COSMOS2020-Pz6.05-5 using method i) and being among the highest dark matter halo mass estimates ($\simeq 2\times 10^{12}\,{\rm \Msolar}$) using the Sh22 method, but also by containing almost twice as many galaxy member candidates (19) as the second-richest overdensity, COSMOS2020-PCz7.17-1 (10). This makes COSMOS2020-Pz6.05-1 the most likely candidate for becoming a truly massive galaxy cluster at the present day.}

\subsubsection{Present day halo masses from spherical collapse}
It is well known that once matter overdensities enter the non-linear regime, their future evolution can no longer be rigorously described analytically. Instead, one has to resort to numerical simulations or adopt analytical approximations. 
\textbf{Following the procedure in \citet{Calvi2021}, we tested the so-called spherical collapse model which is the simplest analytical description of the non-linear evolution of matter overdensities with cosmic time \citep{Gunn1972}. We found that the resulting upper limit on the total mass had a number of uncertainties that ultimately led us to disregard the estimate. An example on the small scale is that if we calculate the overdensity $\delta_{\rm OD}$ using the number of galaxies in the overdensity with the $4\sigma$ area and the number of galaxies in a field with the area of COSMOS, we obtain a higher value of $\delta_{\rm OD}$, which in turn gives a higher estimate to the total mass upper limit.
$\delta_{\rm OD}$ could also be affected by the lack of detections of fainter objects in COSMOS since if more of these objects are part of a given overdensity compared to the field and we had the depth to detect them, $\delta_{\rm OD}$ would have an increased value for the overdensity, in turn resulting in a higher present-day mass. The effect of galaxy downsizing would run counter to this argument, since the supposed less massive and faint galaxies would not have formed at $z\simeq6$. On the other hand if there were more undetected galaxies in the field relative to the overdensity, the $\delta_{\rm OD}$ would be lower, leading to a lower mass estimate.
An example on a larger scale is the effect the volume has on the final result. Due to the uncertainty associated with the photometric redshift, true volume occupied by the protocluster candidate is unclear and determining the extend from the median photometric redshift of the galaxies would overestimate the volume of the protocluster candidate and in turn the the present day total mass. Taking these effect in aggregate, the difference on the upper limit for the protocluster candidate total mass estimates vary more than an order of magnitude.}
\subsection{Overdensity maps using LBG selection methods}\label{subsection:LBG-selection}
In this section, we compare the overdensities found with the WAK and WVT methods described in \S\ref{section:method} with the traditional LBG dropout selection methods. To this end, we create overdensity maps using the relevant dropout selection techniques between $z\sim 6-10$. We use the full {\sc The Farmer} catalog and put no restrictions on the galaxies other than the dropout selections and the use of the same mask as we did for the WAK and WVT methods that masks out bright stars in the field. 

To cover the redshift range $z \sim 5.5-6.3$, we adopt the $i$-dropout criteria from \cite{Ono2018}, which are based on the HSC-bands. The $i$-dropout criteria are:
\begin{eqnarray}
    i-z &>& 1.5\\
    z-y &<& 0.5\\
    i-z &>& 2.0(z-y) + 1.1,    
\end{eqnarray}
with the further requirement that the sources must be detected at a $S/N > 5$ in $z$ and $S/N > 4$ in $y$ \citep{Ono2018}. 
To select sources in the redshift range $z \sim 6.3-7.7$, we use the $z$-dropout criteria specified by \citet{Ono2018}. They require a $S/N > 5$ detection in $y$-band and a color-selection given by $z-y > 1.6$.
Sources at $z \sim 7.4-8.8$ are selected using the $Y$-dropout selection criteria from \cite{Schmidt2014}.
Since {\sc The Farmer} does not include measurements in the $V$-band, which is used for one of the non-detection criteria in  \cite{Schmidt2014}, we use the HSC $g$-band as a substitute. Our adopted selection criteria are: 
\begin{eqnarray}
    \left ( S/N \right )_{g} &<& 1.5 \\
    \left ( S/N \right )_{J} &>& 5\\
    Y-J &>& 1.75\\
    J-H &<& 0.02 + 0.15\times(Y-J-1.75),
\end{eqnarray}
where the $YJHK$ magnitudes are from the UltraVISTA imaging in COSMOS2020.
For the $J$-dropouts, we use the selection criteria from \cite{Oesch2014}, which covers the redshift range $z \sim 9-10$ and above:
\begin{eqnarray}
    \left (S/N \right )_{H} &>& 5\\
    \left ( S/N \right )_{g~{\rm to}~Y} &<& 2\\
    J-H &>& 0.5\\
    H-[4.5] &>& 3.2\\
    \chi^2_{{\rm opt}+Y} &<& 3.2, 
\end{eqnarray}
where $\chi^2_{{\rm opt+}Y}={\rm SGN}(f_{i})\times\sum_{i}(f_{i}/\sigma_{i})^2$, and
$f_{i}$/$\sigma_{i}$ is the band flux divided by the flux error for the optical (griz) and $Y$-bands \citep{Oesch2014}.  The $\chi^2$ criterion accounts for low-$z$ interlopers while only slightly  ($\sim20\%$) decreasing the galaxy sample size of real $z>9$ sources. The $H-[4.5]<3.2$ criterion accounts for intermediate redshift, dusty sources at $z=2-4$. 
\begin{figure*}
\centering
  \includegraphics[width=\linewidth]{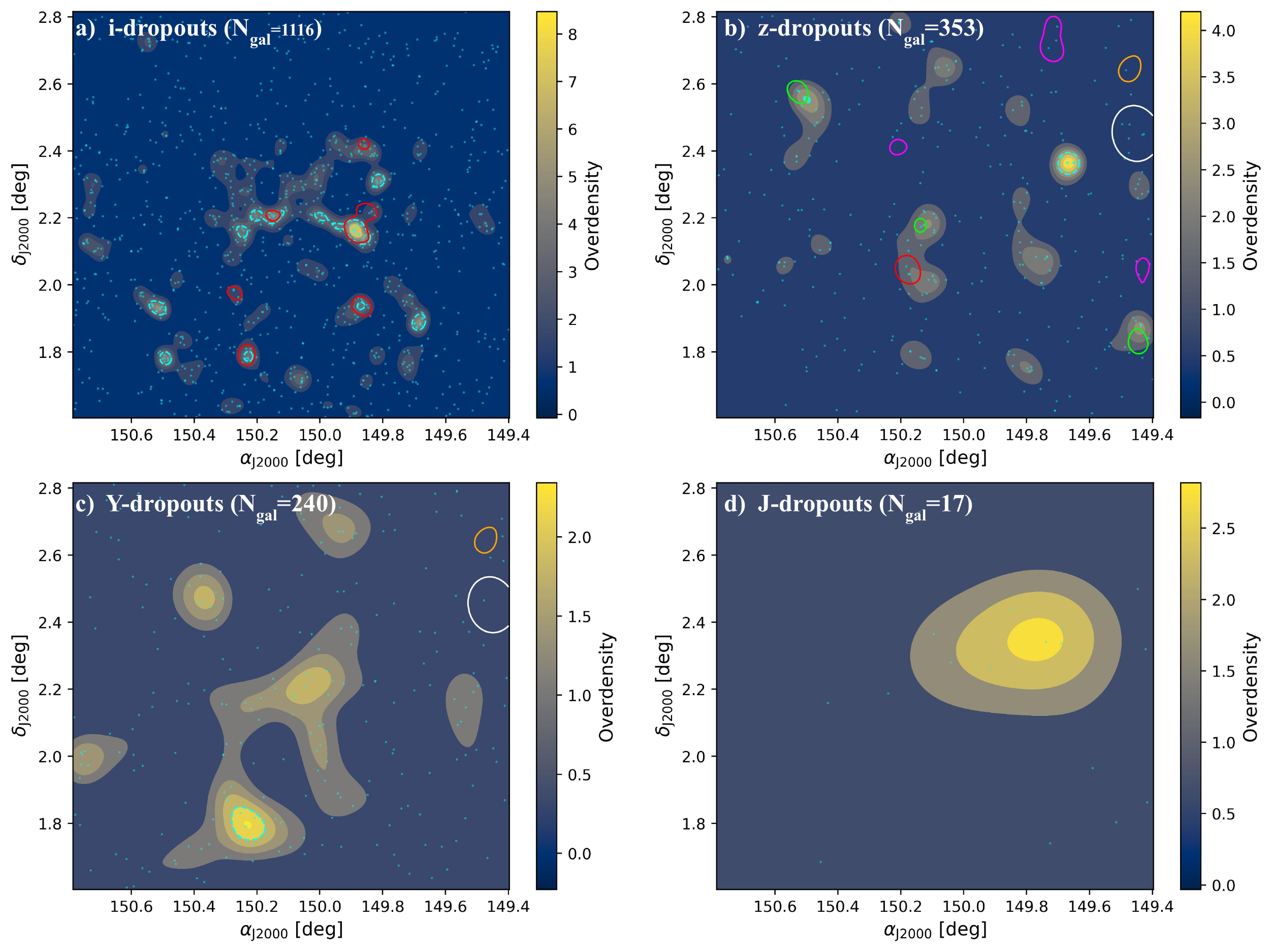}
\caption{Panels a) to d) show overdensity maps of LBGs in COSMOS2020 based on the $i$- (1116 galaxies), $z$-(353 galaxies), $Y$-(240 galaxies), and $J$-(17 galaxies) dropout techniques (see Section \ref{subsection:LBG-selection}), which select galaxies in the redshift ranges $z\sim 5.5-6.3$, $\sim 6.3-7.7$, $\sim 7.4-8.8$, and $\sim 9-10$, respectively. The $4\sigma$ LBG overdensities are shown as cyan dashed contours. For comparison, the other lines show the $4\sigma$ overdensities that fulfill our selection criteria across redshift bins. In figure a) the red lines are for the $z=6.05\pm0.10$ bin. In figure b) the red line is (are) for the $z=6.69\pm0.11$ bin, the green lines for the $z=6.92\pm0.12$ bin, the magenta lines for the $z=7.17\pm0.12$ bin, the orange line for the $z=7.42\pm0.13$ bin and the white line for the $z=7.69\pm0.14$ bin. The orange and white lines in figure c) are the same ones as in figure b).}
\label{fig:dropouts}
\end{figure*}

Fig.~\ref{fig:dropouts} shows the galaxy overdensity maps in COSMOS when applying the above $i$-, $z$-, $Y$-, and $J$- dropout selection criteria to the COSMOS2020 catalog. Comparing these overdensity maps, which corresponds to redshift ranges of $z\sim 5.5-6.3$, $6.3-7.7$, $\sim 7.4-8.8$, and $\sim 9-10$, with the relevant WAK galaxy overdensity maps derived in \S\ref{section:method}, we find multiple overdensities that line up. For $i$-dropouts, Fig.\,\ref{fig:dropouts}a show corresponding overdensities to the ones we have found in the $z=6.05\pm0.1$ bin (Fig.~\ref{fig:WAVT}a), with the exception of COSMOS2020-PCz6.05-08. For $z$-dropouts in Fig.\,\ref{fig:dropouts}b, there are also agreements between the overdensity maps, though less significant ones as seen with COSMOS2020-PCz6.69-01 and COSMOS2020-PCz6.92-01, -04, -05. The $Y$-dropout overdensity map in Fig.~\ref{fig:dropouts}c shows one $\geq4\sigma$ overdensity contour with 4 galaxies inside, though none of our $4\sigma$ contours align with it. For the $J$-dropout overdensity map in Fig.~\ref{fig:dropouts}d, There are no overdensity contours at $4\sigma$ or above and no corresponding contour that fulfill our selection criteria. The small number of significant ($ \geq 4\sigma$) overdensities in the $Y-$ and $J-$dropout overdensity maps is in agreement with the lack of significant overdensities in the WAK overdensity maps in the redshift bins $z \geq 7.4$
(Figs.~\ref{fig:WAVT}m-\ref{fig:WAVT}z).
Comparing the maps, we find little agreement between the WAK and dropout maps, which we ascribe this to the low number statistics in the highest redshift bins. 

\section{Discussion}\label{section:discussion}
\subsection{Dark matter halo mass estimates}\label{subsection:dark-matter-halo}
\begin{figure}
    \centering
    \includegraphics[width=\linewidth]{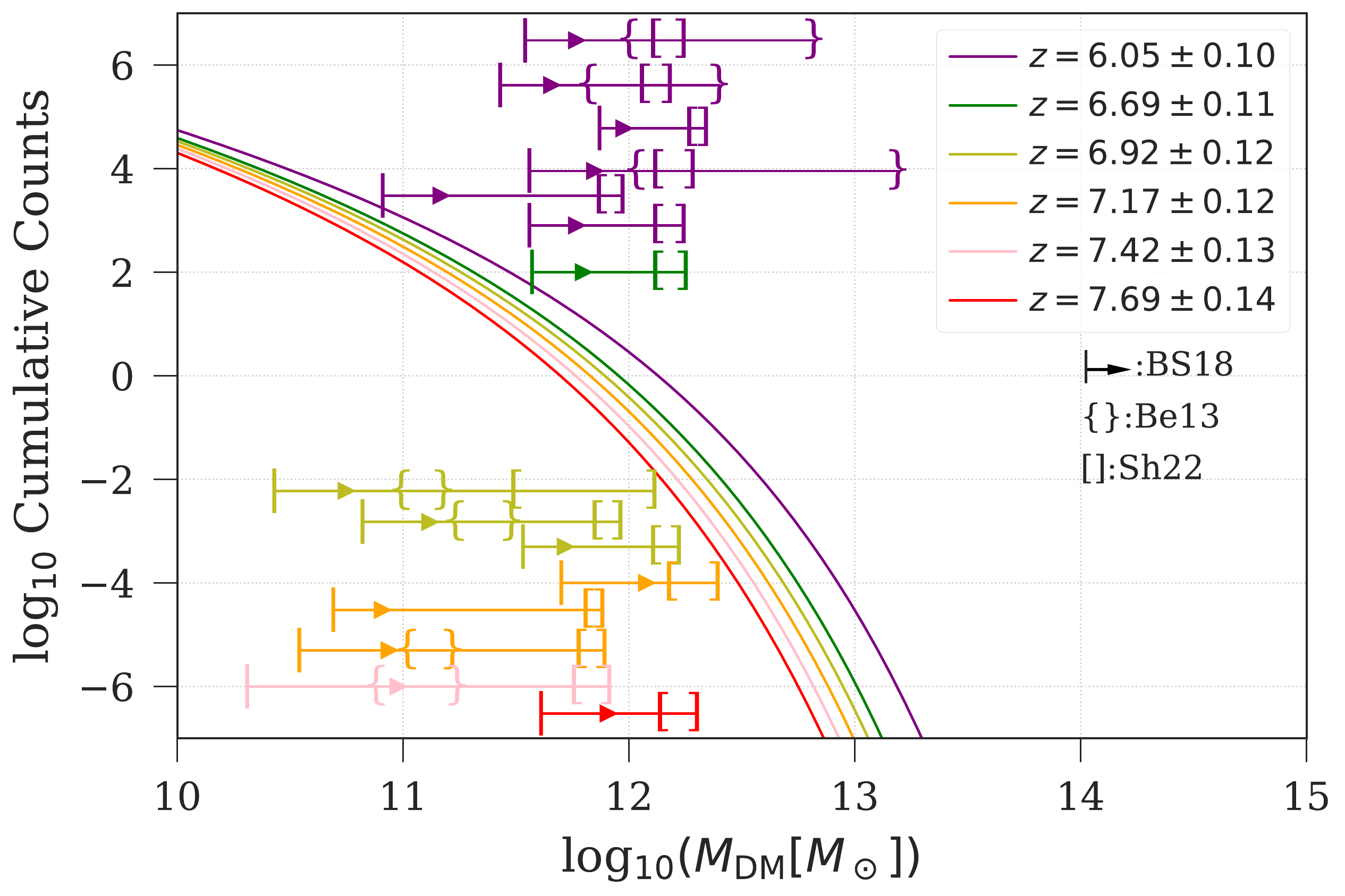}
    \caption{Expected cumulative counts from a COSMOS-like survey following the analytical fit of \cite{Warren2006}. The curves show the expected cumulative counts (number of halos with mass greater than a given mass) for the redshift bins where we have located protocluster candidates. The horizontal lines indicate a range of estimates for a given protocluster candidate using the  different $M_{\rm cent}$ estimates described in section \ref{section:DMHM} and shown in Table \ref{tab:MDM}. The lower bar for each estimate is the value using the \textbf{BS18 method method minus its lower error and the position of the arrow is determined by its value plus the upper error. The "\}" indicates the upper error of the Be13 abundance matching estimates and the "]" marks the upper error for the Sh22 method estimates.} The position of each estimate on the y-axis is chosen arbitrarily for clarity. Estimates from the same bin have the same color, which correspond to the color of each \cite{Warren2006} curve. \textbf{Estimates are shown from top to bottom in the same manner as table \ref{tab:MDM}, so that the top line is for PC$z$6.05-01.}}
    \label{fig:ncumulative}
\end{figure}
\begin{figure}
    \centering
    \includegraphics[width=\linewidth]{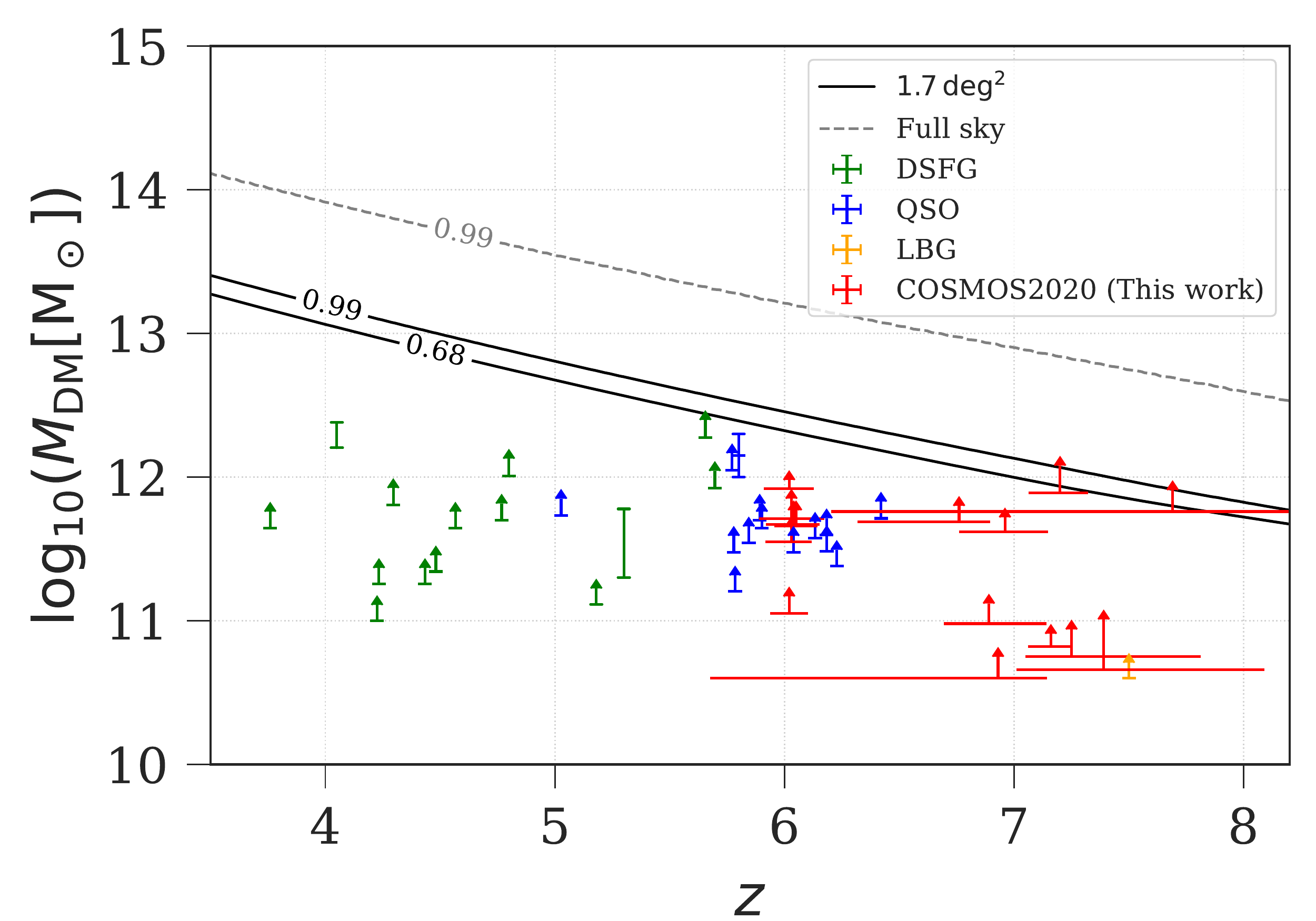}
    \caption{Central protocluster dark matter halo masses ($M_{\rm cent}$) vs redshift. The red points show the central dark matter halo estimates using the BS18 method for the COSMOS2020 protocluster candidates. The bar and arrow are the same as in Fig. \ref{fig:ncumulative}. The exclusion curves are based on the models by \citet{Harrison2013} and show the most massive halos that are expected to be observable within the COSMOS $1.7\,{\rm deg^2}$ area (black curve) at 99\% and 68\% levels and the whole sky (grey dashed curve) at the 99\% level, meaning for an 100$\alpha$\% exclusion curve we should expect to observe a (proto-)cluster above the line only 100(1-$\alpha$)\% of the time \citep{Harrison2013}. \textbf{The green, blue, and orange points are dark matter halo mass estimates based on overdensities associated with DSFG, QSO and LBG, respectively, and taken from \cite{Casey2016,Marrone2018,Overzier2022}.}
    }
    \label{fig:dm-rareness}
\end{figure}
Considering the \textbf{central dark matter halo mass estimates in Table~\,\ref{tab:MDM}}, Fig. \ref{fig:ncumulative} shows the expected cumulative counts for a COSMOS-like survey. We have chosen to use the analytical fit of \cite{Warren2006} because it matches simulations over a wide range of masses well. We use the number density 
of halos with mass greater than a given mass from the \cite{Warren2006} analytical fit. To obtain the cumulative counts (the number of halos with mass greater than a given mass), we then multiply the number density with the volume of each redshift bin where we have found a protocluster candidate, calculated as a truncated pyramid.
\textbf{The range of mass estimate, shown as lines, using the three different methods are between $~10^{11}-10^{13}M_{\odot}$. The estimates using the BS18 method are shown with lower error as the lower bar and the position of the arrow as the upper error. The upper error from the Be13 abundance matching estimates are shown as "\}" and the upper error of the estimates using the Sh22 method are shown as "]".}
We have chosen the cumulative count value for each dark matter halo mass estimate arbitrarily, so it is easier to discern the relation between the estimates and the \cite{Warren2006} curves. 
A comparison with the curves of \cite{Warren2006} shows that our candidates have dark matter halo mass estimates that have corresponding cumulative count values \textbf{that span a wide range. For the number of protocluster candidates (1-6) we have found in each redshift bin, we would expect the cumulative counts to be between $0.0\,{\rm dex}$ to $1.0\,{\rm dex}$.
The BS18 lower limits give high cumulative counts, generally $2.0\,{\rm dex}$ to $4.0\,{\rm dex}$. The Be13 estimates, where available, give lower counts than the BS18 limits. For the $z=6.05\pm0.1$ bin, the estimates range from $1.0\,{\rm dex}$ to $-6.0\,{\rm dex}$, while for the higher bins the range is higher at $2.5\,{\rm dex}$ to $1.5\,{\rm dex}$. The Be13 estimates suggest we would expect to find more halos of similar masses to ours, at least for the higher redshifts bins. This could be an effect of survey depth, as we might be missing out on faint and massive galaxies that would increase the value $M_{\rm DM,cent}$, though at $z>6$ we would expect the most massive galaxies to be the brightest as well.
Finally, for the Sh22 estimates, the cumulative counts ranges from $1.5\,{\rm dex}$ to $-3.0\,{\rm dex}$. With the exception of a couple of outliers, the Sh22 counts generally range from $1.0\,{\rm dex}$ to $-0.5\,{\rm dex}$ for each candidate, similar to what we would expect given the number of protocluster candidates we have found in each bin.} 
The greatest outliers are COSMOS2020-PCz6.05-01 and -05, where their upper bounds using the Be13 method have corresponding cumulative counts of $\sim-3.0\,{\rm dex}$ and $\sim-6.0\,{\rm dex}$ respectively, meaning these appear to be rare structures. \textbf{Another structure that appears to be rare is COSMOS2020-PCz7.69-01, where the Sh22 bounds varies between $-2.0\,{\rm dex}$ to $-3.0\,{\rm dex}$}. These comparisons have to be viewed in light of the fact that both simulation and analytical models have trouble reproducing the most extreme overdensities that we find observationally \citep{Warren2006, Harrison2013}.
\\\\
Another way to investigate if our dark matter halo mass estimates are realistic is to compare the most massive halos that are expected to be observable for a COSMOS-like survey for a given redshift, in a similar fashion to what was done in Fig.\,3 of \cite{Marrone2018}. Figure \ref{fig:dm-rareness} shows how our protocluster candidates compare to the curves showing the expected most massive halo at a given redshift for a both a $2\,{\rm deg^2}$ COSMOS-like survey and a full sky survey, based on the models from \cite{Harrison2013}. We have chosen only to show the BS18 estimates for clarity. We also compare with a number of protoclusters found with different selection methods from \cite{Marrone2018}, as wel as \cite{Casey2016} and \cite{Overzier2022}. In \cite{Harrison2013} they argue that the rareness method of constructing exclusion curves as in Fig. \ref{fig:ncumulative} is biased so that it overestimates the amount of tension a particular observation may cause in relation to a $\Lambda$CDM cosmology.
We observe that our candidates are generally under what is expected to be the most massive halo at their given redshift for a COSMOS-like survey and they have similar DM halo mass estimates as other structures found with different methods. The exceptions are again COSMOS2020-PCz6.05-01 and -05, where their upper estimates using the Be13 method are close to 99\% exclusion curve. We also note that the lower bound of some of the candidates between $z=6.69-7.7$ are close to the expected most massive DM halo mass at their redshift, \textbf{though the large redshift uncertainty on some of the candidate galaxies makes it difficult to conclude if this result is significant.}   
\subsection{Comparison with literature}
\begin{figure*}
\centering
  \includegraphics[width=\textwidth]{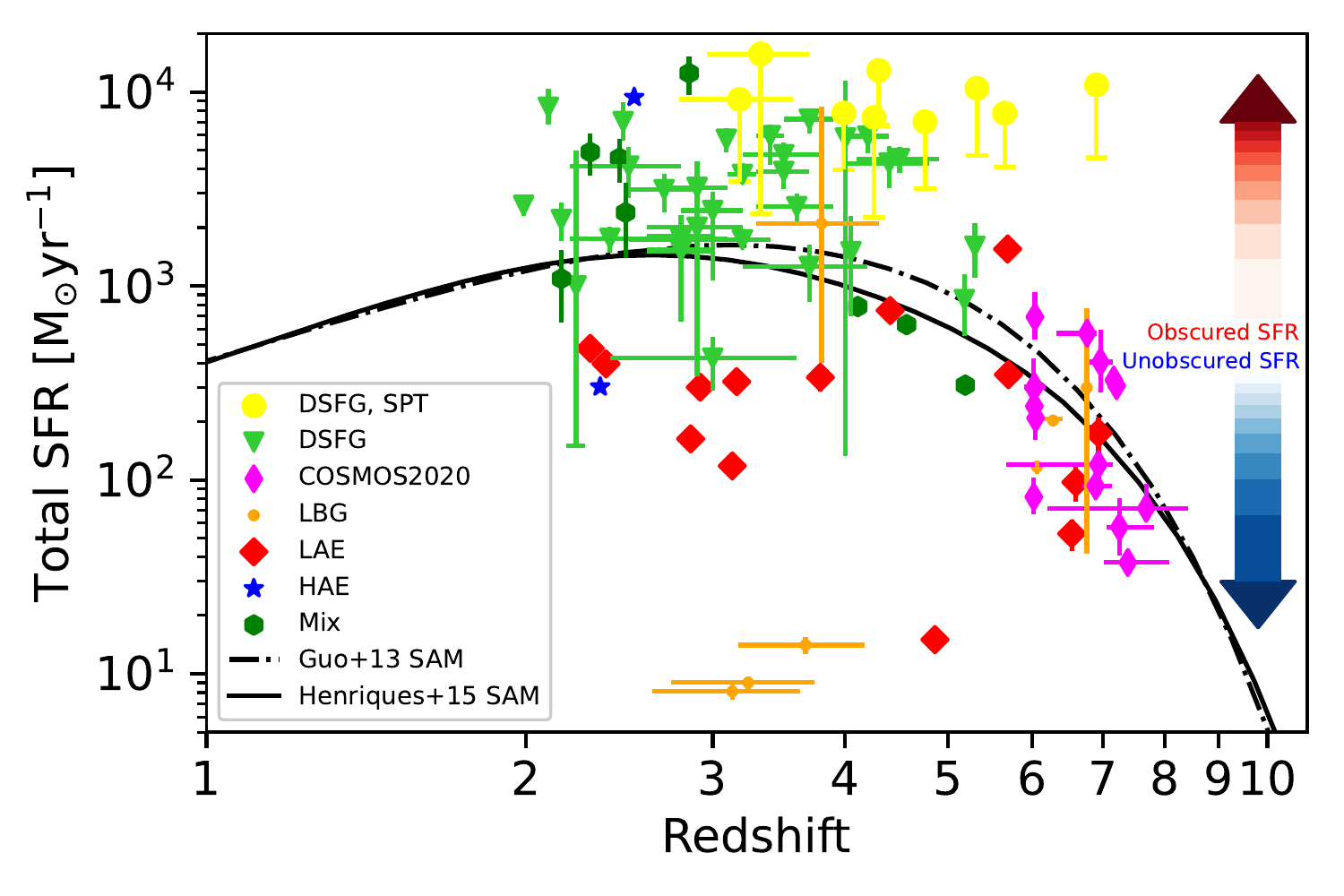}
  \caption{Total SFR vs $z$ for protoclusters using different sample selection methods. Data-points that are (sub-)millimeter-selected protoclusters are from \cite{Wang2020}, while LAEs, LBGs, HAEs and DSFG are from \cite{Harikane2019} (DSFGs are classified as "SMG" in their paper), with the following exceptions: The HAE/SMG protoclusters in \cite{Tadaki2019}, the DSFG protocluster at $2.23$ in \cite{Kato2016}, some of the galaxies from the two COSMOS2015 protoclusters at $z=2.1$, $2.47$ by \cite{Zavala2019}, The LBG protocluster at $z\sim3.8$ from \cite{Kubo2019}, the LBG protocluster at $z=6.31$ by \cite{Mignoli2020}, the LBG protocluster discovered by \cite{Endsley2021} at $z=6.80$ and the LAE protocluster at $z=6.90$ by \cite{Hu2021}. \textbf{Mix refers to the use of a mix of different selection methods for a given protocluster}. The curves show the evolution of the total average SFR per protocluster for different Semi Analytical Models (SAM) from \cite{Chiang2017}.
  }
  \label{fig:SFRoverz}
\end{figure*}
\begin{figure}
\centering
  \includegraphics[width=0.49\textwidth]{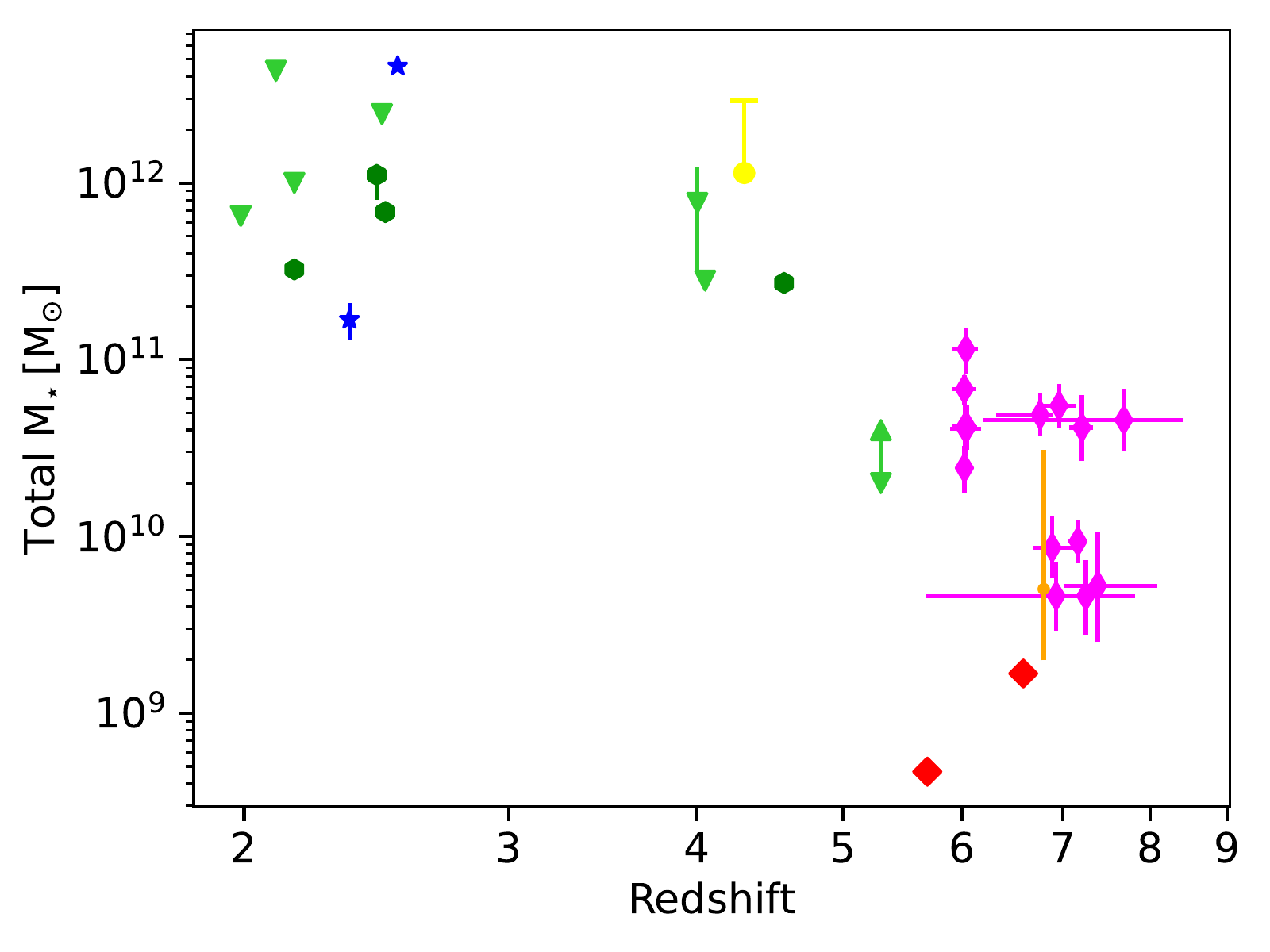}
  \caption{\textbf{Total stellar mass vs $z$ for protoclusters. The physical parameters for these protoclusters can be found in the references of \cite{Casey2016,Harikane2019}, with the exceptions being some of the galaxies in the protoclusters at $z=2.1$ and $z=2.47$ from \cite{Zavala2019}, the protoclusters in \cite{Tadaki2019} at $z=2.16$ and $z=2.49$ and the protocluster at $z=6.8$ from \cite{Endsley2021}. Labels are the same as in Fig. \ref{fig:SFRoverz}.}}
  \label{fig:Mstarvsz}
\end{figure}
Fig.~\ref{fig:SFRoverz} shows the SFR of protoclusters selected via several different methods over a wide redshift range. The different methods can generally be grouped into targeting obscured and unobscured star-formation. The former typically selects protoclusters with higher SFRs than the latter. 

For obscured star-formation and at lower redshifts, we have a combination of protoclusters rich with dusty star-forming galaxies (DSFG) and some with ultra-luminous AGN (Quasars, see \cite{Casey2016}).
These types of objects are rare and have intrinsically short duty cycles ($\sim 100\,{\rm Myr}$). \cite{Casey2016} finds multiple DSFG-rich proto-clusters within a few square degrees and suggests this is evidence that proto-clusters assemble in short-lived, stochastic bursts that likely correspond to the collapse of large-scale filaments on $10\,{\rm Mpc}$ scales. Comparing with our COSMOS2020 candidates, we also have appear to have an extended filament like structure, but none of our galaxies are DSFGs.
This does not exclude the scenario that our candidates had or will have a rapid growth phase of dusty star-formation at some point in time, merely that the different methods we use to select galaxies let us probe structures that are outside of this phase. \textbf{Two of these DSFG-rich protocluster in COSMOS at $z=2.1$ and $z=2.47$ were followed up with ALMA to study their less rare, more typical galaxies by \cite{Zavala2019}. We have included the SFR and stellar masses of those galaxies in Fig. \ref{fig:SFRoverz} and \ref{fig:Mstarvsz}.}
\\\\
\cite{Lewis2018} use ultra red galaxies as "signposts" for dense regions in the early Universe. They find multiple DSFGs near each signpost and posit that with the average total SFR and space density they obtain for their candidate protoclusters (which are similar to the most massive $M_{\rm DM}\sim 10^{15}\,{\rm M_{\odot}}$ galaxy clusters at $z<0.2$), these DSFG systems will evolve into the massive ETGs (relatively passive ellipticals and lenticulars) at the centers of rich galaxy clusters at present. Our descendant mass estimates for PCz6.05-01 are in a similar range as those of \citep{Lewis2018}, though we, as mentioned, have no DSFGs.
\\\\
\textbf{\cite{Ota2018} observes the environment around a quasar at $z=6.6$ and detect multiple LBG and LAE candidates around in the quasar field. They compare with a control field, a general blank sky field where they confirm the non-existence of quasars at similar $z$, any clustering of LAEs and LBGs, or over/underdensities of them. In the southern area of their quasar field, they find the number density of LAEs is almost equivalent to the mean to mean $-1\sigma$ density of LAEs in the control field. Conversely, over this area, LBGs exhibit a filamentary overdensity structure running from east to west. The LBG structure contains several $3\sigma–7\sigma$ high-density excess clumps. Their northern area is very sparse of both LAEs and LBGs. They find that the quasar is in an high-density LBG environment at the $4\sigma$ level, but in an near edge region and not at the center. To our knowledge, none of the galaxies in our overdensities are quasars or have strong AGN emission. \cite{Ota2018} provide no line fluxes, luminosities, stellar masses or SFR estimates (only the Ly$\alpha$ luminosity for one Ly$\alpha$-blob candidate) and we therefore cannot provide further comparisons.}
\\\\
At the extreme end of star-formation, we have the SPT sources of \cite{Wang2020}, compact structures with SFRs at least an order of magnitude higher than COSMOS2020 protocluster candidates. These objects were selected due to their bright $870\mu m$ flux densities and point-source nature. They are classified as protoclusters since they constitute overdensities of {\bf dusty star-forming} galaxies and, therefore, likely the progenitors of present-day clusters. \textbf{See also the SMG selected protocluster of \cite{Oteo2018}.} This highlights the current discrepancy between the SFR of (proto)clusters determined from observations and the SFR expected from simulations, with a tendency for the former to be higher than the models at a given (high) redshift. This discrepancy could come from several factors like simulation volume, sub-grid models not capturing the physics of extreme environments or insufficient resolution due to computational constraints. The difference in SFR for the SPT sources and COSMOS2020-PCz6.05-01 shows the most apparent difference between the methods used to select the objects. The SPT sources are selected with $870\,{\rm \mu m}$ data and are therefore biased towards highly obscured star-formation, and they find all the bright sub-mm galaxies to be in a very active (starburst) phase of their formation where star-formation is at a maximum. The COSMOS2020 protocluster candidates are NIR selected and therefore biased towards unobscured star-formation but are not necessarily undergoing a dusty starburst phase, \textbf{which can be argued due to the lack of mm and far infrared detections. Moreover, the COSMOS2020 candidate galaxies \textbf{(e.g., Fig.\,\ref{fig:mainsequence})} appear to be on the main sequence, although due to the significant scatter we cannot rule out the possibility that some of the galaxies are undergoing starburst.}
\\\\
For unobscured star-formation, we have 
Lyman Alpha Emitters (LAE), which, similarly to the COSMOS2020 candidates, are biased towards unobscured star-formation. The search for LAEs at high-$z$ is typically done with narrow-band filters, which restricts the search to small ranges at specific redshifts ($z=5.7$, $z=6.6$ as seen in \cite{Ouchi2005,Ouchi2008,Jiang2018,Harikane2019}). We see from Fig. \ref{fig:SFRoverz} that the SFRs of LAE selected protoclusters are the ones most similar to ours, though their structures contain objects with lower masses compared to the COSMOS2020 candidates (see Fig.~\ref{fig:mainsequence}).

\textbf{\cite{Higuchi2019} looks for LAE overdensities in a number of fields, including COSMOS, and they find a $z=6.6$ overdensity HSC-z7PCC17 containing a handful of galaxies close to our protocluster candidate PCz6.69-02 and at a similar peak overdensity ($\delta_{\rm max}=7.0^{+6.1}_{-3.1}$ for theirs and $\delta_{\rm max}=8.5^{+4.6}_{-2.4}$ for ours), though their overdensity have no spectroscopically confirmed LAEs within $10\,{\rm cMpc}$ from the center of the protocluster candidates and is not marked as a high density region by their standard (see \S5.1.2. in \cite{Higuchi2019} for more detail). One of their other overdensities HSC-z7PCC26 also at $z=6.6$ is highlighted in \cite{Harikane2019}, and is shown in Fig.\,\ref{fig:SFRoverz} with the SFR taken from \cite{Calvi2019, Espinosa2020}, as well as HSC-z6PCC5 with the SFR taken from \cite{Pavesi2018}}.
\\\\
Lyman Break Galaxy (LBG) selected protoclusters are most similar to our selection of candidates, at least in method and are also biased towards unobscured star-formation. LBGs are selected using color selection criteria applicable to specific redshifts. These color selections use 2-4 observational bands, and as a result, the redshifts of the galaxies that fulfil these criteria are only known inside a broad range (see the LBG selected protoclusters at $z\approx3-4$ from \cite{Toshikawa2016} in Fig.~\ref{fig:SFRoverz}). 

Because of the redshift uncertainty associated with LBG selection, the method is often used to select objects which can be followed up with other methods, such as searching for LAEs. This is the case for the LBG selected protocluster at $z=6.01$ from \citet{Toshikawa2012,Toshikawa2014}, where followup on an LBG overdensity was done by searching for LAEs. \textbf{Another example of is the protocluster at $z\sim3.8$ from \cite{Kubo2019}, where LBG overdensities have been followed up with Planck and Herschel data. This followup gives IR SFRs and places the protocluster next to the DSFGs in Fig. \ref{fig:SFRoverz}.}
\textbf{\cite{Toshikawa2016} found a number of dropout overdensities, specifically 5 i-band dropouts overdensities. The 5 dropout overdensities have peak OD values between $4.1\sigma-7.6\sigma$, similar to the peak overdensities for our protocluster candidates (see table \ref{tab:fulldata}). Only two of the \cite{Toshikawa2016} i-dropout overdensities have galaxies with spec-z, and of those there are only 2-3 galaxies with spec-z in each. \cite{Toshikawa2016} marks one as unclear and the other possible as to whether or not they are protoclusters. Since there is only data available for the galaxies with spec-z and it is unclear whether they are actual protoclusters, we have chosen to not include these overdensities in our plots and further analysis. the lower redshift dropout overdensities with followup from \cite{Toshikawa2016} (D1UD01, D4UD01 and D4GD01 as seen in \cite{Harikane2019}) are included in Fig.~\ref{fig:SFRoverz}.}
For the $z=6.8$ protocluster found in \citet{Endsley2021}, two of their galaxies with strong ($>7\sigma$) Ly$\alpha$ detections are inside one of our $4\sigma$ overdensity contours in the $z=6.69\pm0.11$ bin and another galaxy which has no Ly$\alpha$ detection is inside a $4\sigma$ contour in the $z=6.92\pm0.12$. Both of these contours are not included in our results, since they individually have less than 5 galaxies inside them, but they are clearly part of an overdensity in both bins 
(see Fig. \ref{fig:WAVT}g,h and \ref{fig:WAVT}i,j).
Combining the galaxies from the contours in both bins would give us 5 galaxies in total, thereby classifying it as a candidate using our selection criteria. This highlights the trade-off when choosing a specific bin size, as there is the possibility of missing overdensities where the galaxies fall between the bin edges. Referencing the positions of the 12 candidate galaxies in \citet{Endsley2021}, we see that they are generally more spread out than our candidates, but their overdensity value is also lower than ours at $\delta=3$, whereas our candidates have an overdensity value of $\delta\sim5$ or above. decreasing the overdensity value of the contours in the two bins to $\delta=3$ does not include any more of the \citet{Endsley2021} galaxies.   
\\\\
We also compare the total average protocluster SFR from the semi-analytical models (SAM) in \cite{Chiang2017} (see the black curves in Fig. \ref{fig:SFRoverz}). From a comparison with the models, we see that for our candidates at $z\approx6-8$, our selection seems to target protoclusters that are average in terms of their SFR, as indicated by the candidates being close to the values predicted by the SAMs.
\\\\
A subsample of the galaxies in Fig. \ref{fig:SFRoverz} also has stellar mass estimates, making it possible to compare the total stellar mass for these protoclusters with the COSMOS2020 candidates, as shown in Fig.\,\ref{fig:Mstarvsz}. \textbf{We observe a general trend where the lower redshift protoclusters have higher total stellar masses. This is in line with what we would expect given that the lower redshift protoclusters have had more time to build up mass.}
\subsection{Candidate stellar mass, SFR, sSFR evolution}
\begin{figure*}
\centering
  \centering
  \includegraphics[width=0.49\linewidth]{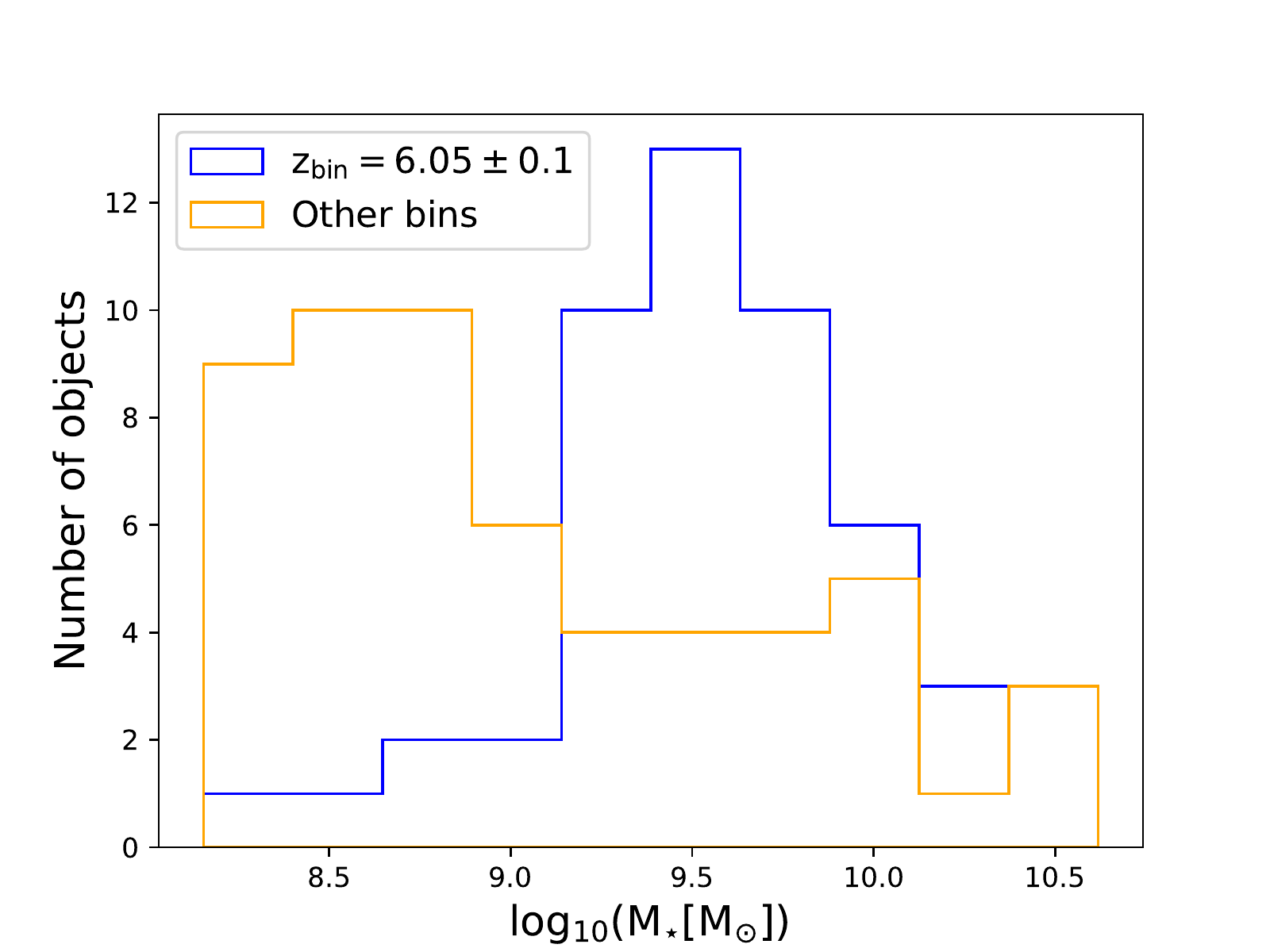}
  \label{fig:masshistograms}
  \includegraphics[width=0.49\linewidth]{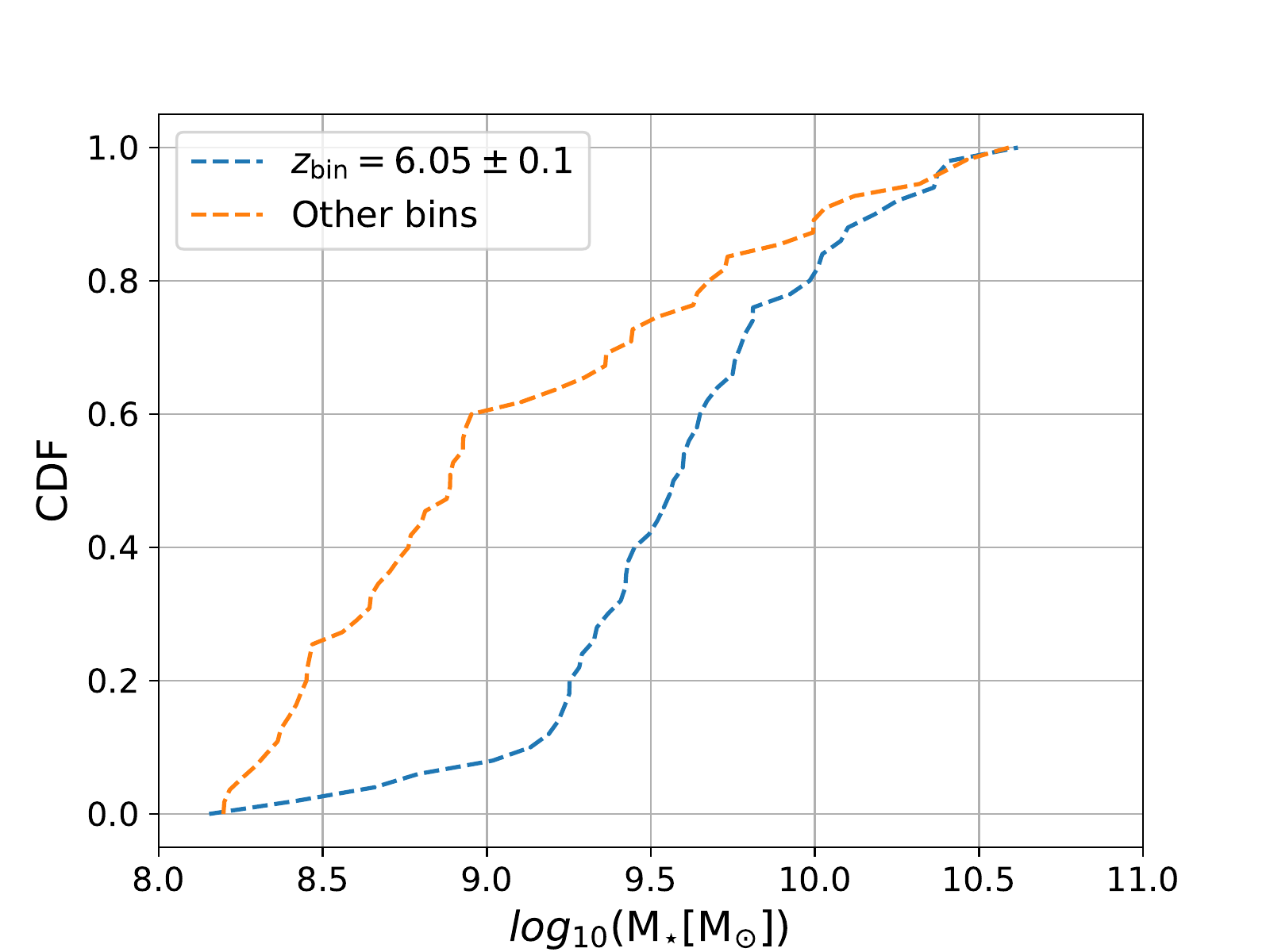}
  \label{fig:CDF_allmass}
  \includegraphics[width=0.49\linewidth]{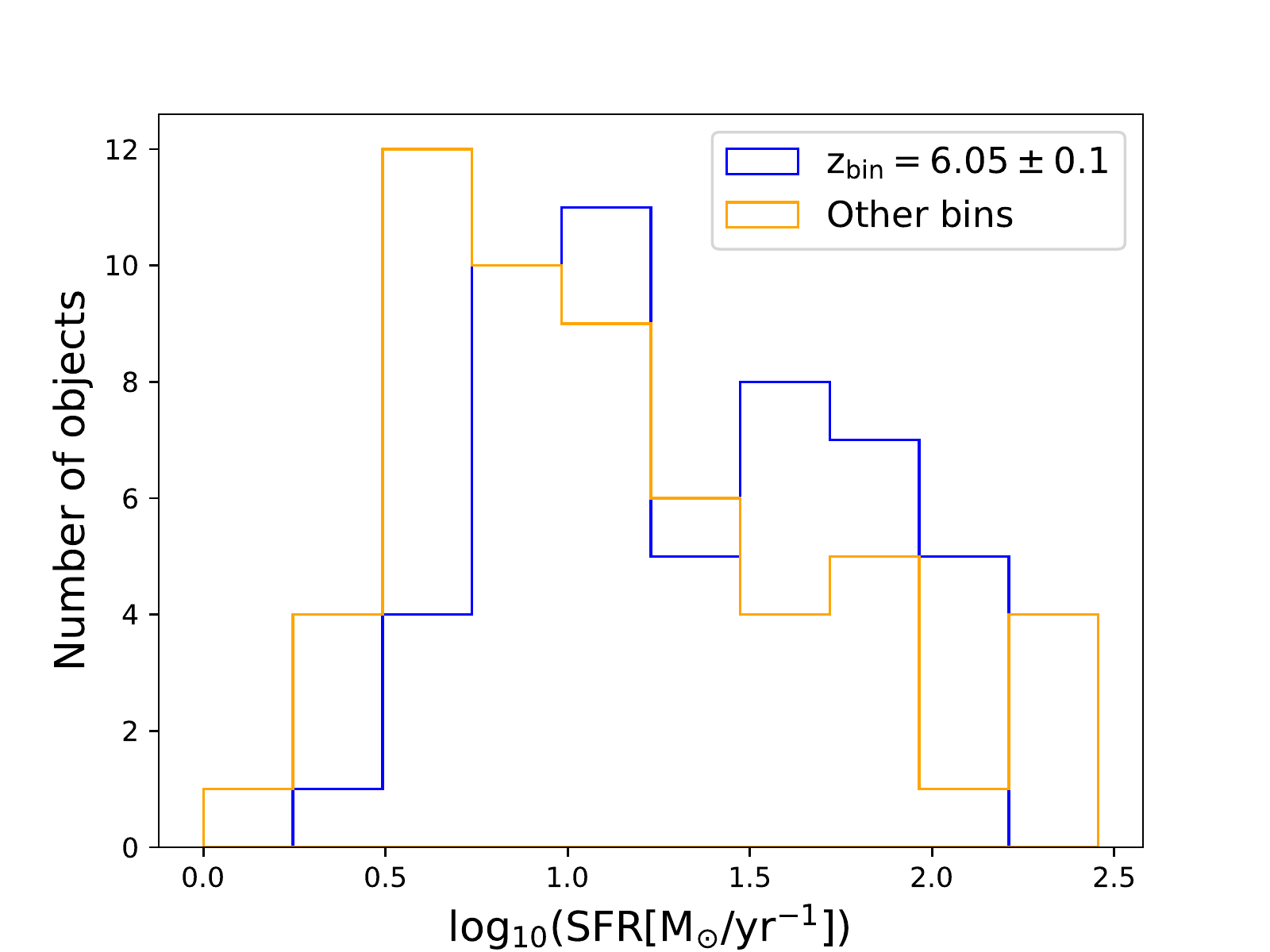}
  \label{fig:SFRhistograms}
  \includegraphics[width=0.49\linewidth]{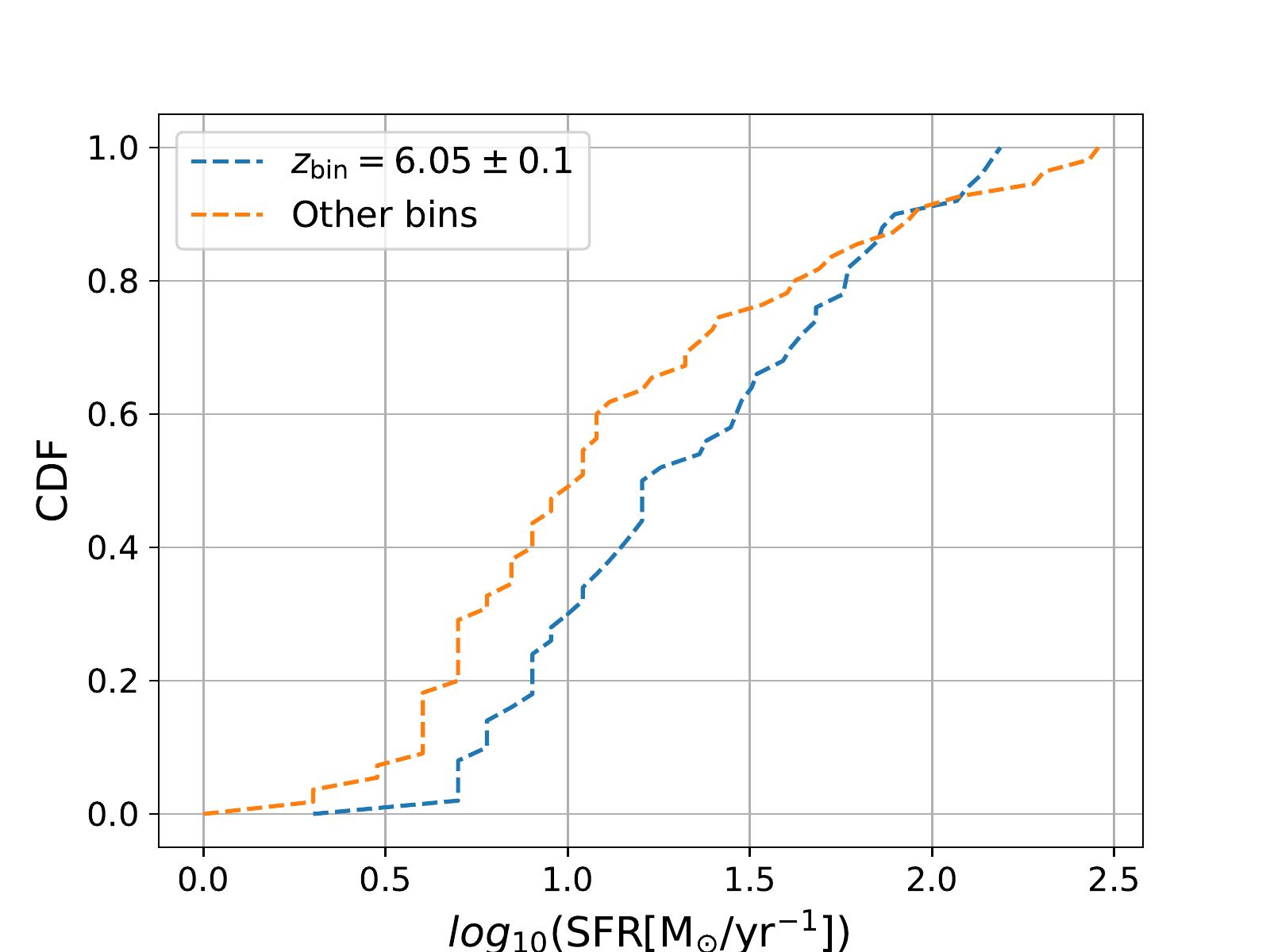}
  \label{fig:CDF_allSFR}
  \includegraphics[width=0.49\linewidth]{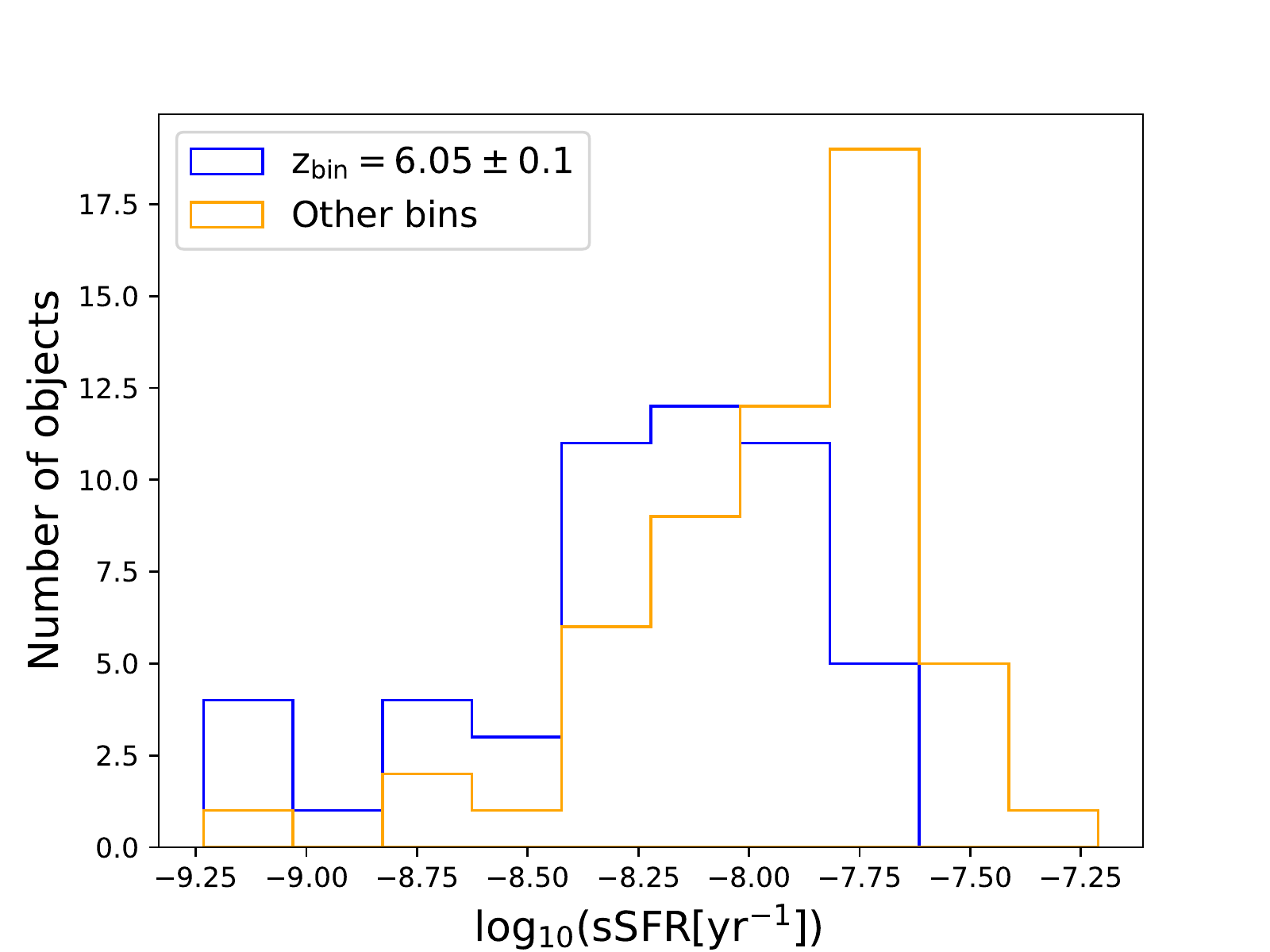}
  \label{fig:sSFRhistograms}
  \includegraphics[width=0.49\linewidth]{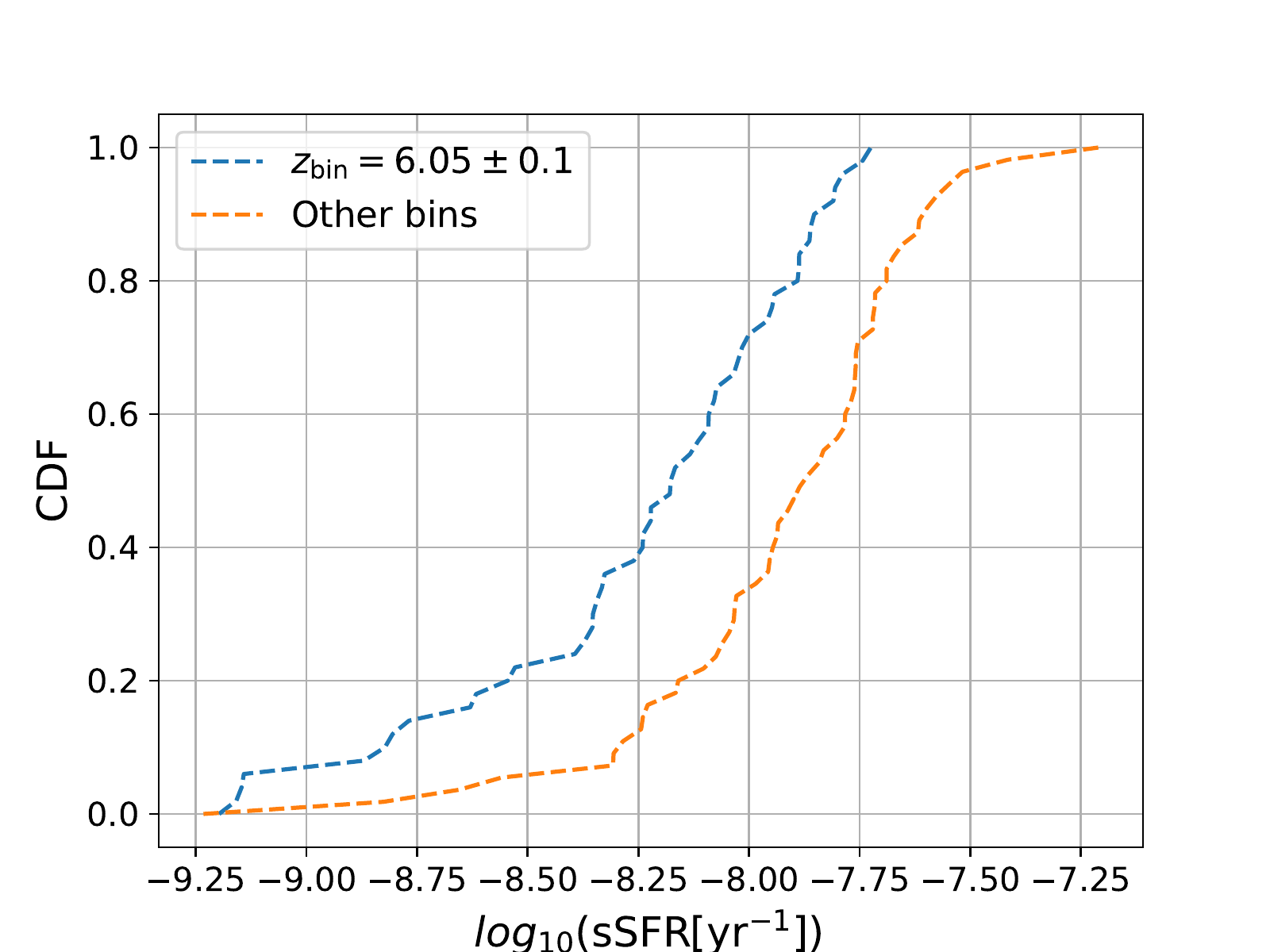}
  \label{fig:CDF_allsSFR}
\caption{\textbf{Upper left:} Comparison of stellar mass histograms of all the protocluster candidate galaxies inside the z=6.05 bin (blue) and the other bins (orange). \textbf{Upper right:} Stellar mass CDF showing the difference in distribution between the protocluster candidate galaxies of the z=6.05 bin and the other bins. \textbf{Middle left:} Same as upper left plot, but for SFR. \textbf{Middle right:} Same as upper right plot, but for SFR. \textbf{Lower left:} Same as upper left plot, but for sSFR. \textbf{Lower right:} Same as upper right plot, but for sSFR.}
\label{fig:comparison}
\end{figure*}

To investigate the differences between the $z=6.05$ bin candidate galaxies and the higher $z$-bins, we show their stellar mass, SFR and specific SFR (sSFR) histograms and CDFs in Fig.\,\ref{fig:comparison}. It is clear that the distribution at $z=6.05$ is skewed towards higher masses with a median of $M_{\rm \star}=10^{9.6}\,{\rm M_{\rm \odot}}$ for the $z=6.05$ bin galaxies and $M_{\rm \star}=10^{8.9}\,{\rm M_{\rm \odot}}$ for the galaxies in the other bins. \textbf{Running a KS-test on the two distributions gives a KS-value of $0.53$ and a p-value of $2.22\times10^{-7}$, suggesting that these are stellar mass distributions distinct from each other.
To test whether this difference is mainly due to our most significant overdensity PCz6.05-01, we removed its galaxies from the mass distribution and found the change in the median mass negligible ($\sim 0.005\,{\rm dex}$), and the stellar mass CDF skewed towards even higher masses, since we loose some $M_{\rm \star}<10^{8.5}\,{\rm M_{\rm \odot}}$ galaxies. For the SFR histogram and CDF, the difference between the bins is less noticeable. the $z=6.05$  bin protocluster candidates skew towards higher SFR, though the highest values are found in the higher $z$ bins. Running a KS test for these distributions give a KS-value of $0.23$ and a p-value of $0.09$, meaning we cannot reject the null-hypothesis and the $9\%$ level. Finally for the sSFR histogram and CDF, we see a noticeable difference between the bins. The higher $z$ bins are skewed towards high sSFR, with a median of $7.88\,{\rm dex}$, as opposed to the $z=6.05$ bin candidate galaxies with a median of $8.18\,{\rm dex}$. The sSFR CDF shows that the higher bin candidates always skew towards higher sSFR than the ones in the $z=6.05$ bin. A KS test of the two distributions give a KS-value of $0.40$ and a p-value of $2.66\times10^{-4}$, the two distribution appear to be distinct from one another.}

\textbf{Combining the information from the stellar mass, SFR and sSFR distributions, we see a tentative evolution of the protocluster candidates from having higher sSFR at higher redshift (i.e. more efficient at star-formation) to higher stellar masses at lower redshift.} The difference we are seeing could be because we are probing larger volumes at lower redshift, thereby taking in more galaxies. However, comparing the expected sizes of protoclusters from \cite{Chiang2017}, the difference between $z=6$ and $z=7.7$ is small, suggesting the effects we see in Fig. \ref{fig:comparison} are not due to volume selection effects.

\section{Conclusions}\label{section:conclusions}
In this paper, we have analyzed galaxies that are part of The Cosmic Evolution Survey between $z=6-10$, utilizing the new {\sc The Farmer} version of the COSMOS2020 catalog \citep{Weaver2021}. The photometric redshifts are determined by the state-of-the-art SED fitting codes LePhare and EAZY. The goal was to probe the onset of protocluster formation by locating high-redshift structures in redshift bins between $z=6$ and $10$. We perform a galaxy number density analysis in $80\,{\rm cMpc}$ bins between $z=6-10$ using a 2D Weighted Adaptive Kernel and 2D Voronoi tessellation.
The main findings of this paper are as follows:
\begin{enumerate}
  \item We locate 15 significant ($> 4\sigma$) overdensities between $z=6.0-7.7$, each containing at least five galaxies. Of the 15 overdensities, 14 are recovered at $\geq 4\sigma$ by both galaxy number density estimators employed here. 
   The most prominent of these, COSMOS2020-PCz6.05-01, is an overdensity at $z\simeq 6.05\pm 0.10$ consisting of 19 galaxies, with 14 being $i$-band dropouts. The median photometric redshifts of the 19 galaxies from LePhare spans a range of $5.97\leq z_{\rm gal}\leq6.13$. Above $z=7.7$, we do
  not find any galaxy overdensities at a significance $\geq 4\sigma$.
  \item To compare with traditional dropout selection techniques, we construct overdensity maps
  consisting of $i$-, $z$-, $Y$- and $J$-band dropouts in the COSMOS2020 catalog. For redshift bins between $z=6.0-7.7$, we find
  excellent correspondence between the overdensities found using the WAK and WVT estimators and the overdensities of $i$- and $z$-dropouts, which correspond to $z\sim 6$ and $\sim 7$, respectively. A similar agreement is not found, however, for the $z > 7.7$ redshift bins, where the low number of sources in our sample hinders a rigorous comparison with the $Y$- and $J$-band dropout maps. 
  \item We estimate the \textbf{total} dark matter halo mass associated with the 15 protocluster candidates using three different techniques and find a range of total halo masses of $M_{\rm DM}\approx 3.5\times 10^{11}-7.2\times 10^{12}\,\Msolar$ using the abundance matching of \cite{Behroozi2013} and $M_{\rm DM}\approx 7.8\times 10^{11}-3.4\times 10^{12}\,\Msolar$ using the method outlined in \cite{Shuntov2022}. Considering the simulation of \citet{Chiang2013}, \textbf{and the central halo mass estimates from \cite{Behroozi2013,Shuntov2022}}, we find the descendant halo mass of COSMOS2020-PCz6.05-01 to be similar to a Virgo-/Coma-like cluster ($\sim10^{14-15}\, M_{\odot}$), \textbf{and we expect the other candidates from the $z=6.05\pm0.10$ bin to end up with a similar halo mass at $z=0$. For the higher $z$ candidates the \cite{Behroozi2013} abundance matching estimates are closer to $M_{\rm DM}\approx 10^{11}\,\Msolar$, whereas the \cite{Shuntov2022} estimates are still $M_{\rm DM}\approx 10^{12}\,\Msolar$, meaning they could evolve to be even more massive than a Virgo-/Coma-like cluster at present.}
  These mass estimates are likely to be underestimated due to faint low mass galaxies and quiescent high mass galaxies not detected in COSMOS. 
  \item We compare the 15 candidates to the number of protoclusters located with different selection methods 
  and find that they occupy a unique position as a NIR-selected structure biased towards \textbf{massive} blue stars and unobscured star-formation \textbf{with galaxies} that appears to be on the galaxy main sequence. \textbf{combined with the lack of mm and far infrared detections, the candidate galaxies appear to not be in a starbursting phase}. Compared with the semi-analytical models of \cite{Chiang2017}, we find that our candidates agree with the total average star-formation rate predicted by those models. 
  \item We show a trend in the evolution of the stellar masses between the $z=6.05$ bin and the higher z-bins by comparing their histograms and CDFs, with more massive galaxies at lower redshift. Combined with a higher specific SFR for the candidates in the higher redshift bins, we appear to trace the evolution of the protocluster galaxies from being more efficiently star-forming at $z>6$ to more massive at $z=6$. 
\end{enumerate}

Many of the protocluster candidates presented here will be covered by COSMOS-Web (Cycle 1, ID. \#1727), a large
program on the {\it James Webb Space Telescope} (JWST),
which will survey the COSMOS field using the MIRI and NIRCam instruments.
Our results emulate what will be possible 
with {\it Euclid} combined with large ground-based optical 
surveys, which, via deep near-IR imaging of tens of square degrees, will uncover dozens of 
$z \geq 6$ protocluster-targets. Following these up with spectroscopic campaigns using 8-10m class telescopes on the ground, the JWST, and the Atacama Large Millimeter/sub-millimeter Array (ALMA), will offer uniquely detailed studies 
of the assembly processes of protoclusters and their galaxies during reionization.

\section{Acknowledgements}
The Cosmic Dawn Center (DAWN) is funded by the Danish National Research Foundation under grant No. 140. We gratefully acknowledge the contributions of the entire COSMOS collaboration consisting of more than 100 scientists in constructing the COSMOS2020 catalog as well as for the feedback and discussion. The research with the COSMOS survey is also partly supported by the Centre National d'Etudes Spatiales (CNES).
TRG and MB are grateful for support from the Carlsberg Foundation via grant no.~CF20-0534.

\bibliography{bib}{}
\bibliographystyle{aasjournal}

\begin{appendix}
\section{COSMOS galaxy overdensity maps 
}\label{sec:maps} 
\begin{figure}[h]
\centering
\includegraphics[width=\linewidth]{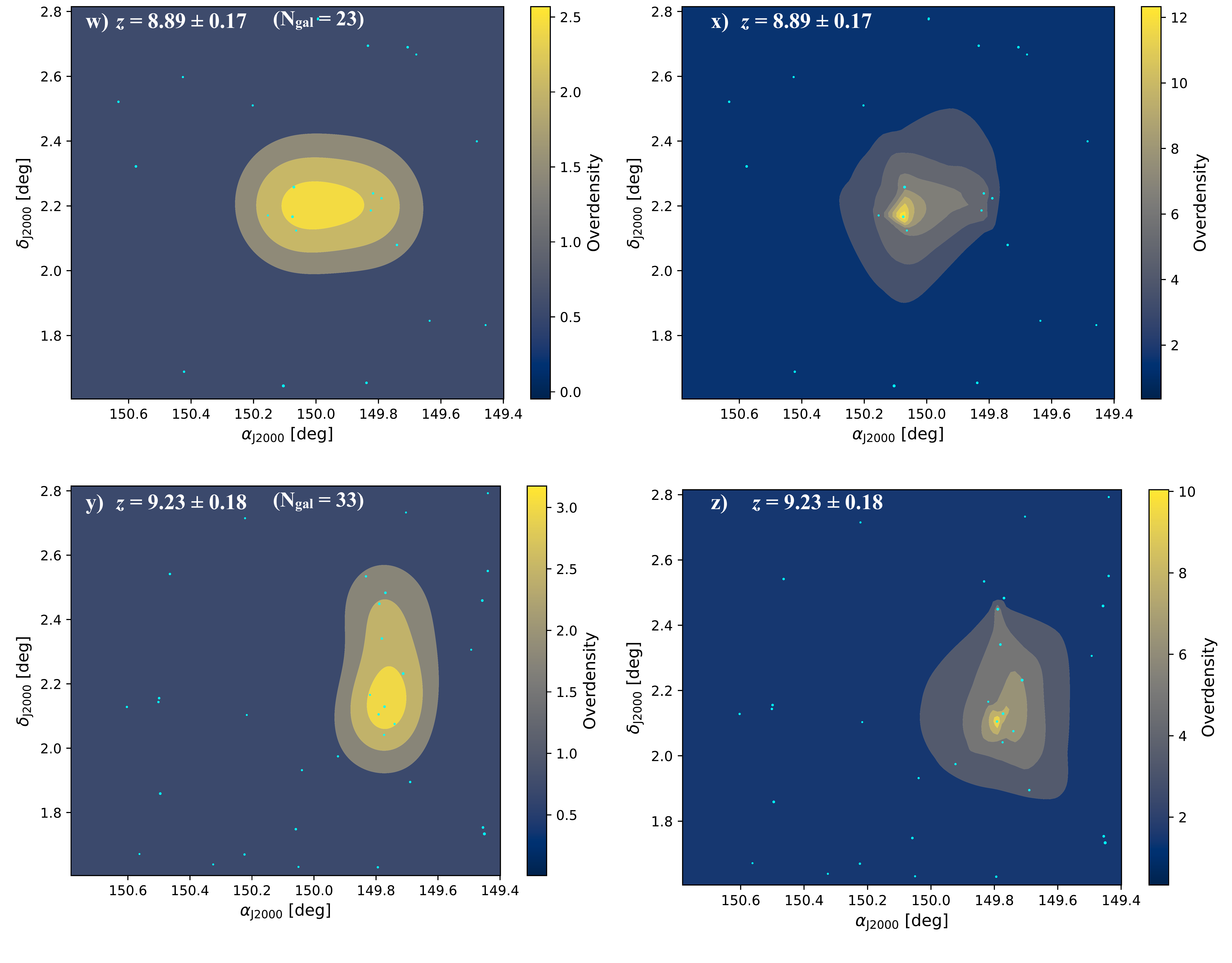}
\caption{Galaxy overdensity maps (colored contours in steps of $2\sigma$) using the WAK (left panels) and WVT estimators (right panels) as described in Section \ref{section:method}. The overdensity maps are for the indicated redshift bins. Red contours indicate regions where the galaxy overdensity is significant by $4\sigma$ or higher and meet our selection criteria.
These $\geq 4\sigma$ overdensities are labeled by numbers that refer to their IDs in Table \ref{tab:fulldata}. The location of the individual galaxies are marked by black dots, and the number of galaxies within each redshift bin is indicated in parentheses.
}
\label{fig:WAVT}
\end{figure}

Figs.~\ref{fig:WAVT}a-\ref{fig:WAVT}z show galaxy overdensity maps (colored contours) using the WAK (left panels) and WVT estimators (right panels).
Starting from the top row 
the overdensity maps shown are for the following redshift bins: a-b) $z=6.05\pm0.10$ (595 galaxies), c-d) $z=6.25\pm0.10$ (226 galaxies), e-f) $z=6.47\pm0.11$ (95 galaxies), g-h) $z=6.69\pm0.11$ (123 galaxies), i-j) $z=6.92\pm0.12$ (187 galaxies), k-l) $z=7.17\pm0.12$ (193 galaxies), m-n) $z=7.42\pm0.13$ (102 galaxies), o-p) $z=7.69\pm0.14$ (26 galaxies), q-r) $z=7.97\pm0.14$ (22 galaxies), s-t) $8.26\pm0.15$ (20 galaxies), u-v) $8.57\pm0.16$ (21 galaxies), w-x) $8.89\pm0.17$ (23 galaxies), y-z)  $9.23\pm0.18$ (33 galaxies). Contours indicate regions of increasing overdensity values in steps of $1\sigma$. Red contours indicate regions where the galaxy overdensity is significant by $4\sigma$ or higher and meet our selection criteria. These are indicated by numbers that refer to their name in Table \ref{tab:fulldata}. The location of the individual galaxies are marked by cyan dots, and the number of galaxies within each redshift bin is indicated in the parentheses above.
\\\\
\includegraphics[width=\linewidth]{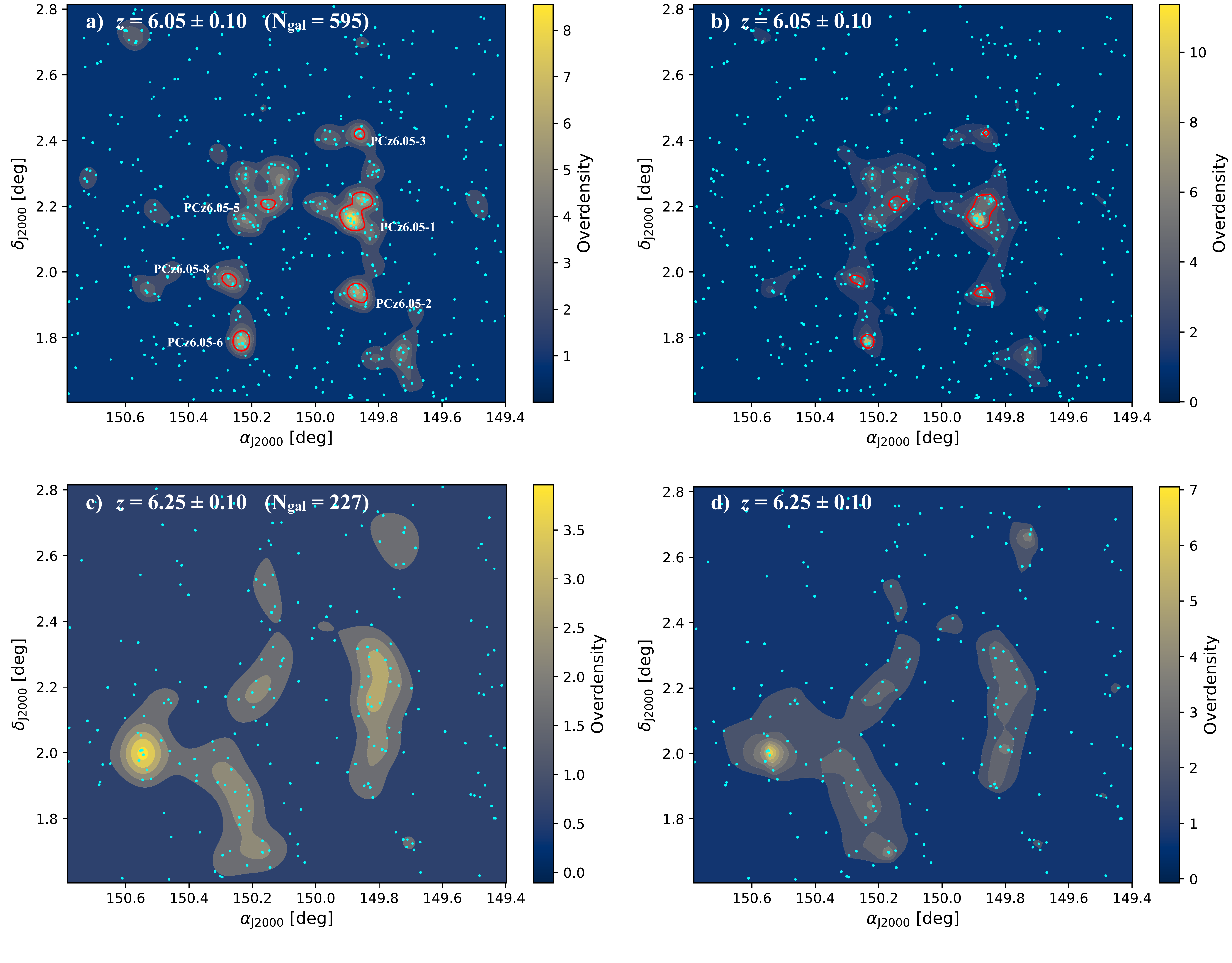}
\newpage
\includegraphics[width=\linewidth]{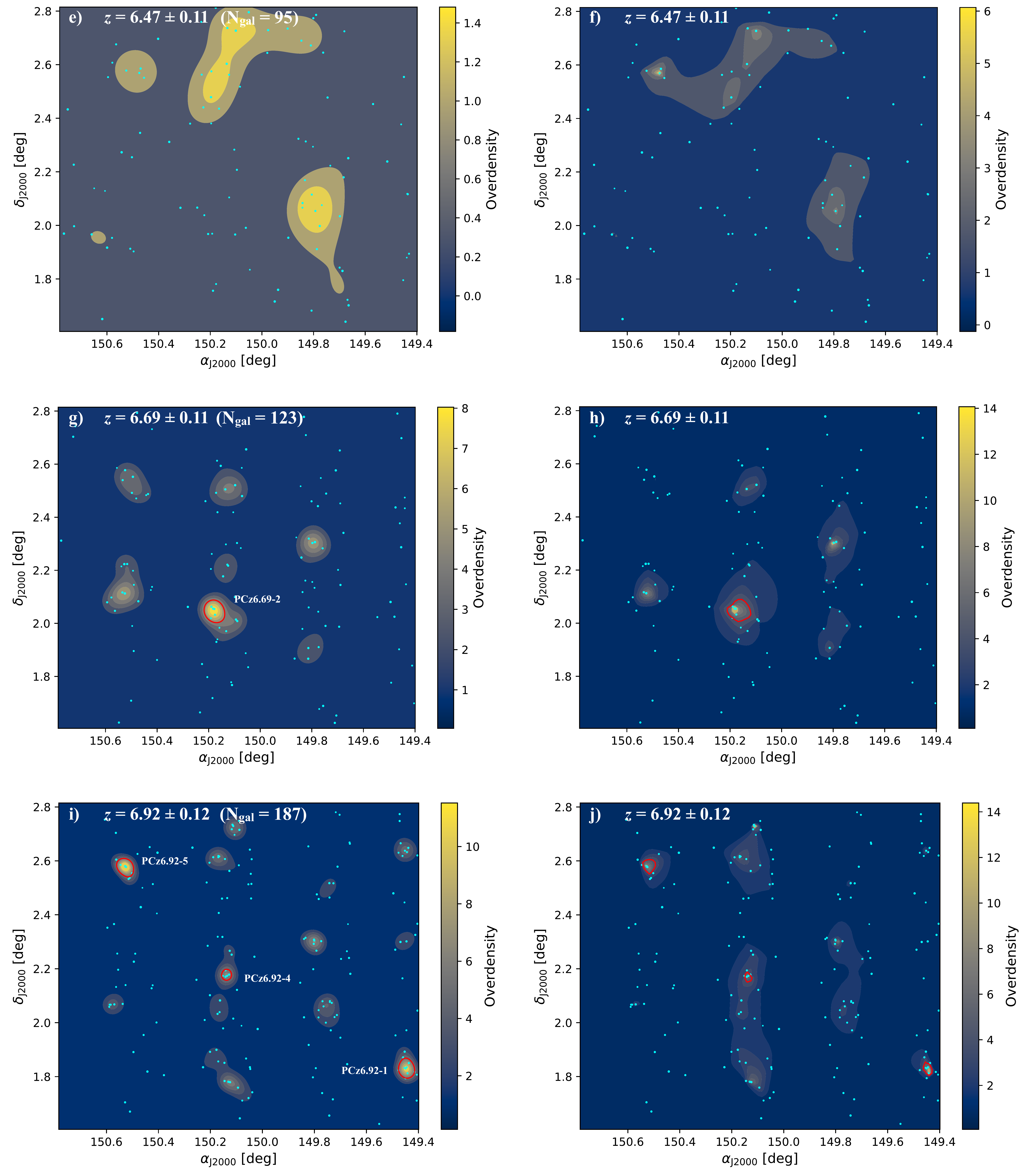}
\newpage
\includegraphics[width=\linewidth]{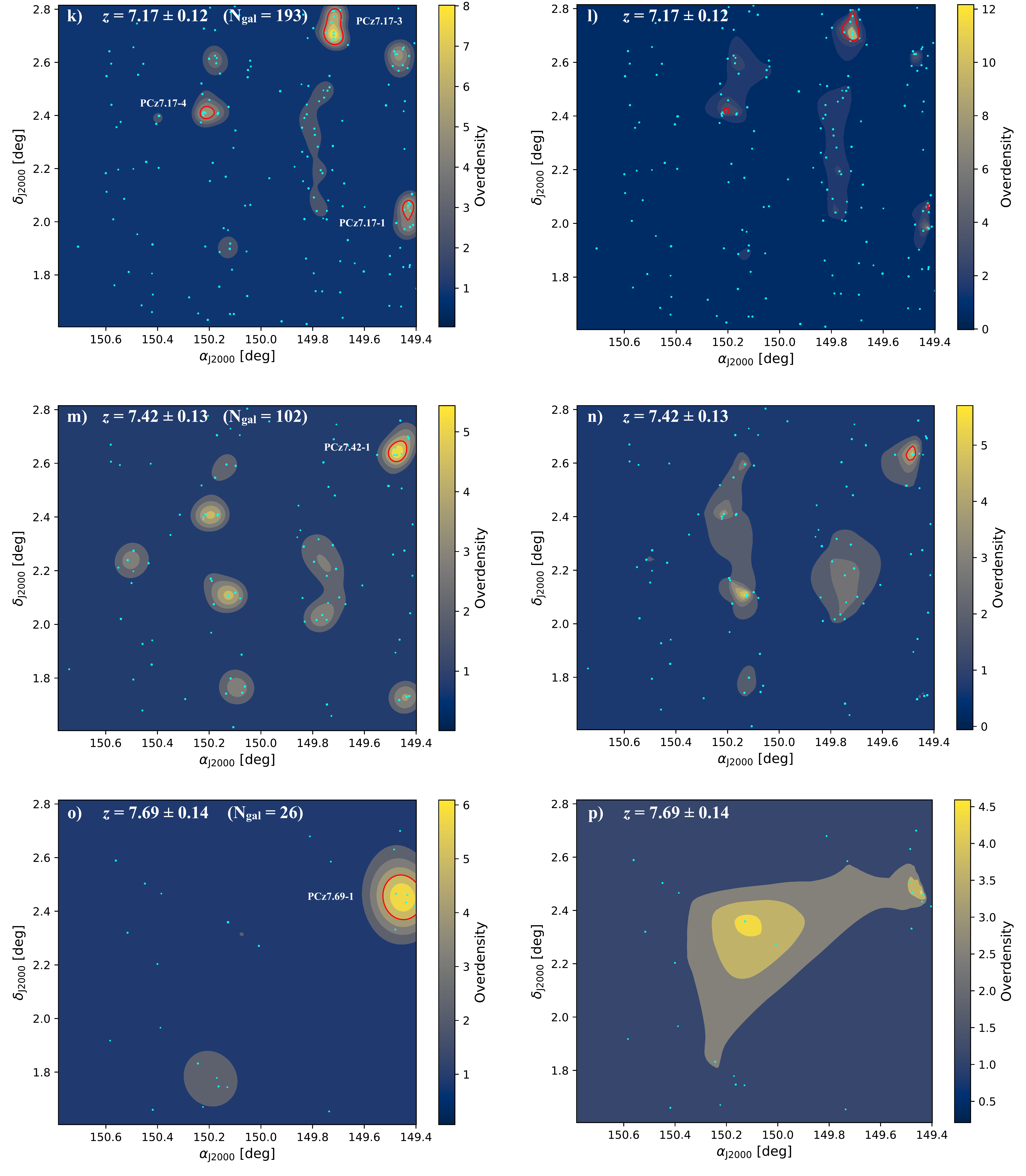}
\newpage
\includegraphics[width=\linewidth]{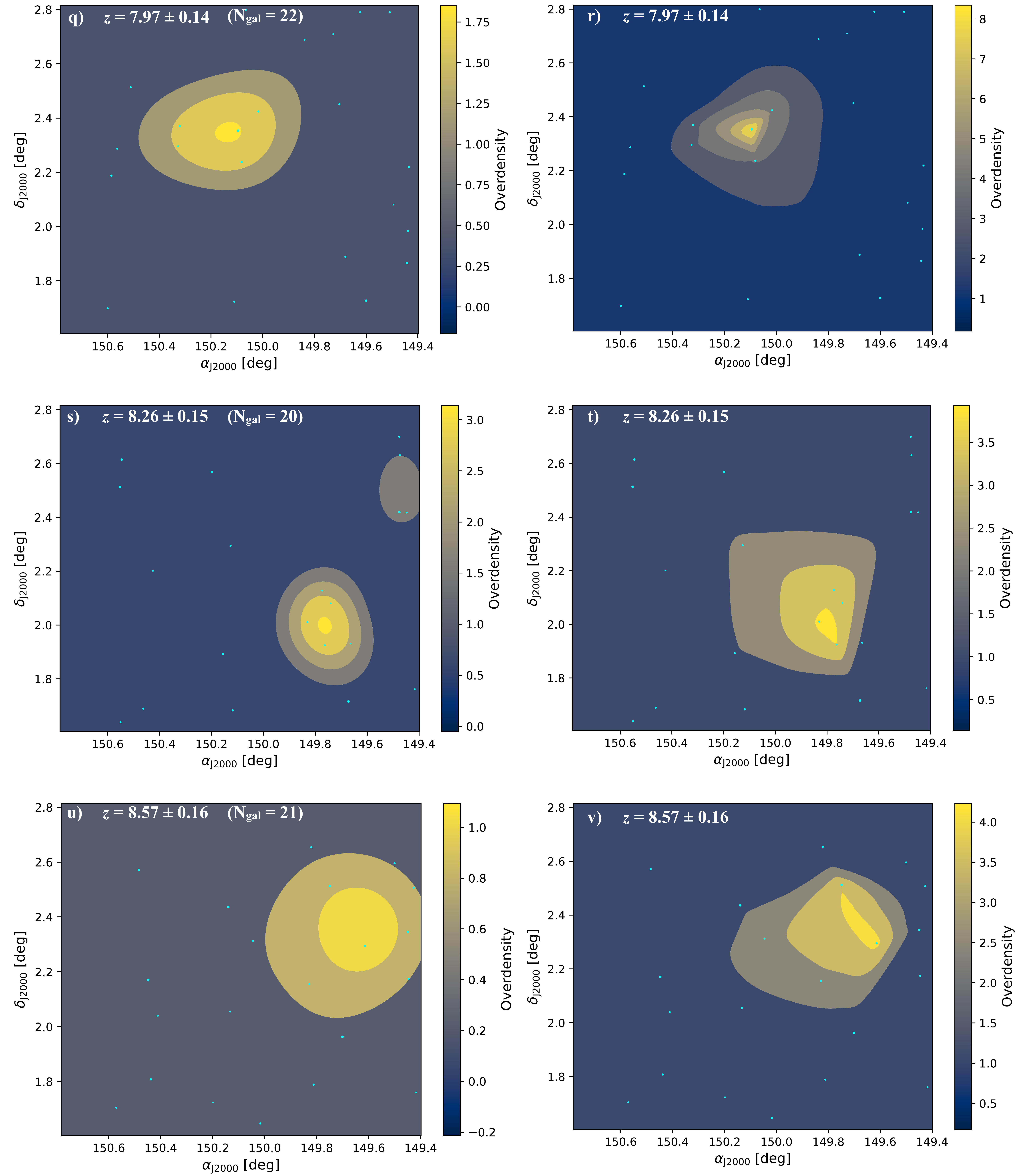}
\newpage

\pagebreak

\section{Weight distribution for galaxy sample of all bins}\label{sec:weights}

\begin{figure}[h]
    \centering
    \includegraphics[width=0.95\textwidth]{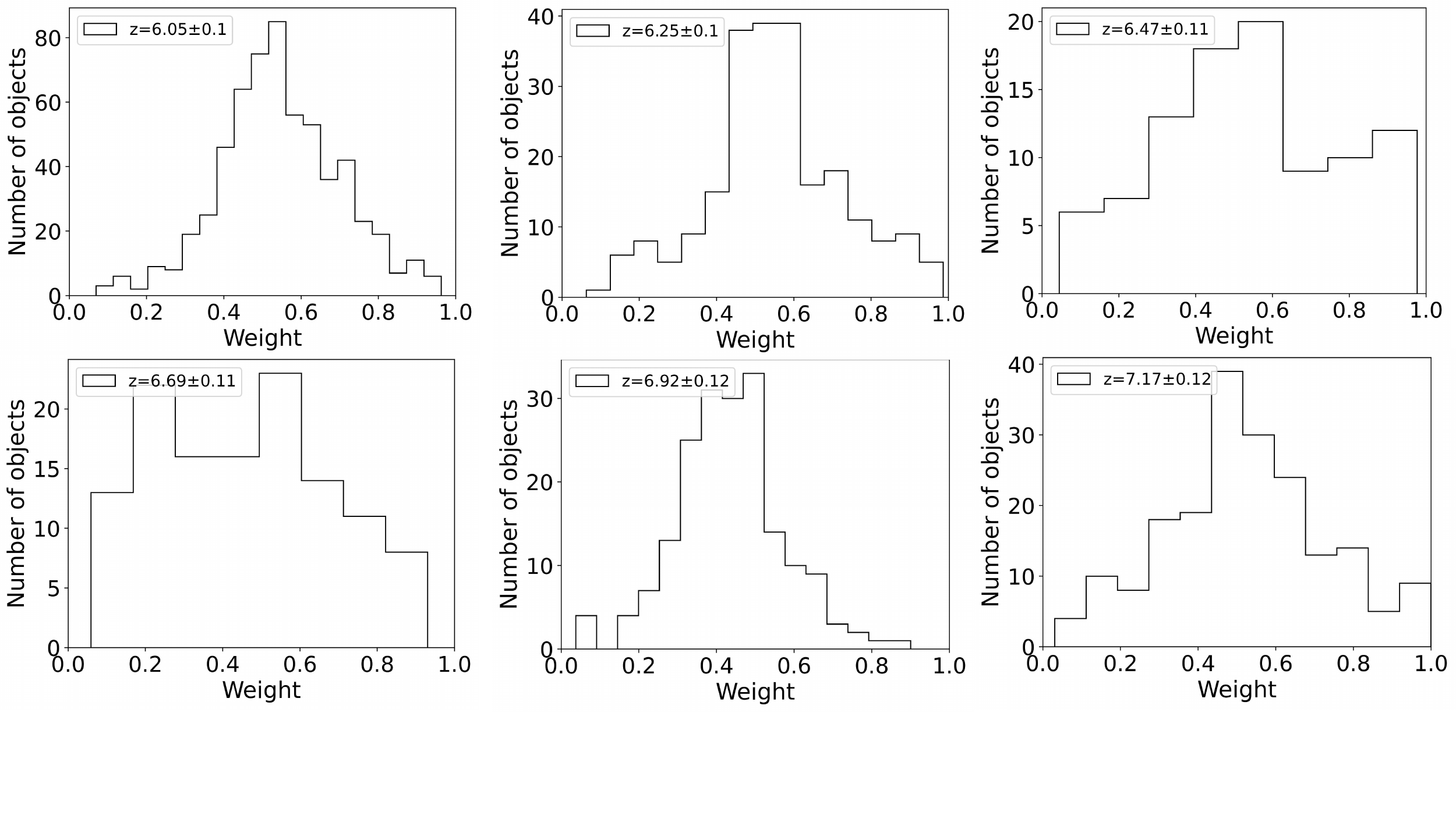}
\end{figure}
\begin{figure}[h]
    \centering
    \vspace{-2.675cm}
    \includegraphics[width=0.95\textwidth]{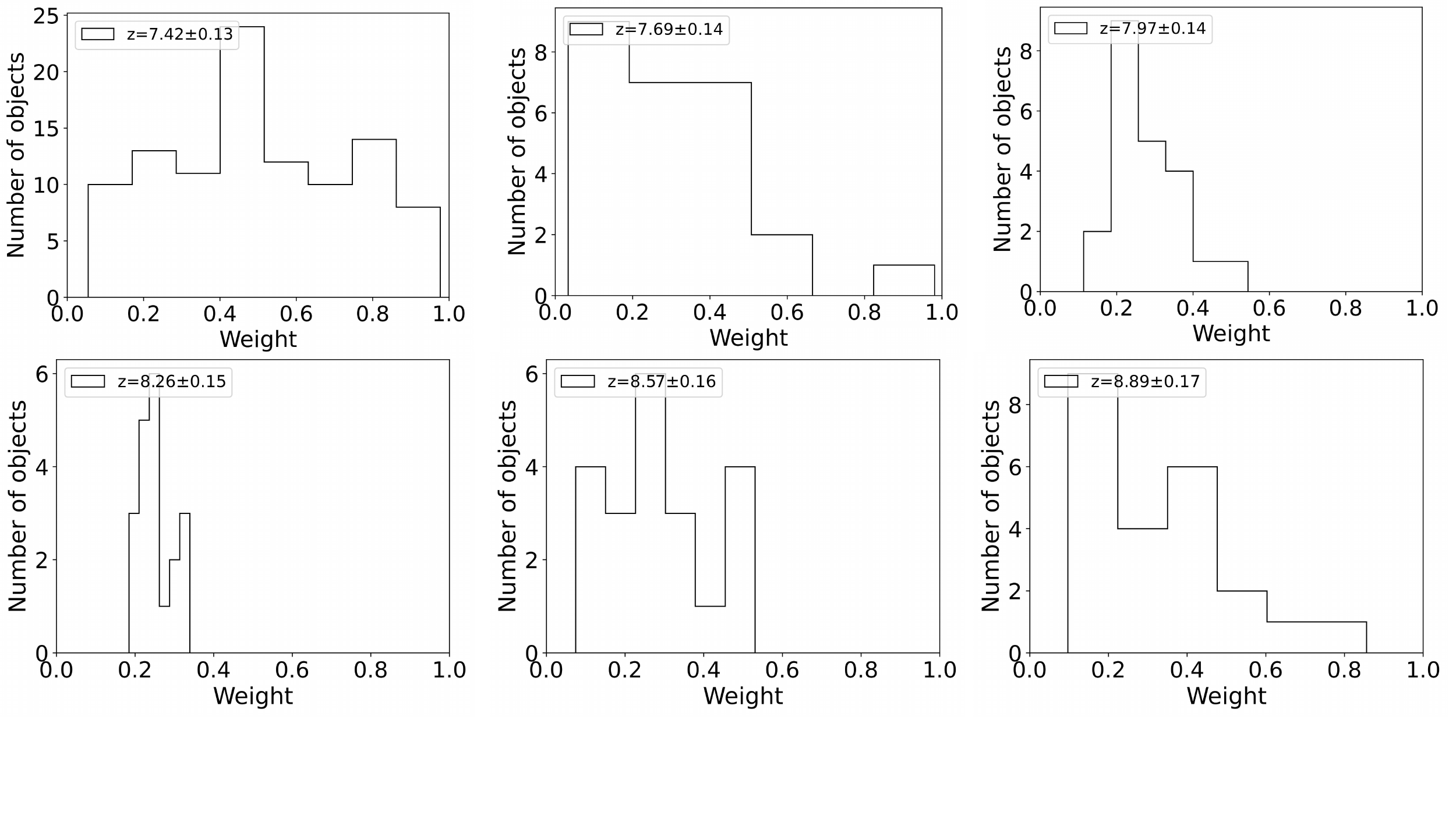}
\end{figure}
\begin{figure}[h]
    \centering
    \vspace{-2.615cm}
    \includegraphics[width=0.95\textwidth]{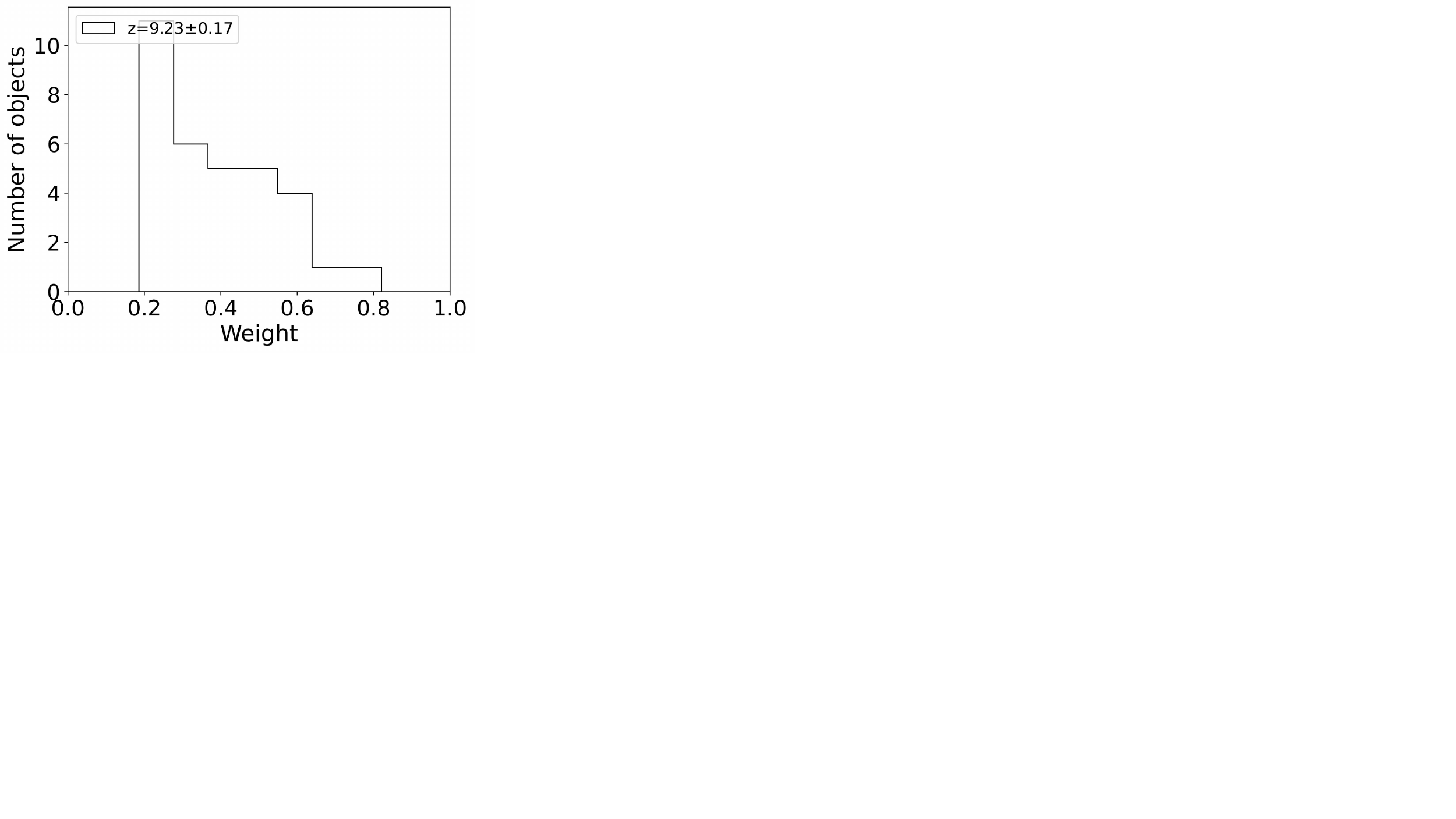}
    \vspace{-5.9cm}
    \caption{Weight distribution of every $z$-bin using LePhare}
    \label{fig:weights}

\end{figure}
\pagebreak

\section{Properties of galaxies associated with $z=6-10$ overdensities in COSMOS}\label{sec:allOD}
In this appendix we list the properties of the galaxies associated with each of the 15 $\geq 4\sigma$ overdensities given in Table \ref{tab:fulldata}.
In Tables \ref{tab:PCz6.05-01data} to \ref{tab:PCz7.69-01data} given below, * indicates galaxies that fullfil the LBG dropout criteria expected for their photometric redshift.  

\bigskip

\begin{figure}[b]
    \centering
    \includegraphics[width=0.8\textwidth]{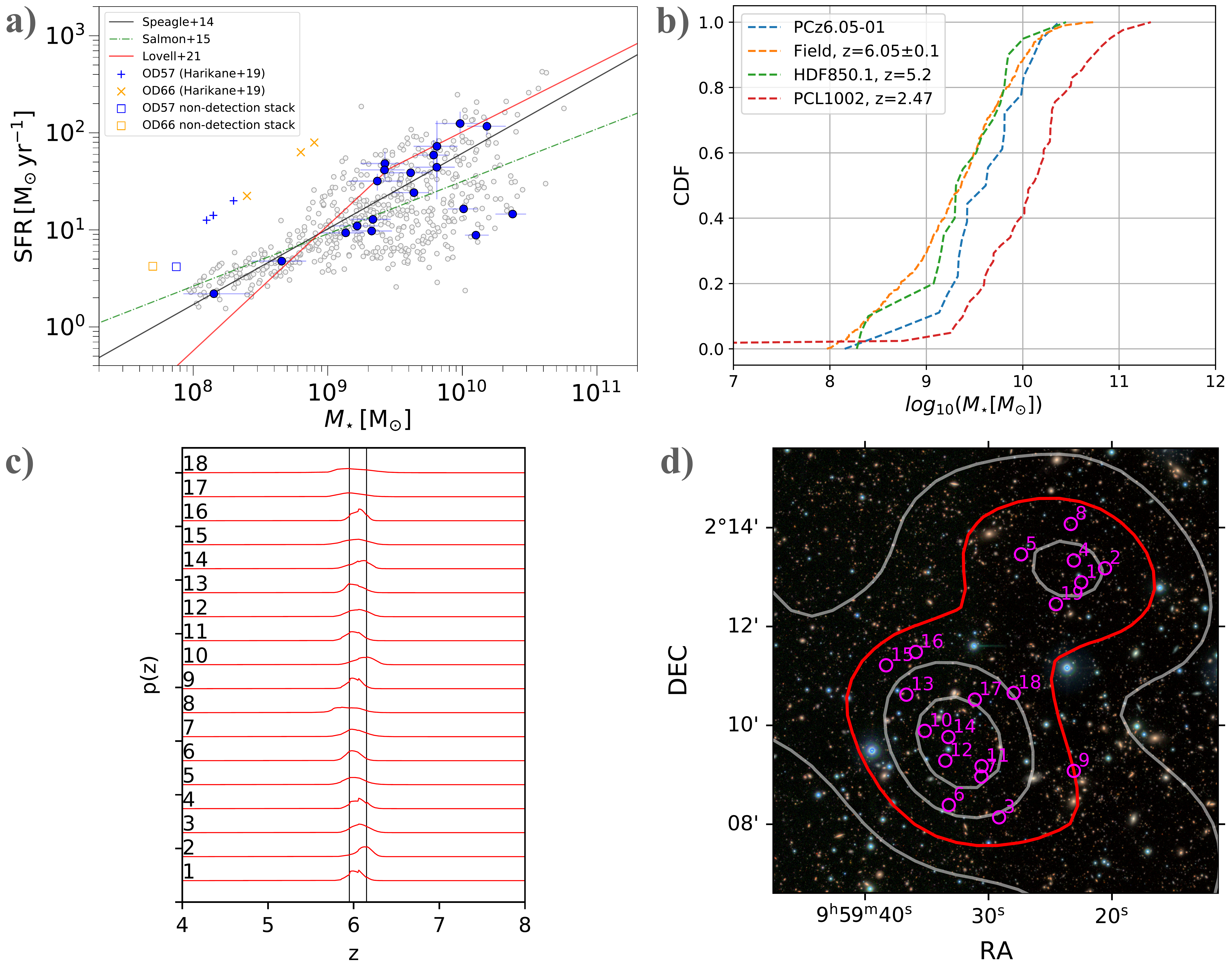}
         \caption{Same figures as {\bf a)} \ref{fig:mainsequence}, {\bf b)} \ref{fig:CDF}, {\bf c)} \ref{pcz6.05-01_pz}, {\bf d)} \ref{fig:adaptive kernel}b, but for PCz6.05-01.}
        \label{fig:PCz6.05-01}
\end{figure}

\noindent {\bf COSMOS2020-PCz6.05-01:}
We identify 19 sources inside the $4\sigma$ contour of this overdensity, of which 14 are $i$-band dropouts (Table \ref{tab:PCz6.05-01data}). The galaxies span a volume of $10.7\times14.3\times66.3\,{\rm cMpc}$ (Fig.~\ref{fig:PCz6.05-01}c). We infer a dark matter halo mass of $M_{\rm DM}\approx 2-13\times 10^{11}\,M_{\odot}$ using the methods described in \ref{subsection:dark-matter-halo}. The stellar mass CDF for the 19 galaxies is shown in Fig.~\ref{fig:PCz6.05-01}b along with the CDF of the field population. A two-sample K-S test of the two CDFs yields a statistic of $0.28$ and a $p$-value of $0.10$.
\begin{table*}[h]
\begin{center}
\caption{Same as table \ref{tab:show19data}}
\label{tab:PCz6.05-01data}
\centering
\begin{tabular}{l l l l l l l l l}
\hline
\hline
\# & ID & R.A.\, (J2000) & Decl.\, (J2000) & $z_{\rm Lephare}$ & $z_{\rm EAZY}$ & $\log(M_{\rm \star, LePhare}/\Msolar)$ & $SFR_{\rm UV}$ & $K_{\rm S}$ \\ 
   &    & {\it hh:mm:ss.ss} & {\it dd:mm:ss.ss} &                   &                &                                                ${\rm dex}$        &                        ${\rm [M_{\odot}\,yr^{-1}]}$                          &     ${\rm [AB\,mag]}$                         \\ \hline
1* & 127337 & 09:59:22.49 & 02:12:53.54 & 6.02$^{+0.09}_{-0.08}$ & 6.06$^{+0.05}_{-0.05}$ & 9.4$^{+0.1}_{-0.2}$ & 48$^{+16}_{-15}$ & 26.57\\
2* & 142959 & 09:59:20.56 & 02:13:10.62 & 6.12$^{+0.08}_{-0.10}$ & 6.12$^{+0.06}_{-0.05}$ & 9.1$^{+0.1}_{-0.2}$ & 9$^{+1}_{-1}$ & 26.17\\
3 & 156780 & 09:59:29.14 & 02:08:08.38 & 6.08$^{+0.10}_{-0.10}$ & 6.08$^{+0.05}_{-0.04}$ & 9.3$^{+0.1}_{-0.2}$ & 13$^{+1}_{-1}$ & 26.68\\
4 & 187778 & 09:59:23.08 & 02:13:20.10 & 6.06$^{+0.09}_{-0.09}$ & 6.09$^{+0.04}_{-0.04}$ & 8.7$^{+0.2}_{-0.2}$ & 5$^{+1}_{-1}$ & 26.28\\
5 & 220530 & 09:59:27.36 & 02:13:27.56 & 6.01$^{+0.12}_{-0.11}$ & 6.03$^{+0.05}_{-0.06}$ & 8.2$^{+0.3}_{-0.2}$ & 2$^{+1}_{-1}$ & 28.08\\
6* & 225263 & 09:59:33.21 & 02:08:23.24 & 6.07$^{+0.14}_{-0.18}$ & 5.91$^{+0.05}_{-0.06}$ & 9.4$^{+0.2}_{-0.2}$ & 32$^{+6}_{-7}$ & 25.47\\
7* & 361608 & 09:59:30.58 & 02:08:57.78 & 6.00$^{+0.09}_{-0.08}$ & 5.99$^{+0.05}_{-0.04}$ & 10.4$^{+0.1}_{-0.1}$ & 15$^{+1}_{-1}$ & 25.46\\
8 & 369661 & 09:59:23.33 & 02:14:04.42 & 6.01$^{+0.13}_{-0.11}$ & 6.05$^{+0.06}_{-0.06}$ & 9.4$^{+0.2}_{-0.2}$ & 41$^{+4}_{-6}$ & 27.05\\
9* & 413243 & 09:59:23.09 & 02:09:04.73 & 6.01$^{+0.08}_{-0.08}$ & 5.99$^{+0.05}_{-0.05}$ & 9.8$^{+0.1}_{-0.2}$ & 44$^{+44}_{-24}$ & 25.80\\
10* & 441761 & 09:59:35.16 & 02:09:53.31 & 5.98$^{+0.12}_{-0.12}$ & 5.90$^{+0.05}_{-0.06}$ & 10.0$^{+0.1}_{-0.2}$ & 125$^{+41}_{-26}$ & 25.43\\
11* & 444487 & 09:59:30.56 & 02:09:10.48 & 6.13$^{+0.11}_{-0.12}$ & 6.08$^{+0.05}_{-0.05}$ & 10.0$^{+0.1}_{-0.1}$ & 16$^{+1}_{-1}$ & 25.89\\
12* & 482804 & 09:59:33.50 & 02:09:17.17 & 6.00$^{+0.09}_{-0.09}$ & 6.00$^{+0.06}_{-0.06}$ & 9.3$^{+0.2}_{-0.1}$ & 10$^{+1}_{-1}$ & 25.32\\
13* & 573604 & 09:59:36.65 & 02:10:37.28 & 6.06$^{+0.15}_{-0.20}$ & 5.94$^{+0.20}_{-0.20}$ & 10.1$^{+0.1}_{-0.1}$ & 9$^{+1}_{-1}$ & 26.57\\
14* & 582186 & 09:59:33.25 & 02:09:45.81 & 6.05$^{+0.11}_{-0.12}$ & 6.10$^{+0.07}_{-0.07}$ & 10.2$^{+0.1}_{-0.1}$ & 117$^{+27}_{-21}$ & 25.88\\
15* & 694706 & 09:59:38.29 & 02:11:12.90 & 6.00$^{+0.11}_{-0.09}$ & 5.98$^{+0.05}_{-0.04}$ & 9.8$^{+0.2}_{-0.2}$ & 73$^{+61}_{-37}$ & 25.83\\
16* & 742465 & 09:59:35.88 & 02:11:29.06 & 6.08$^{+0.10}_{-0.13}$ & 6.05$^{+0.05}_{-0.05}$ & 9.6$^{+0.1}_{-0.2}$ & 24$^{+1}_{-1}$ & 25.08\\
17* & 759747 & 09:59:31.11 & 02:10:31.32 & 5.99$^{+0.15}_{-0.22}$ & 6.04$^{+0.06}_{-0.07}$ & 9.2$^{+0.2}_{-0.2}$ & 11$^{+1}_{-1}$ & 25.97\\
18* & 783817 & 09:59:27.96 & 02:10:39.35 & 6.06$^{+0.07}_{-0.09}$ & 6.03$^{+0.07}_{-0.06}$ & 9.8$^{+0.1}_{-0.1}$ & 59$^{+25}_{-15}$ & 25.48\\
19 & 958367 & 09:59:24.54 & 02:12:27.15 & 5.97$^{+0.23}_{-0.19}$ & 5.96$^{+0.16}_{-0.09}$ & 9.6$^{+0.2}_{-0.2}$ & 39$^{+4}_{-4}$ & 27.17\\
\hline
\end{tabular}
\end{center}
\end{table*}
\clearpage

\bigskip

\begin{figure}[b]
    \centering
    \includegraphics[width=0.8\textwidth]{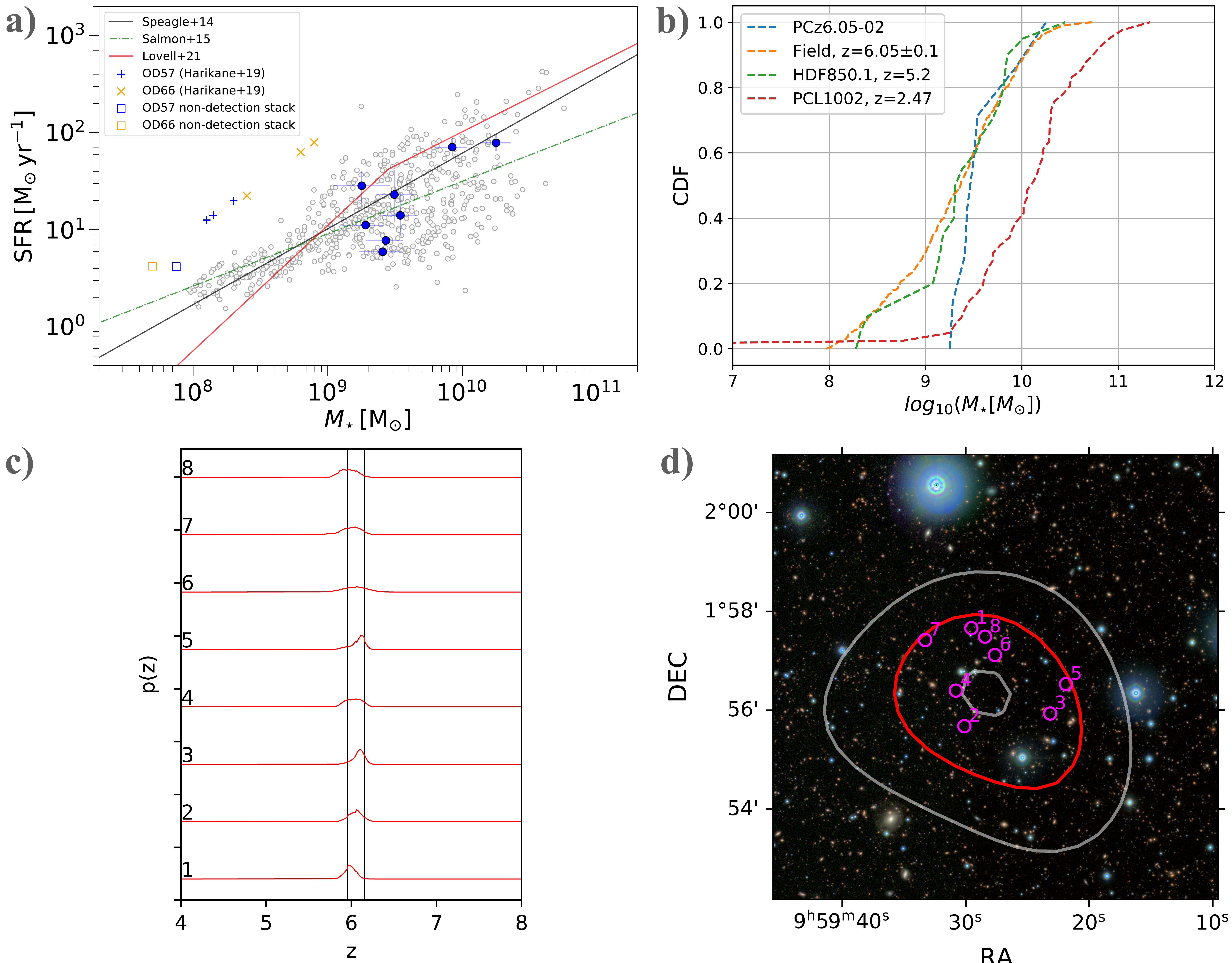}
    \caption{}
    \label{fig:PCz6.05-02}
\end{figure}

\noindent{\bf COSMOS2020-PCz6.05-02:}
We identify 8 sources inside the $4\sigma$ contour of this overdensity, of which 7 are $i$-band dropouts and 1 has no $i$-band data. (Table \ref{tab:PCz6.05-02data}). The galaxies span a volume of $6.9\times4.8\times54.4\,{\rm cMpc}$. (Fig.~\ref{fig:PCz6.05-02}c). We infer a dark matter halo mass of $M_{\rm DM}\approx 8-26\times 10^{11}\,M_{\odot}$ using the methods described in \ref{subsection:dark-matter-halo}. The stellar mass CDF for the 8 galaxies is shown in Fig.~\ref{fig:PCz6.05-02}b along with the CDF of the field population. A two-sample K-S test of the two CDFs yields a statistic of $0.45$ and a $p$-value of $0.05$.
\begin{center}
\begin{table*}[h]
\caption{Same as table \ref{tab:show19data}, but for PCz6.05-02}
\label{tab:PCz6.05-02data}
\centering
\begin{tabular}{l l l l l l l l l}
\hline
\hline
\# & ID & R.A.\, (J2000) & Decl.\, (J2000) & $z_{\rm Lephare}$ & $z_{\rm EAZY}$ & $\log(M_{\rm \star, LePhare}/\Msolar)$ & $SFR_{\rm UV}$ & $K_{\rm S}$ \\ 
   &    & {\it hh:mm:ss.ss} & {\it dd:mm:ss.ss} &                   &                &                                                ${\rm dex}$        &                        ${\rm [M_{\odot}\,yr^{-1}]}$                          &     ${\rm [AB\,mag]}$                         \\ \hline
1 & 28317 & 09:59:29.55 & 01:57:39.75 & 5.99$^{+0.07}_{-0.06}$ & 6.03$^{+0.03}_{-0.04}$ & 9.4$^{+0.1}_{-0.2}$ & 6$^{+1}_{-1}$ & 27.56\\
2* & 644571 & 09:59:30.12 & 01:55:40.81 & 6.05$^{+0.08}_{-0.09}$ & 6.05$^{+0.05}_{-0.04}$ & 9.4$^{+0.1}_{-0.1}$ & 8$^{+1}_{-1}$ & 25.90\\
3* & 683055 & 09:59:23.16 & 01:55:56.08 & 6.09$^{+0.06}_{-0.07}$ & 5.99$^{+0.07}_{-0.04}$ & 9.9$^{+0.1}_{-0.1}$ & 71$^{+19}_{-14}$ & 25.18\\
4* & 743713 & 09:59:30.80 & 01:56:23.88 & 6.03$^{+0.10}_{-0.10}$ & 6.03$^{+0.09}_{-0.06}$ & 9.3$^{+0.2}_{-0.2}$ & 28$^{+12}_{-10}$ & 26.00\\
5* & 763833 & 09:59:21.88 & 01:56:31.77 & 6.10$^{+0.05}_{-0.10}$ & 5.89$^{+0.04}_{-0.04}$ & 10.3$^{+0.1}_{-0.1}$ & 79$^{+17}_{-14}$ & 24.87\\
6* & 828608 & 09:59:27.63 & 01:57:07.11 & 6.01$^{+0.15}_{-0.27}$ & 6.06$^{+0.05}_{-0.06}$ & 9.3$^{+0.1}_{-0.2}$ & 11$^{+1}_{-1}$ & 25.80\\
7* & 862479 & 09:59:33.29 & 01:57:25.31 & 6.00$^{+0.10}_{-0.12}$ & 6.02$^{+0.06}_{-0.09}$ & 9.5$^{+0.1}_{-0.1}$ & 14$^{+10}_{-5}$ & 26.02\\
8* & 871368 & 09:59:28.44 & 01:57:29.36 & 5.97$^{+0.11}_{-0.10}$ & 5.93$^{+0.07}_{-0.06}$ & 9.5$^{+0.1}_{-0.1}$ & 23$^{+18}_{-11}$ & 26.11\\
\hline
\end{tabular}
\end{table*}
\end{center}
\clearpage

\bigskip

\begin{figure}[b]
    \centering
    \includegraphics[width=0.8\textwidth]{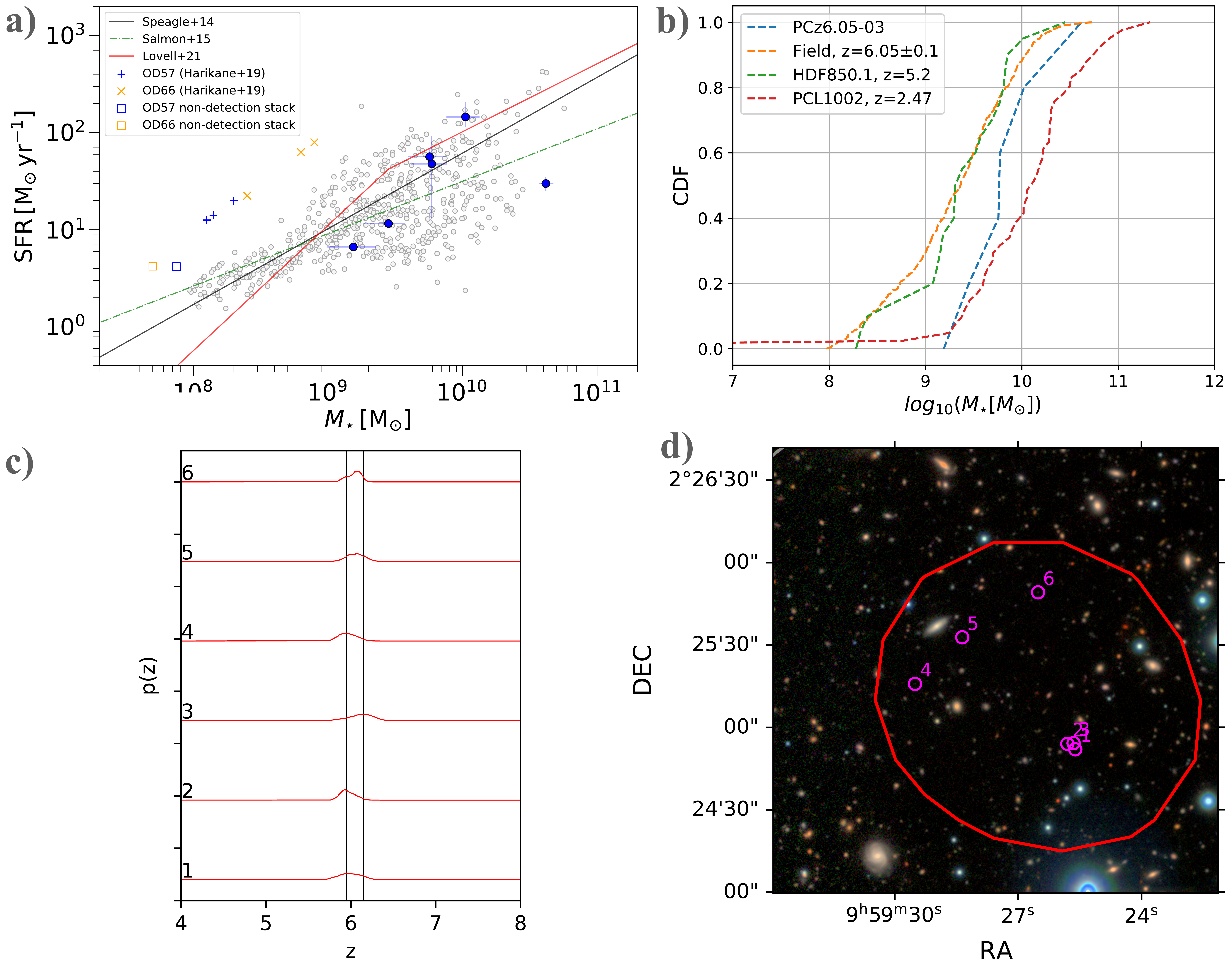}
    \caption{}
        \label{fig:PCz6.05-03}
\end{figure}

\noindent{\bf COSMOS2020-PCz6.05-03:}
We identify 6 sources inside the $4\sigma$ contour of this overdensity, of which 3 are $i$-band dropouts (Table \ref{tab:PCz6.05-03data}). The galaxies span a volume of $2.3\times2.3\times71.4\,{\rm cMpc}$. (Fig.~\ref{fig:PCz6.05-03}c). We infer a dark matter halo mass of 
$M_{\rm DM}\approx 1\times 10^{12}\,M_{\odot}$ using the methods described in \ref{subsection:dark-matter-halo}. The stellar mass CDF for the 8 galaxies is shown in Fig.~\ref{fig:PCz6.05-03}b along with the CDF of the field population. A two-sample K-S test of the two CDFs yields a statistic of $0.43$ and a $p$-value of $0.17$.
\begin{center}
\begin{table*}[h]
\caption{Same as table \ref{tab:show19data}, but for PCz6.05-03}
\label{tab:PCz6.05-03data}
\centering
\begin{tabular}{l l l l l l l l l}
\hline
\hline
\# & ID & R.A.\, (J2000) & Decl.\, (J2000) & $z_{\rm Lephare}$ & $z_{\rm EAZY}$ & $\log(M_{\rm \star, LePhare})$ & $SFR_{\rm UV}$ & $K_{\rm S}$ \\ 
   &    & {\it hh:mm:ss.ss} & {\it dd:mm:ss.ss} &                   &                &                                                ${\rm dex}$        &                        ${\rm [M_{\odot}\,yr^{-1}]}$                          &     ${\rm [AB\,mag]}$                         \\ \hline
1 & 460349 & 09:59:25.61 & 02:24:51.99 & 5.99$^{+0.13}_{-0.13}$ & 6.03$^{+0.08}_{-0.09}$ & 9.8$^{+0.1}_{-0.2}$ & 48$^{+45}_{-35}$ & 26.68\\
2* & 465581 & 09:59:25.80 & 02:24:53.99 & 5.96$^{+0.10}_{-0.08}$ & 5.96$^{+0.04}_{-0.04}$ & 10.0$^{+0.1}_{-0.1}$ & 145$^{+61}_{-31}$ & 25.76\\
3 & 465988 & 09:59:25.65 & 02:24:54.26 & 6.13$^{+0.12}_{-0.15}$ & 6.10$^{+0.07}_{-0.09}$ & 10.6$^{+0.1}_{-0.0}$ & 30$^{+5}_{-5}$ & 25.50\\
4* & 545707 & 09:59:29.51 & 02:25:15.82 & 5.96$^{+0.11}_{-0.10}$ & 5.90$^{+0.06}_{-0.05}$ & 9.2$^{+0.2}_{-0.2}$ & 7$^{+1}_{-1}$ & 26.50\\
5 & 591296 & 09:59:28.36 & 02:25:32.79 & 6.06$^{+0.11}_{-0.11}$ & 6.08$^{+0.06}_{-0.05}$ & 9.8$^{+0.1}_{-0.2}$ & 57$^{+13}_{-12}$ & 26.18\\
6* & 650067 & 09:59:26.52 & 02:25:49.17 & 6.05$^{+0.07}_{-0.11}$ & 5.99$^{+0.06}_{-0.05}$ & 9.4$^{+0.1}_{-0.2}$ & 12$^{+1}_{-1}$ & 25.31\\
\hline
\end{tabular}
\end{table*}
\end{center}
\clearpage

\bigskip

\begin{figure}[b]
    \centering
    \includegraphics[width=0.8\textwidth]{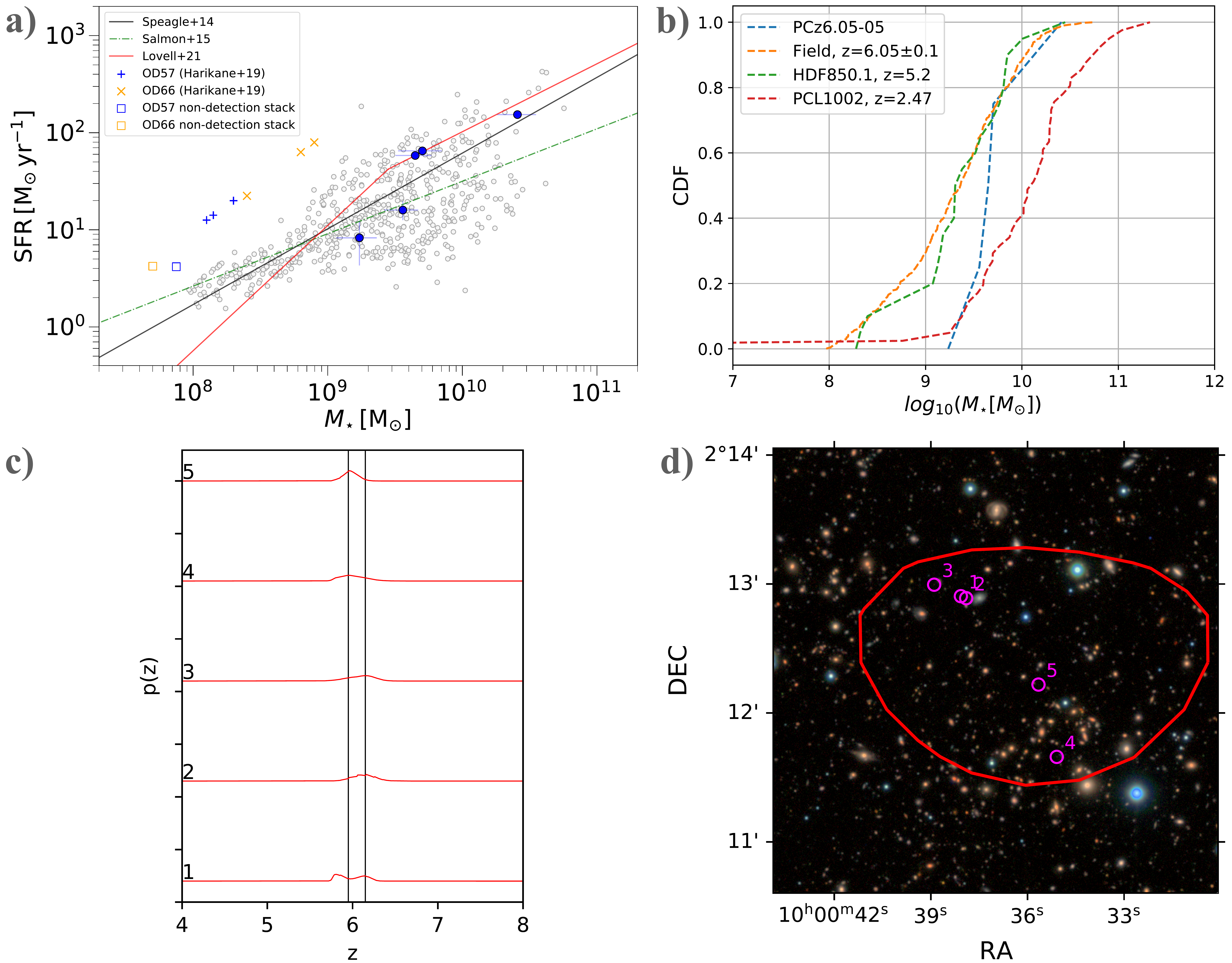}
    \caption{}
        \label{fig:PCz6.05-05}
\end{figure}

\noindent{\bf COSMOS2020-PCz6.05-05:}
We identify 6 sources inside the $4\sigma$ contour of this overdensity, of which 2 are $i$-band dropouts (Table \ref{tab:PCz6.05-05data}). The galaxies span a volume of $2.3\times3.2\times62.7\,{\rm cMpc}$. (Fig.~\ref{fig:PCz6.05-05}c). We infer a dark matter halo mass of 
$M_{\rm DM}\approx 8-58\times 10^{11}\,M_{\odot}$ using the methods described in \ref{subsection:dark-matter-halo}. The stellar mass CDF for the 8 galaxies is shown in Fig.~\ref{fig:PCz6.05-05}b along with the CDF of the field population. A two-sample K-S test of the two CDFs yields a statistic of $0.45$ and a $p$-value of $0.20$.
\begin{center}
\begin{table*}[h]
\caption{Same as table \ref{tab:show19data}, but for PCz6.05-05}
\label{tab:PCz6.05-05data}
\centering
\begin{tabular}{l l l l l l l l l}
\hline
\hline
\# & ID & R.A.\, (J2000) & Decl.\, (J2000) & $z_{\rm Lephare}$ & $z_{\rm EAZY}$ & $\log(M_{\rm \star, LePhare})$ & $SFR_{\rm UV}$ & $K_{\rm S}$ \\ 
   &    & {\it hh:mm:ss.ss} & {\it dd:mm:ss.ss} &                   &                &                                                ${\rm dex}$        &                        ${\rm [M_{\odot}\,yr^{-1}]}$                          &     ${\rm [AB\,mag]}$                         \\ \hline
1* & 84673 & 10:00:38.07 & 02:12:54.33 & 5.98$^{+0.19}_{-0.16}$ & 5.79$^{+0.04}_{-0.04}$ & 9.6$^{+0.1}_{-0.1}$ & 16$^{+4}_{-3}$ & 26.76\\
2 & 90989 & 10:00:37.90 & 02:12:53.40 & 6.13$^{+0.13}_{-0.14}$ & 6.18$^{+0.07}_{-0.09}$ & 9.6$^{+0.1}_{-0.1}$ & 58$^{+5}_{-5}$ & 25.98\\
3 & 102863 & 10:00:38.90 & 02:12:59.67 & 6.10$^{+0.14}_{-0.18}$ & 5.77$^{+0.18}_{-0.49}$ & 9.7$^{+0.1}_{-0.2}$ & 65$^{+6}_{-5}$ & 25.51\\
4 & 862755 & 10:00:35.09 & 02:11:39.56 & 5.99$^{+0.17}_{-0.14}$ & 5.88$^{+0.08}_{-0.08}$ & 10.4$^{+0.1}_{-0.2}$ & 154$^{+5}_{-5}$ & 25.96\\
5* & 936376 & 10:00:35.66 & 02:12:13.21 & 5.98$^{+0.09}_{-0.08}$ & 5.97$^{+0.05}_{-0.05}$ & 9.2$^{+0.1}_{-0.2}$ & 8$^{+4}_{-4}$ & 27.24\\
\hline
\end{tabular}
\end{table*}
\end{center}

\clearpage

\bigskip

\begin{figure}[b]
    \centering
    \includegraphics[width=0.8\textwidth]{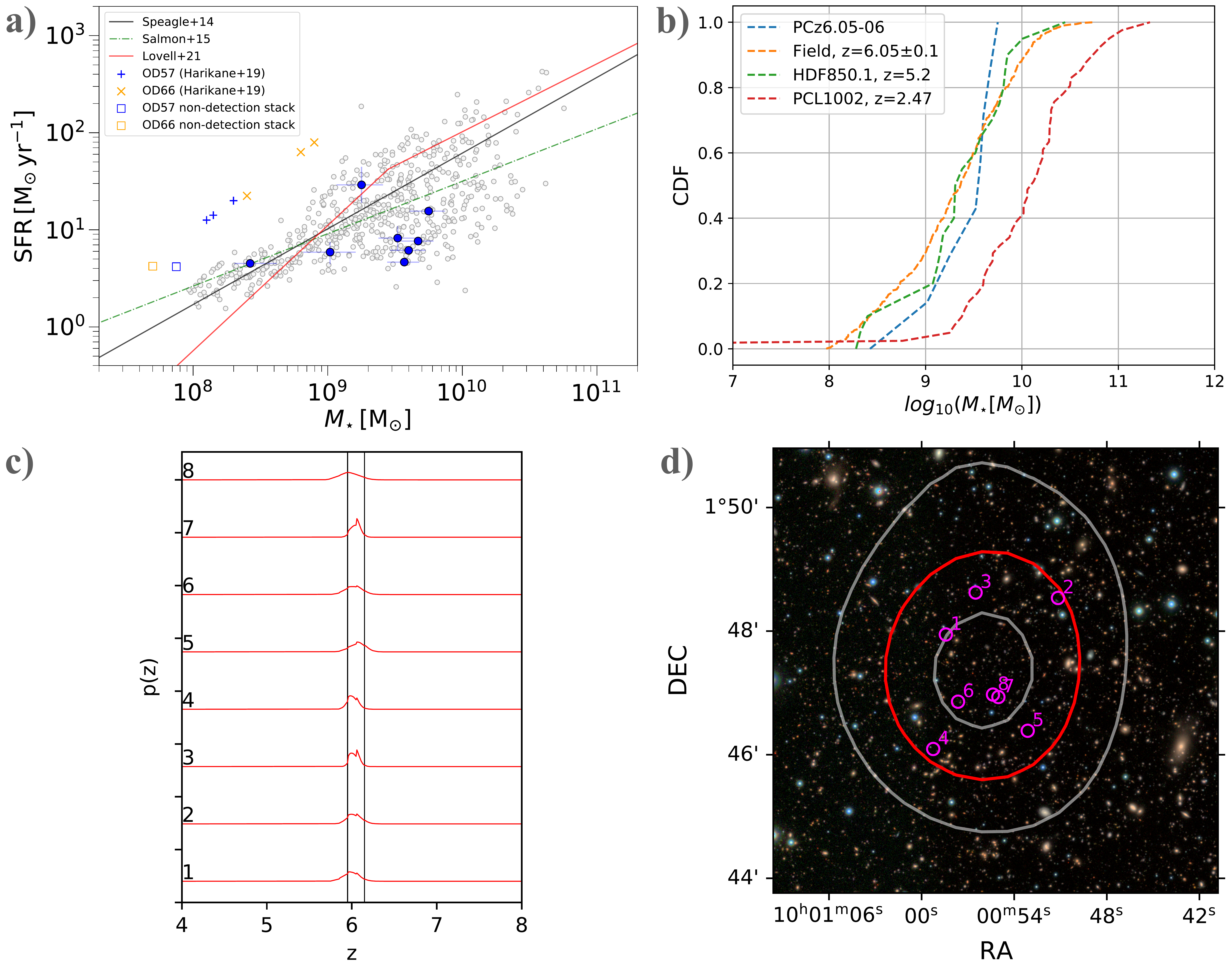}
    \caption{}
        \label{fig:PCz6.05-06}
\end{figure}

\noindent{\bf COSMOS2020-PCz6.05-06:}
We identify 8 sources inside the $4\sigma$ contour of this overdensity, of which 5 are $i$-band dropouts (Table \ref{tab:PCz6.05-06data}). The galaxies span a volume of $4.9\times6.1\times43.0\,{\rm cMpc}$. (Fig.~\ref{fig:PCz6.05-06}c). We infer a dark matter halo mass of 
$M_{\rm DM}\approx 5\times 10^{11}\,M_{\odot}$ using the methods described in \ref{subsection:dark-matter-halo}. The stellar mass CDF for the 8 galaxies is shown in Fig.~\ref{fig:PCz6.05-06}b along with the CDF of the field population. A two-sample K-S test of the two CDFs yields a statistic of $0.24$ and a $p$-value of $0.66$.
\begin{center}
\begin{table*}[h]
\caption{Same as table \ref{tab:show19data}, but for PCz6.05-06}
\label{tab:PCz6.05-06data}
\centering
\begin{tabular}{l l l l l l l l l}
\hline
\hline
\# & ID & R.A.\, (J2000) & Decl.\, (J2000) & $z_{\rm Lephare}$ & $z_{\rm EAZY}$ & $\log(M_{\rm \star, LePhare})$ & $SFR_{\rm UV}$ & $K_{\rm S}$ \\ 
   &    & {\it hh:mm:ss.ss} & {\it dd:mm:ss.ss} &                   &                &                                                ${\rm dex}$        &                        ${\rm [M_{\odot}\,yr^{-1}]}$                          &     ${\rm [AB\,mag]}$                         \\ \hline
1 & 117712 & 10:00:58.43 & 01:47:56.73 & 6.00$^{+0.09}_{-0.09}$ & 6.05$^{+0.04}_{-0.04}$ & 9.6$^{+0.1}_{-0.1}$ & 5$^{+1}_{-1}$ & 31.44\\
2 & 199719 & 10:00:51.17 & 01:48:32.19 & 6.03$^{+0.09}_{-0.08}$ & 6.05$^{+0.04}_{-0.05}$ & 9.5$^{+0.1}_{-0.1}$ & 8$^{+3}_{-2}$ & 26.14\\
3* & 305408 & 10:00:56.50 & 01:48:37.47 & 6.03$^{+0.06}_{-0.05}$ & 6.02$^{+0.08}_{-0.04}$ & 9.3$^{+0.2}_{-0.2}$ & 29$^{+16}_{-10}$ & 25.36\\
4* & 839753 & 10:00:59.25 & 01:46:05.58 & 6.01$^{+0.07}_{-0.06}$ & 6.05$^{+0.03}_{-0.03}$ & 9.7$^{+0.1}_{-0.1}$ & 8$^{+1}_{-1}$ & 26.76\\
5* & 912236 & 10:00:53.13 & 01:46:23.26 & 6.07$^{+0.09}_{-0.10}$ & 6.06$^{+0.04}_{-0.04}$ & 8.4$^{+0.2}_{-0.1}$ & 5$^{+1}_{-1}$ & 26.32\\
6* & 944694 & 10:00:57.64 & 01:46:51.49 & 6.02$^{+0.10}_{-0.10}$ & 6.06$^{+0.05}_{-0.05}$ & 9.6$^{+0.1}_{-0.1}$ & 6$^{+1}_{-1}$ & nan\\
7* & 946072 & 10:00:55.03 & 01:46:56.03 & 6.05$^{+0.05}_{-0.06}$ & 6.04$^{+0.04}_{-0.04}$ & 9.7$^{+0.1}_{-0.1}$ & 16$^{+1}_{-1}$ & 25.56\\
8 & 947720 & 10:00:55.38 & 01:46:58.40 & 5.97$^{+0.12}_{-0.11}$ & 5.96$^{+0.08}_{-0.06}$ & 9.0$^{+0.2}_{-0.2}$ & 6$^{+2}_{-1}$ & 26.21\\
\hline
\end{tabular}
\end{table*}
\end{center}

\clearpage

\bigskip

\begin{figure}[h]
    \centering
    \includegraphics[width=0.8\textwidth]{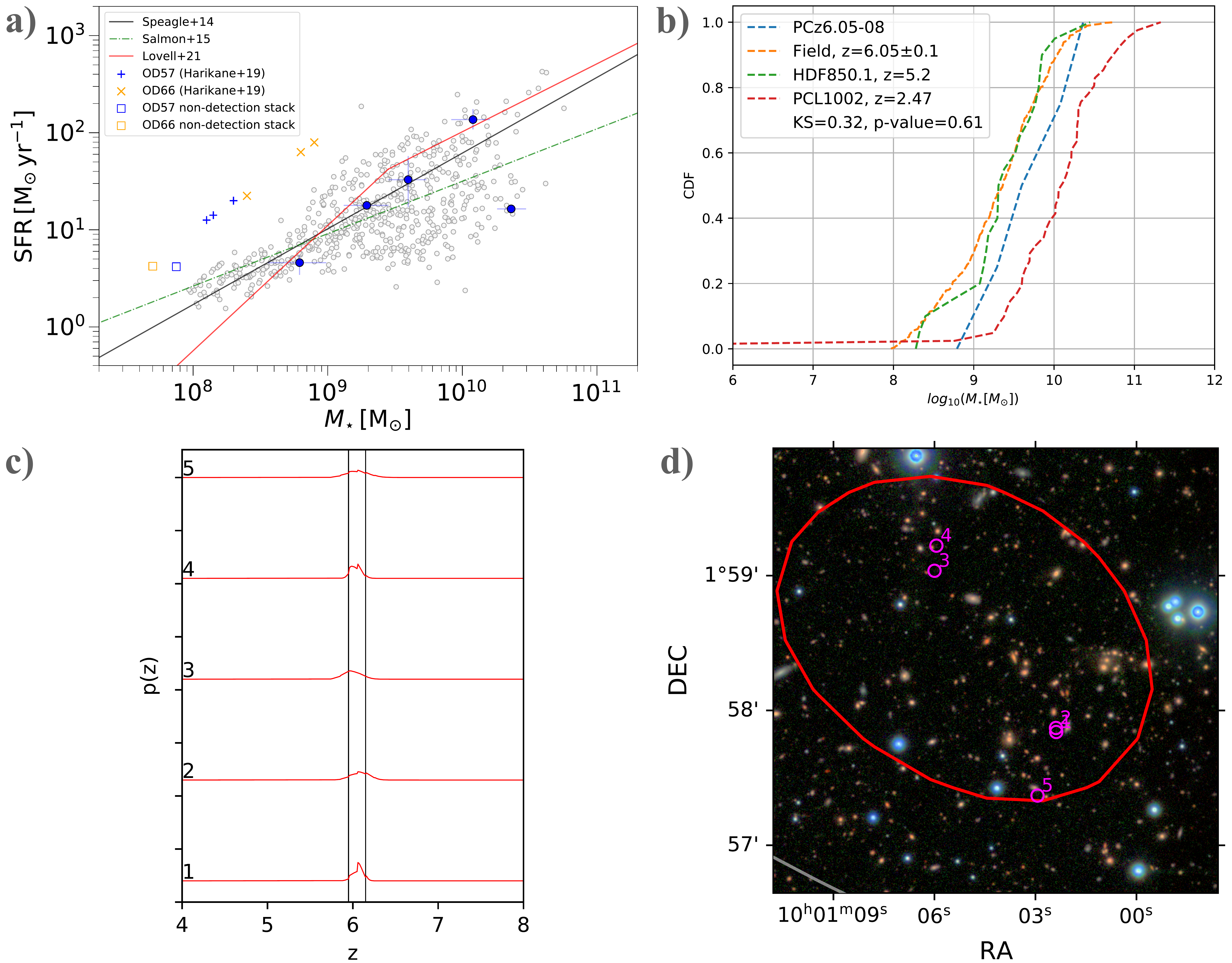}
    \caption{}
        \label{fig:PCz6.05-08}
\end{figure}

\noindent{\bf COSMOS2020-PCz6.05-08:}
We identify 5 sources inside the $4\sigma$ contour of this overdensity, of which 3 are $i$-band dropouts (Table \ref{tab:PCz6.05-08data}). The galaxies span a volume of $2.2\times4.5\times39.2\,{\rm cMpc}$. (Fig.~\ref{fig:PCz6.05-08}c). We infer a dark matter halo mass of 
$M_{\rm DM}\approx 8\times 10^{11}\,M_{\odot}$ using the methods described in \ref{subsection:dark-matter-halo}. The stellar mass CDF for the 5 galaxies is shown in Fig.~\ref{fig:PCz6.05-08}b along with the CDF of the field population. A two-sample K-S test of the two CDFs yields a statistic of $0.32$ and a $p$-value of $0.61$.
\begin{center}
\begin{table*}[h]
\caption{Same as table \ref{tab:show19data}, but for PCz6.05-08}
\label{tab:PCz6.05-08data}
\centering
\begin{tabular}{l l l l l l l l l}
\hline
\hline
\# & ID & R.A.\, (J2000) & Decl.\, (J2000) & $z_{\rm Lephare}$ & $z_{\rm EAZY}$ & $\log(M_{\rm \star, LePhare})$ & $SFR_{\rm UV}$ & $K_{\rm S}$ \\ 
   &    & {\it hh:mm:ss.ss} & {\it dd:mm:ss.ss} &                   &                &                                                ${\rm dex}$        &                        ${\rm [M_{\odot}\,yr^{-1}]}$                          &     ${\rm [AB\,mag]}$                         \\ \hline
1* & 85122 & 10:01:02.38 & 01:57:50.42 & 6.07$^{+0.05}_{-0.07}$ & 6.05$^{+0.06}_{-0.04}$ & 10.1$^{+0.1}_{-0.2}$ & 137$^{+39}_{-29}$ & 24.93\\
2 & 89385 & 10:01:02.39 & 01:57:52.21 & 6.09$^{+0.11}_{-0.11}$ & 6.15$^{+0.06}_{-0.06}$ & 9.6$^{+0.1}_{-0.1}$ & 33$^{+25}_{-16}$ & nan\\
3* & 333962 & 10:01:06.00 & 01:59:02.17 & 5.99$^{+0.11}_{-0.10}$ & 5.94$^{+0.06}_{-0.06}$ & 9.3$^{+0.2}_{-0.2}$ & 18$^{+1}_{-2}$ & 26.03\\
4* & 368321 & 10:01:05.94 & 01:59:13.33 & 6.03$^{+0.07}_{-0.06}$ & 6.03$^{+0.05}_{-0.06}$ & 10.4$^{+0.1}_{-0.1}$ & 16$^{+1}_{-1}$ & 25.02\\
5 & 916028 & 10:01:02.94 & 01:57:22.12 & 6.05$^{+0.13}_{-0.13}$ & 6.09$^{+0.07}_{-0.06}$ & 8.8$^{+0.2}_{-0.2}$ & 5$^{+1}_{-1}$ & 26.59\\
\hline
\end{tabular}
\end{table*}
\end{center}

\clearpage

\bigskip

\begin{figure}[h]
    \centering
    \includegraphics[width=0.8\textwidth]{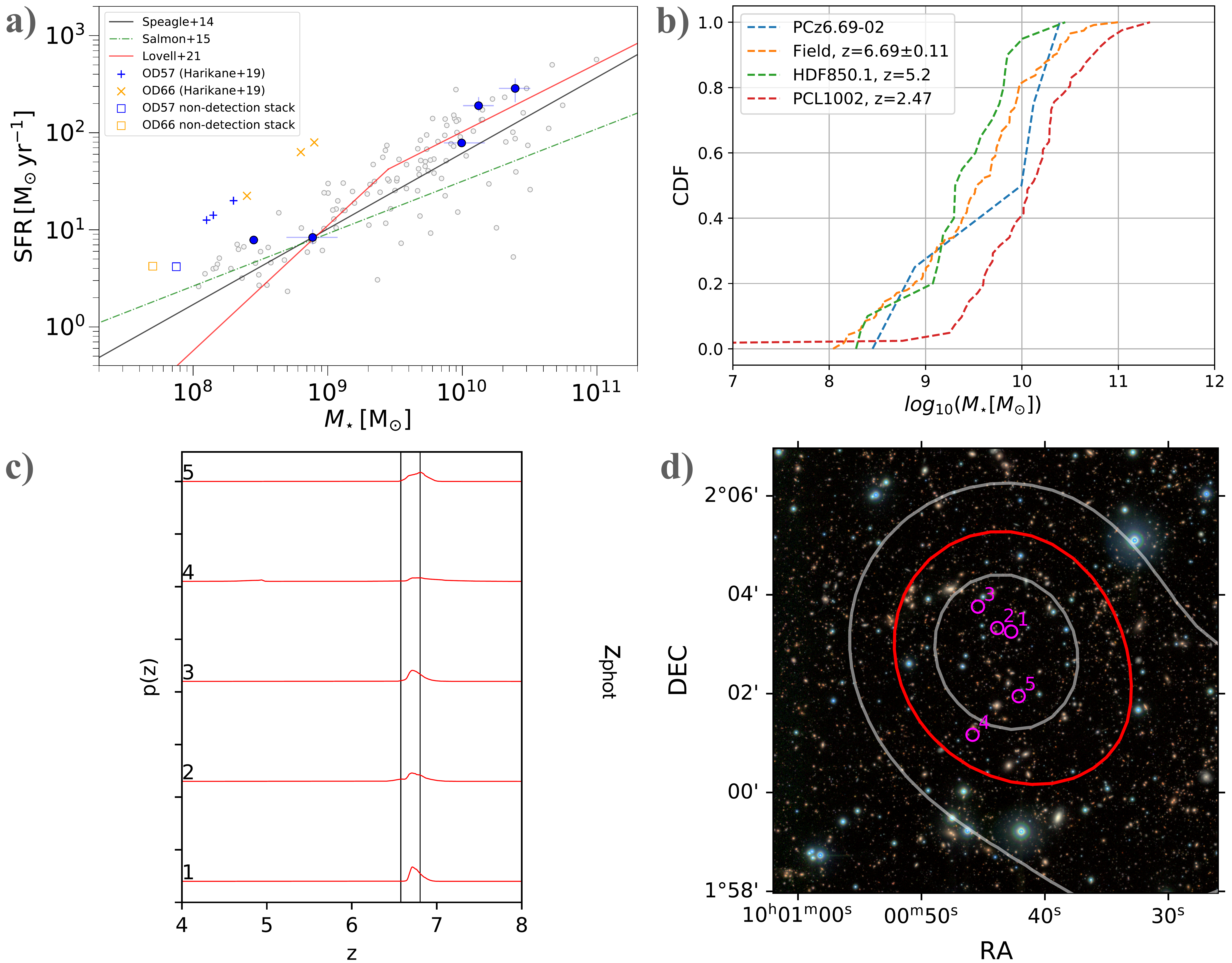}
    \caption{}
        \label{fig:PCz6.69-02}
\end{figure}

\noindent{\bf COSMOS2020-PCz6.69-02:}
We identify 5 sources inside the $4\sigma$ contour of this overdensity, of which 2 are $z$-band dropouts (Table \ref{tab:PCz6.69-02data}). The galaxies span a volume of $2.3\times6.4\times15.0\,{\rm cMpc}$. (Fig.~\ref{fig:PCz6.69-02}c). We infer a dark matter halo mass of 
$M_{\rm DM}\approx 1\times 10^{12}\,M_{\odot}$ using the methods described in \ref{subsection:dark-matter-halo}. The stellar mass CDF for the 5 galaxies is shown in Fig.~\ref{fig:PCz6.69-02}b along with the CDF of the field population. A two-sample K-S test of the two CDFs yields a statistic of $0.41$ and a $p$-value of $0.30$.
\begin{center}
\begin{table*}[h]
\caption{Same as table \ref{tab:show19data}, but for PCz6.69-02}
\label{tab:PCz6.69-02data}
\centering
\begin{tabular}{l l l l l l l l l}
\hline
\hline
\# & ID & R.A.\, (J2000) & Decl.\, (J2000) & $z_{\rm Lephare}$ & $z_{\rm EAZY}$ & $\log(M_{\rm \star, LePhare})$ & $SFR_{\rm UV}$ & $K_{\rm S}$ \\ 
   &    & {\it hh:mm:ss.ss} & {\it dd:mm:ss.ss} &                   &                &                                                ${\rm dex}$        &                        ${\rm [M_{\odot}\,yr^{-1}]}$                          &     ${\rm [AB\,mag]}$                         \\ \hline
1 & 188158 & 10:00:42.73 & 02:03:15.34 & 6.75$^{+0.08}_{-0.05}$ & 6.91$^{+0.01}_{-0.02}$ & 8.5$^{+0.1}_{-0.1}$ & 8$^{+1}_{-1}$ & 25.27\\
2* & 231335 & 10:00:43.87 & 02:03:19.66 & 6.74$^{+0.12}_{-0.10}$ & 6.74$^{+0.04}_{-0.04}$ & 10.4$^{+0.1}_{-0.1}$ & 286$^{+76}_{-80}$ & 25.48\\
3 & 318344 & 10:00:45.43 & 02:03:45.64 & 6.75$^{+0.10}_{-0.07}$ & 6.74$^{+0.04}_{-0.03}$ & 8.9$^{+0.2}_{-0.2}$ & 8$^{+2}_{-1}$ & 26.16\\
4 & 750274 & 10:00:45.86 & 02:01:10.13 & 6.78$^{+0.29}_{-1.88}$ & 6.75$^{+0.02}_{-0.02}$ & 10.0$^{+0.2}_{-0.1}$ & 78$^{+6}_{-8}$ & nan\\
5* & 894738 & 10:00:42.13 & 02:01:56.92 & 6.78$^{+0.09}_{-0.10}$ & 6.70$^{+0.07}_{-0.03}$ & 10.1$^{+0.1}_{-0.1}$ & 190$^{+43}_{-24}$ & 25.10\\
\hline
\end{tabular}
\end{table*}
\end{center}

\clearpage

\bigskip

\begin{figure}[h]
    \centering
    \includegraphics[width=0.8\textwidth]{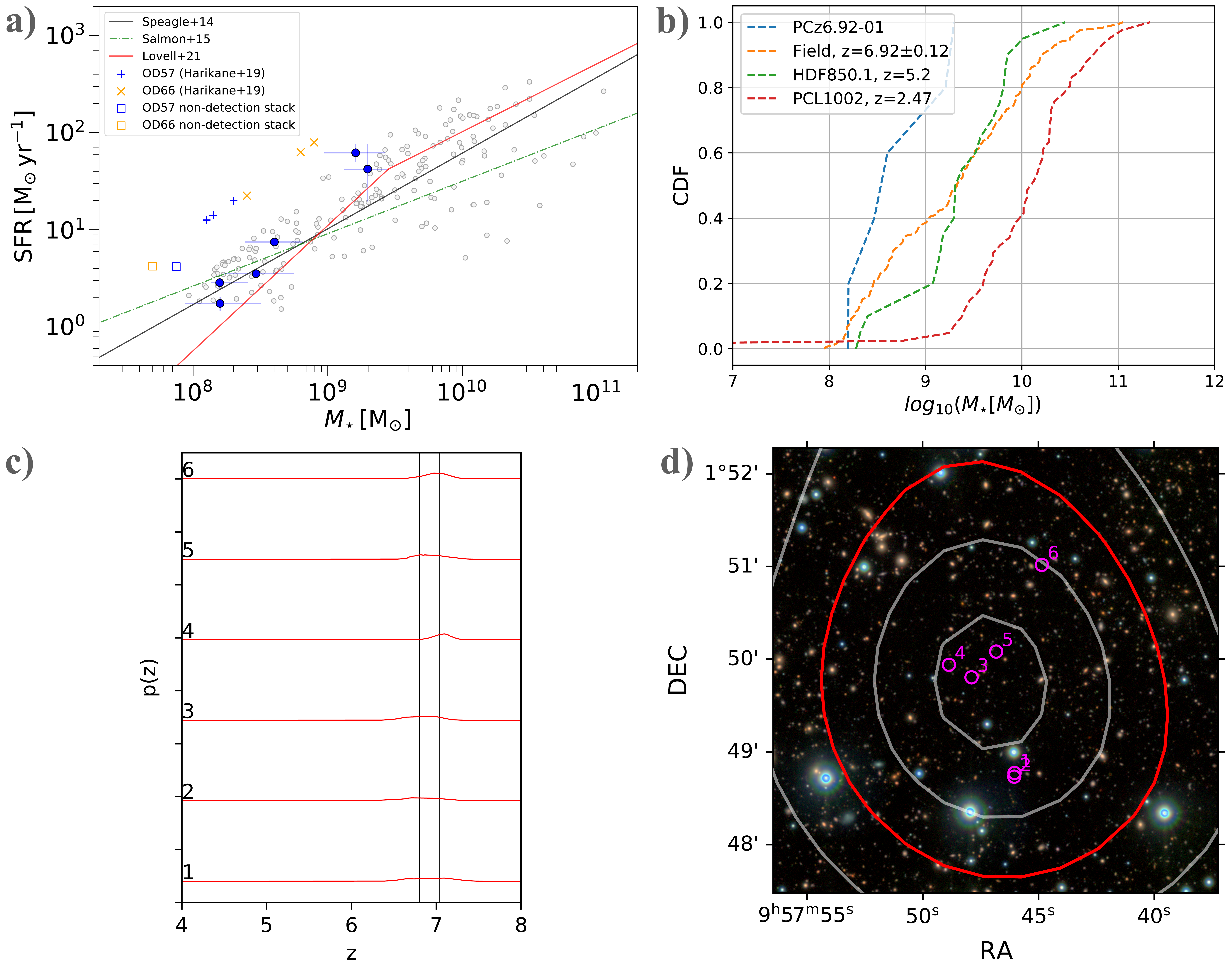}
    \caption{}
        \label{fig:PCz6.92-01}
\end{figure}

\noindent{\bf COSMOS2020-PCz6.92-01:}
We identify 6 sources inside the $4\sigma$ contour of this overdensity, of which 0 are $z$-band dropouts (Table \ref{tab:PCz6.92-01data}). The galaxies span a volume of $2.5\times5.7\times56.8\,{\rm cMpc}$. (Fig.~\ref{fig:PCz6.92-01}c). We infer a dark matter halo mass of 
$M_{\rm DM}\approx 9-39\times 10^{10}\,M_{\odot}$ using the methods described in \ref{subsection:dark-matter-halo}. The stellar mass CDF for the 6 galaxies is shown in Fig.~\ref{fig:PCz6.92-01}b along with the CDF of the field population. A two-sample K-S test of the two CDFs yields a statistic of $0.51$ and a $p$-value of $0.06$.
\begin{center}
\begin{table*}[h]
\caption{Same as table \ref{tab:show19data}, but for PCz6.92-01}
\label{tab:PCz6.92-01data}
\begin{tabular}{l l l l l l l l l}
\hline
\hline
\# & ID & R.A.\, (J2000) & Decl.\, (J2000) & $z_{\rm Lephare}$ & $z_{\rm EAZY}$ & $\log(M_{\rm \star, LePhare})$ & $SFR_{\rm UV}$ & $K_{\rm S}$ \\ 
   &    & {\it hh:mm:ss.ss} & {\it dd:mm:ss.ss} &                   &                &                                                ${\rm dex}$        &                        ${\rm [M_{\odot}\,yr^{-1}]}$                          &     ${\rm [AB\,mag]}$                         \\ \hline
1 & 184953 & 09:57:46.04 & 01:48:46.31 & 6.93$^{+0.24}_{-0.28}$ & 7.10$^{+0.06}_{-0.08}$ & 9.2$^{+0.2}_{-0.2}$ & 62$^{+14}_{-12}$ & 26.03\\
2 & 185324 & 09:57:46.05 & 01:48:43.93 & 6.87$^{+0.38}_{-0.27}$ & 6.98$^{+4.73}_{-0.13}$ & 8.2$^{+0.3}_{-0.3}$ & 2$^{+1}_{-1}$ & nan\\
3 & 383012 & 09:57:47.89 & 01:49:48.22 & 6.84$^{+0.19}_{-0.22}$ & 7.04$^{+0.05}_{-0.06}$ & 8.5$^{+0.3}_{-0.2}$ & 4$^{+1}_{-1}$ & nan\\
4 & 426584 & 09:57:48.87 & 01:49:56.33 & 7.01$^{+0.13}_{-6.44}$ & 7.11$^{+0.05}_{-0.05}$ & 8.6$^{+0.2}_{-0.2}$ & 7$^{+1}_{-1}$ & nan\\
5 & 450133 & 09:57:46.82 & 01:50:04.89 & 6.93$^{+0.20}_{-0.17}$ & 6.76$^{+0.03}_{-0.02}$ & 9.3$^{+0.1}_{-0.2}$ & 42$^{+35}_{-23}$ & 27.92\\
6 & 651664 & 09:57:44.85 & 01:51:00.99 & 6.99$^{+0.14}_{-0.15}$ & 6.91$^{+0.03}_{-0.05}$ & 8.2$^{+0.2}_{-0.1}$ & 3$^{+1}_{-1}$ & 26.01\\
\hline
\end{tabular}
\end{table*}
\end{center}

\clearpage

\bigskip

\begin{figure}[h]
    \centering
    \includegraphics[width=0.8\textwidth]{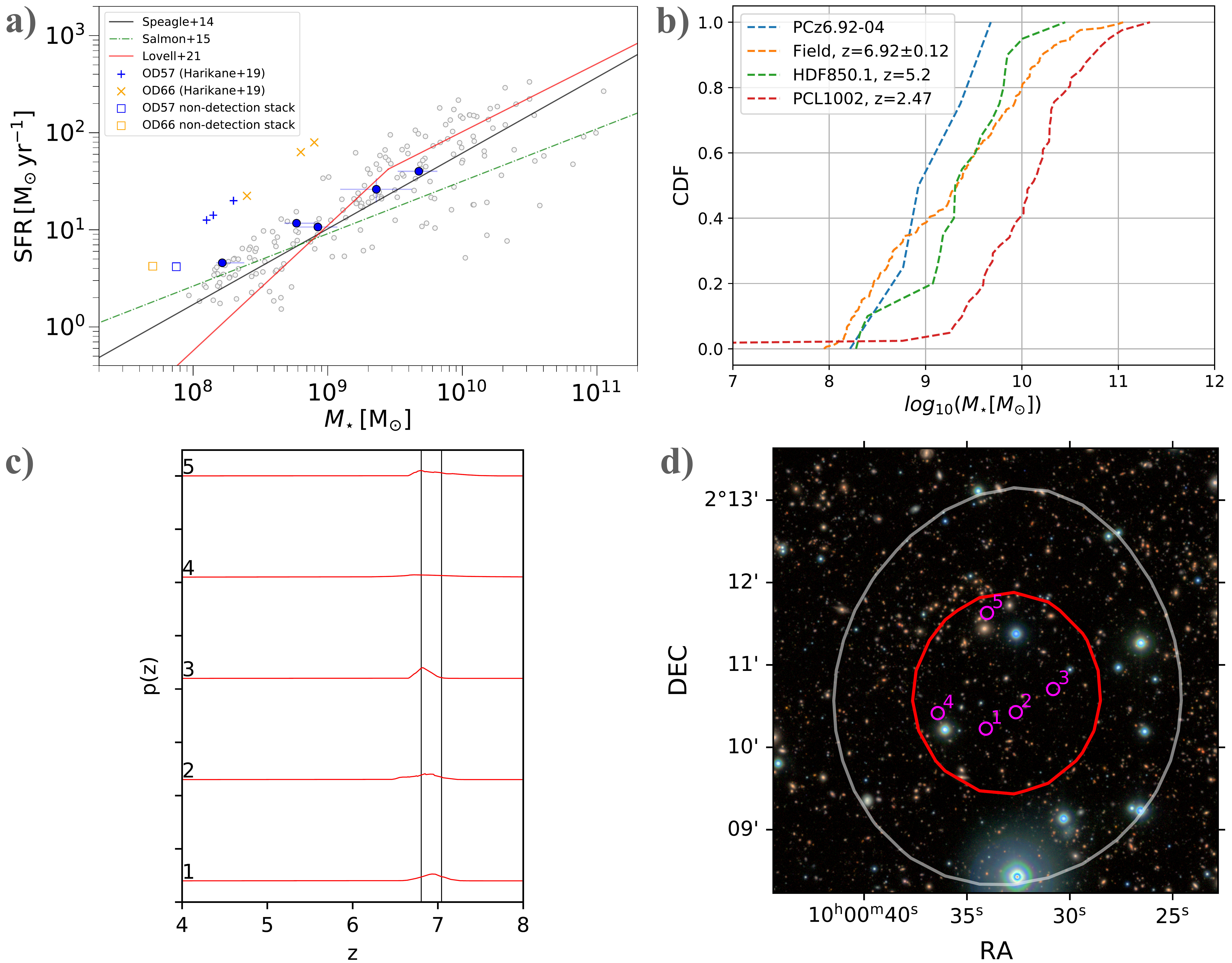}
    \caption{}
        \label{fig:PCz6.92-04}
\end{figure}

\noindent{\bf COSMOS2020-PCz6.92-04}
We identify 5 sources inside the $4\sigma$ contour of this overdensity, of which 2 are $z$-band dropouts (Table \ref{tab:PCz6.92-04data}). The galaxies span a volume of $3.5\times3.5\times37.4\,{\rm cMpc}$. (Fig.~\ref{fig:PCz6.92-04}c). We infer a dark matter halo mass of 
$M_{\rm DM}\approx 2-5\times 10^{11}\,M_{\odot}$ using the methods described in \ref{subsection:dark-matter-halo}. The stellar mass CDF for the 5 galaxies is shown in Fig.~\ref{fig:PCz6.92-04}b along with the CDF of the field population. A two-sample K-S test of the two CDFs yields a statistic of $0.34$ and a $p$-value of $0.51$.
\begin{table*}[h]
\caption{Same as table \ref{tab:show19data}, but for PCz6.92-04}
\label{tab:PCz6.92-04data}
\centering
\begin{tabular}{l l l l l l l l l}
\hline
\hline
\# & ID & R.A.\, (J2000) & Decl.\, (J2000) & $z_{\rm Lephare}$ & $z_{\rm EAZY}$ & $\log(M_{\rm \star, LePhare})$ & $SFR_{\rm UV}$ & $K_{\rm S}$ \\ 
   &    & {\it hh:mm:ss.ss} & {\it dd:mm:ss.ss} &                   &                &                                                ${\rm dex}$        &                        ${\rm [M_{\odot}\,yr^{-1}]}$                          &     ${\rm [AB\,mag]}$                         \\ \hline
1 & 445086 & 10:00:34.08 & 02:10:13.61 & 6.93$^{+0.13}_{-0.13}$ & 6.94$^{+0.03}_{-0.03}$ & 8.2$^{+0.2}_{-0.1}$ & 5$^{+1}_{-1}$ & 27.62\\
2* & 493215 & 10:00:32.63 & 02:10:25.58 & 6.85$^{+0.14}_{-0.19}$ & 6.91$^{+0.03}_{-0.03}$ & 8.9$^{+0.1}_{-0.2}$ & 11$^{+1}_{-1}$ & 25.66\\
3* & 567863 & 10:00:30.81 & 02:10:42.46 & 6.83$^{+0.09}_{-0.08}$ & 6.76$^{+0.03}_{-0.02}$ & 8.8$^{+0.2}_{-0.1}$ & 12$^{+1}_{-1}$ & 25.67\\
4 & 610764 & 10:00:36.42 & 02:10:24.91 & 6.89$^{+0.64}_{-0.41}$ & 6.75$^{+0.05}_{-0.05}$ & 9.4$^{+0.3}_{-0.3}$ & 26$^{+8}_{-7}$ & nan\\
5 & 750581 & 10:00:34.03 & 02:11:37.84 & 6.94$^{+0.26}_{-0.17}$ & 6.76$^{+0.02}_{-0.02}$ & 9.7$^{+0.1}_{-0.2}$ & 40$^{+3}_{-3}$ & 26.63\\
\hline
\end{tabular}
\end{table*}

\clearpage

\bigskip

\begin{figure}[h]
    \centering
    \includegraphics[width=0.8\textwidth]{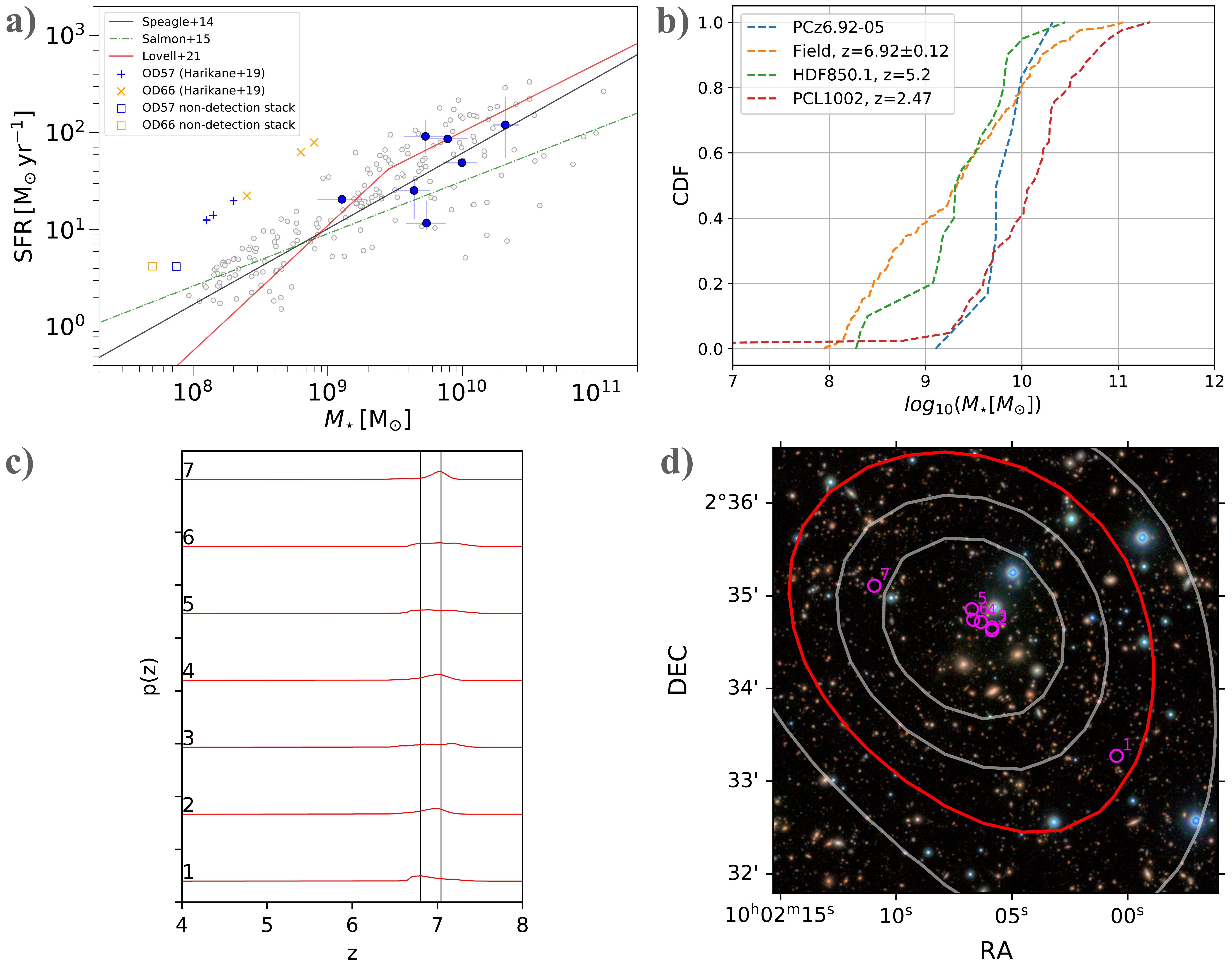}
    \caption{}
        \label{fig:PCz6.92-05}
\end{figure}

\noindent{\bf COSMOS2020-PCz6.92-05:}
We identify 7 sources inside the $4\sigma$ contour of this overdensity, of which 3 are $z$-band dropouts (Table \ref{tab:PCz6.92-05data}). The galaxies span a volume of $6.6\times4.6\times55.9\,{\rm cMpc}$. (Fig.~\ref{fig:PCz6.92-05}c). We infer a dark matter halo mass of 
$M_{\rm DM}\approx 1\times 10^{12}\,M_{\odot}$ using the methods described in \ref{subsection:dark-matter-halo}. The stellar mass CDF for the 7 galaxies is shown in Fig.~\ref{fig:PCz6.92-05}b along with the CDF of the field population. A two-sample K-S test of the two CDFs yields a statistic of $0.50$ and a $p$-value of $0.05$.
\begin{table*}[h]
\caption{Same as table \ref{tab:show19data}, but for PCz6.92-05}
\label{tab:PCz6.92-05data}
\centering
\begin{tabular}{l l l l l l l l l}
\hline
\hline
\# & ID & R.A.\, (J2000) & Decl.\, (J2000) & $z_{\rm Lephare}$ & $z_{\rm EAZY}$ & $\log(M_{\rm \star, LePhare})$ & $SFR_{\rm UV}$ & $K_{\rm S}$ \\ 
   &    & {\it hh:mm:ss.ss} & {\it dd:mm:ss.ss} &                   &                &                                                ${\rm dex}$        &                        ${\rm [M_{\odot}\,yr^{-1}]}$                          &     ${\rm [AB\,mag]}$                         \\ \hline
1* & 134434 & 10:02:00.48 & 02:33:16.52 & 6.87$^{+0.25}_{-0.14}$ & 6.83$^{+0.03}_{-0.07}$ & 9.7$^{+0.1}_{-0.2}$ & 92$^{+45}_{-32}$ & 25.63\\
2 & 462038 & 10:02:05.87 & 02:34:37.81 & 6.93$^{+0.13}_{-0.20}$ & 6.93$^{+0.09}_{-0.11}$ & 9.7$^{+0.1}_{-0.2}$ & 12$^{+8}_{-2}$ & 26.60\\
3* & 507497 & 10:02:05.88 & 02:34:39.53 & 6.96$^{+0.24}_{-0.26}$ & 6.89$^{+0.04}_{-0.05}$ & 9.1$^{+0.2}_{-0.2}$ & 21$^{+2}_{-2}$ & 25.49\\
4 & 509093 & 10:02:06.34 & 02:34:43.20 & 6.96$^{+0.13}_{-0.19}$ & 6.51$^{+0.06}_{-0.05}$ & 10.0$^{+0.1}_{-0.1}$ & 49$^{+4}_{-4}$ & 25.44\\
5* & 554918 & 10:02:06.74 & 02:34:51.33 & 6.97$^{+0.25}_{-0.21}$ & 6.94$^{+0.03}_{-0.02}$ & 9.9$^{+0.2}_{-0.2}$ & 86$^{+6}_{-6}$ & 25.07\\
6 & 555718 & 10:02:06.68 & 02:34:44.61 & 7.03$^{+0.22}_{-0.22}$ & 6.76$^{+0.04}_{-0.03}$ & 10.3$^{+0.1}_{-0.1}$ & 120$^{+114}_{-65}$ & 26.02\\
7 & 592899 & 10:02:10.95 & 02:35:06.62 & 7.00$^{+0.10}_{-0.17}$ & 7.04$^{+0.05}_{-0.05}$ & 9.6$^{+0.1}_{-0.1}$ & 25$^{+8}_{-12}$ & 26.89\\
\hline
\end{tabular}
\end{table*}

\clearpage

\begin{figure}[h]
    \centering
    \includegraphics[width=0.8\textwidth]{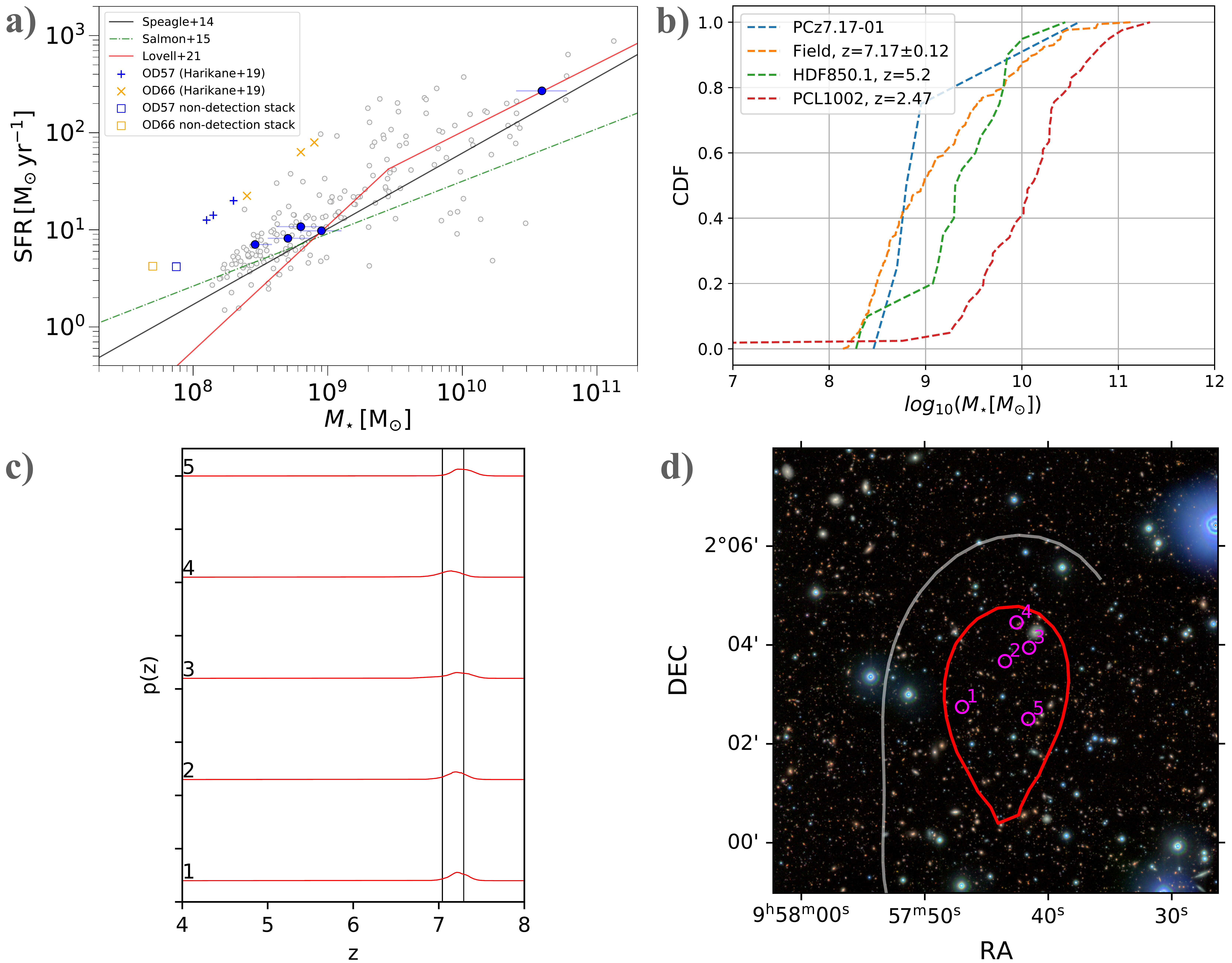}
    \caption{}
        \label{fig:PCz7.17-01}
\end{figure}

\noindent{\bf COSMOS2020-PCz7.17-01:}
We identify 5 sources inside the $4\sigma$ contour of this overdensity, of which 0 are $z$-band dropouts (Table \ref{tab:PCz7.17-01data}). The galaxies span a volume of $3.4\times4.9\times52.6\,{\rm cMpc}$. (Fig.~\ref{fig:PCz7.17-01}c). We infer a dark matter halo mass of 
$M_{\rm DM}\approx 8\times 10^{11}\,M_{\odot}$ using the methods described in \ref{subsection:dark-matter-halo}. The stellar mass CDF for the 5 galaxies is shown in Fig.~\ref{fig:PCz7.17-01}b along with the CDF of the field population. A two-sample K-S test of the two CDFs yields a statistic of $0.31$ and a $p$-value of $0.64$.
\begin{table*}[h]
\caption{Same as table \ref{tab:show19data}, but for PCz7.17-01}
\label{tab:PCz7.17-01data}
\centering
\begin{tabular}{l l l l l l l l l}
\hline
\hline
\# & ID & R.A.\, (J2000) & Decl.\, (J2000) & $z_{\rm Lephare}$ & $z_{\rm EAZY}$ & $\log(M_{\rm \star, LePhare})$ & $SFR_{\rm UV}$ & $K_{\rm S}$ \\ \hline
   &    & {\it hh:mm:ss.ss} & {\it dd:mm:ss.ss} &                   &                &                                                ${\rm dex}$        &                        ${\rm [M_{\odot}\,yr^{-1}]}$                          &     ${\rm [AB\,mag]}$                         \\ \hline
1 & 43737 & 09:57:46.98 & 02:02:44.79 & 7.23$^{+0.11}_{-0.11}$ & 7.24$^{+0.05}_{-0.04}$ & 9.0$^{+0.1}_{-0.2}$ & 10$^{+1}_{-1}$ & 25.46\\
2 & 204754 & 09:57:43.51 & 02:03:40.26 & 7.20$^{+0.11}_{-0.12}$ & 7.20$^{+0.04}_{-0.04}$ & 8.7$^{+0.2}_{-0.2}$ & 8$^{+1}_{-1}$ & 26.23\\
3 & 258823 & 09:57:41.55 & 02:03:56.50 & 7.22$^{+0.14}_{-0.18}$ & 6.86$^{+0.05}_{-0.06}$ & 8.8$^{+0.2}_{-0.2}$ & 11$^{+2}_{-2}$ & 26.28\\
4 & 368191 & 09:57:42.57 & 02:04:27.35 & 7.12$^{+0.13}_{-0.17}$ & 7.15$^{+0.05}_{-0.06}$ & 10.6$^{+0.2}_{-0.2}$ & 270$^{+27}_{-26}$ & 25.46\\
5 & 871193 & 09:57:41.62 & 02:02:30.18 & 7.28$^{+0.12}_{-0.11}$ & 7.17$^{+0.05}_{-0.22}$ & 8.5$^{+0.1}_{-0.1}$ & 7$^{+1}_{-1}$ & 26.22\\
\hline
\end{tabular}
\end{table*}

\clearpage

\bigskip

\begin{figure}[h]
    \centering
    \includegraphics[width=0.8\textwidth]{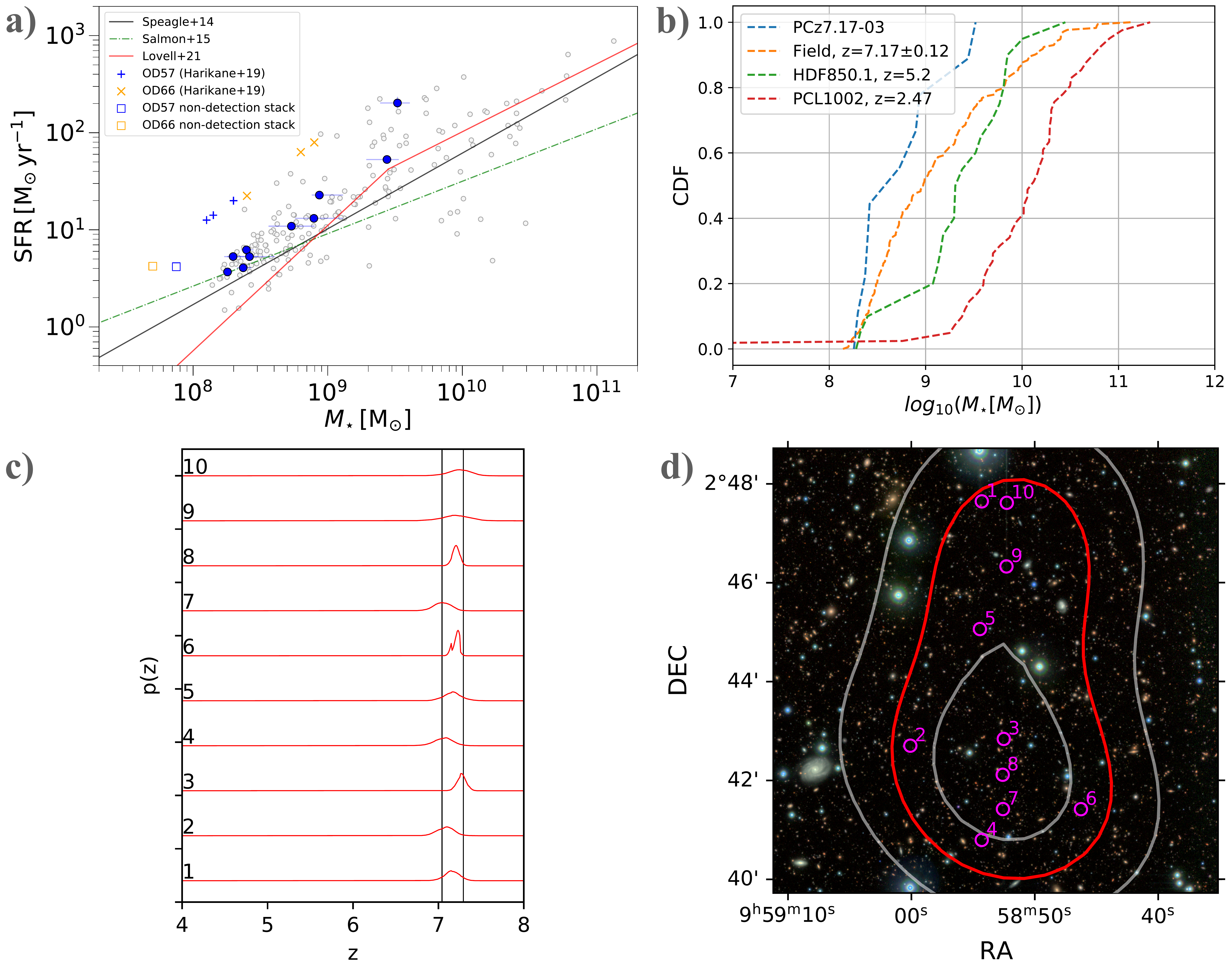}
    \caption{}
        \label{fig:PCz7.17-03}
\end{figure}

\noindent{\bf COSMOS2020-PCz7.17-03:}
We identify 10 sources inside the $4\sigma$ contour of this overdensity, of which 0 are $z$-band dropouts (Table \ref{tab:PCz7.17-03data}). The galaxies span a volume of $8.7\times17.3\times76.0\,{\rm cMpc}$. (Fig.~\ref{fig:PCz7.17-03}c). We infer a dark matter halo mass of 
$M_{\rm DM}\approx 2-7\times 10^{11}\,M_{\odot}$ using the methods described in \ref{subsection:dark-matter-halo}. The stellar mass CDF for the 10 galaxies is shown in Fig.~\ref{fig:PCz7.17-03}b along with the CDF of the field population. A two-sample K-S test of the two CDFs yields a statistic of $0.36$ and a $p$-value of $0.13$.
\begin{table*}[h]
\caption{Same as table \ref{tab:show19data}, but for PCz7.17-03}
\label{tab:PCz7.17-03data}
\centering
\begin{tabular}{l l l l l l l l l}
\hline
\hline
\# & ID & R.A.\, (J2000) & Decl.\, (J2000) & $z_{\rm Lephare}$ & $z_{\rm EAZY}$ & $\log(M_{\rm \star, LePhare})$ & $SFR_{\rm UV}$ & $K_{\rm S}$ \\ 
   &    & {\it hh:mm:ss.ss} & {\it dd:mm:ss.ss} &                   &                &                                                ${\rm dex}$        &                        ${\rm [M_{\odot}\,yr^{-1}]}$                          &     ${\rm [AB\,mag]}$                         \\ \hline
1 & 12090 & 09:58:54.28 & 02:47:38.65 & 7.16$^{+0.09}_{-0.09}$ & 7.17$^{+0.04}_{-0.04}$ & 8.7$^{+0.2}_{-0.2}$ & 11$^{+1}_{-1}$ & 25.77\\
2 & 15707 & 09:59:00.08 & 02:42:42.19 & 7.09$^{+0.09}_{-0.11}$ & 6.96$^{+0.05}_{-0.03}$ & 8.3$^{+0.1}_{-0.1}$ & 4$^{+1}_{-1}$ & 27.11\\
3 & 65232 & 09:58:52.51 & 02:42:50.36 & 7.27$^{+0.05}_{-0.05}$ & 7.21$^{+0.03}_{-0.03}$ & 8.9$^{+0.2}_{-0.1}$ & 23$^{+1}_{-1}$ & 24.89\\
4 & 425359 & 09:58:54.30 & 02:40:47.76 & 7.06$^{+0.10}_{-0.12}$ & 7.08$^{+0.05}_{-0.08}$ & 8.3$^{+0.2}_{-0.1}$ & 5$^{+1}_{-1}$ & 26.05\\
5 & 561655 & 09:58:54.44 & 02:45:03.72 & 7.17$^{+0.10}_{-0.10}$ & 6.94$^{+0.03}_{-0.03}$ & 8.4$^{+0.1}_{-0.1}$ & 4$^{+1}_{-1}$ & 25.45\\
6 & 572331 & 09:58:46.26 & 02:41:25.08 & 7.21$^{+0.03}_{-0.06}$ & 7.14$^{+0.01}_{-0.01}$ & 9.5$^{+0.1}_{-0.1}$ & 203$^{+32}_{-25}$ & 24.62\\
7 & 573770 & 09:58:52.58 & 02:41:25.24 & 7.05$^{+0.10}_{-0.10}$ & 6.94$^{+0.03}_{-0.02}$ & 8.4$^{+0.1}_{-0.1}$ & 6$^{+1}_{-1}$ & 26.39\\
8 & 720685 & 09:58:52.60 & 02:42:06.86 & 7.21$^{+0.04}_{-0.04}$ & 7.20$^{+0.02}_{-0.02}$ & 9.4$^{+0.1}_{-0.2}$ & 53$^{+1}_{-1}$ & 24.01\\
9 & 748702 & 09:58:52.29 & 02:46:19.69 & 7.20$^{+0.16}_{-0.16}$ & 7.16$^{+0.06}_{-0.08}$ & 8.4$^{+0.2}_{-0.2}$ & 5$^{+1}_{-1}$ & 25.9\\
10 & 931635 & 09:58:52.26 & 02:47:36.93 & 7.24$^{+0.13}_{-0.17}$ & 7.23$^{+0.06}_{-0.05}$ & 8.9$^{+0.2}_{-0.1}$ & 13$^{+1}_{-1}$ & 26.57\\
\hline
\end{tabular}
\end{table*}

\clearpage

\bigskip

\begin{figure}[h]
    \centering
    \includegraphics[width=0.8\textwidth]{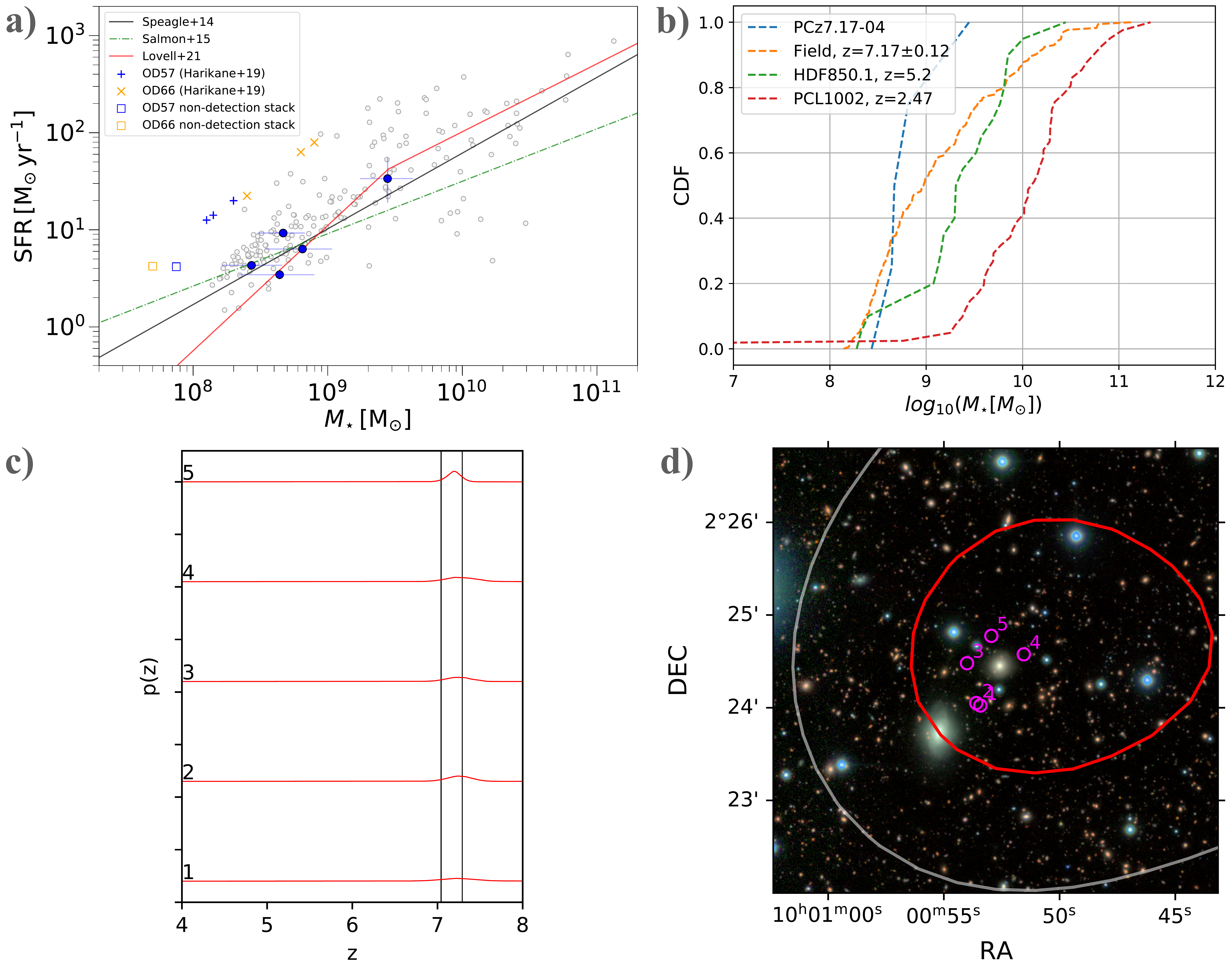}
    \caption{}
       \label{fig:PCz7.17-04}
\end{figure}

\noindent{\bf COSMOS2020-PCz7.17-04:}
We identify 5 sources inside the $4\sigma$ contour of this overdensity, of which 0 are $z$-band dropouts (Table \ref{tab:PCz7.17-04data}). The galaxies span a volume of $1.6\times1.9\times31.7\,{\rm cMpc}$. (Fig.~\ref{fig:PCz7.17-04}c). We infer a dark matter halo mass of 
$M_{\rm DM}\approx 9-35\times 10^{10}\,M_{\odot}$ using the methods described in \ref{subsection:dark-matter-halo}. The stellar mass CDF for the 5 galaxies is shown in Fig.~\ref{fig:PCz7.17-04}b along with the CDF of the field population. A two-sample K-S test of the two CDFs yields a statistic of $0.34$ and a $p$-value of $0.52$.
\begin{table*}[h]
\caption{Same as table \ref{tab:show19data}, but for PCz7.17-04}
\label{tab:PCz7.17-04data}
\centering
\begin{tabular}{l l l l l l l l l}
\hline
\hline
\# & ID & R.A.\, (J2000) & Decl.\, (J2000) & $z_{\rm Lephare}$ & $z_{\rm EAZY}$ & $\log(M_{\rm \star, LePhare})$ & $SFR_{\rm UV}$ & $K_{\rm S}$ \\ 
   &    & {\it hh:mm:ss.ss} & {\it dd:mm:ss.ss} &                   &                &                                                ${\rm dex}$        &                        ${\rm [M_{\odot}\,yr^{-1}]}$                          &     ${\rm [AB\,mag]}$                         \\ \hline
1 & 300458 & 10:00:53.42 & 02:24:01.42 & 7.25$^{+0.68}_{-0.35}$ & 7.26$^{+0.09}_{-0.08}$ & 8.6$^{+0.3}_{-0.3}$ & 3$^{+1}_{-1}$ & 27.51\\
2 & 308815 & 10:00:53.60 & 02:24:03.06 & 7.24$^{+0.15}_{-0.20}$ & 7.28$^{+0.06}_{-0.06}$ & 8.8$^{+0.2}_{-0.3}$ & 6$^{+1}_{-1}$ & 26.98\\
3 & 354781 & 10:00:54.00 & 02:24:28.83 & 7.29$^{+1.68}_{-0.19}$ & 7.29$^{+0.07}_{-0.07}$ & 9.4$^{+0.2}_{-0.2}$ & 34$^{+21}_{-15}$ & 25.95\\
4 & 368364 & 10:00:51.55 & 02:24:34.59 & 7.29$^{+0.22}_{-0.18}$ & 7.26$^{+0.09}_{-0.07}$ & 8.4$^{+0.2}_{-0.2}$ & 4$^{+1}_{-1}$ & 26.39\\
5 & 409511 & 10:00:52.95 & 02:24:46.54 & 7.19$^{+0.08}_{-0.09}$ & 7.17$^{+0.04}_{-0.04}$ & 8.7$^{+0.2}_{-0.2}$ & 9$^{+1}_{-1}$ & 25.89\\
\hline
\end{tabular}
\end{table*}

\clearpage

\bigskip

\begin{figure}[h]
    \centering
    \includegraphics[width=0.8\textwidth]{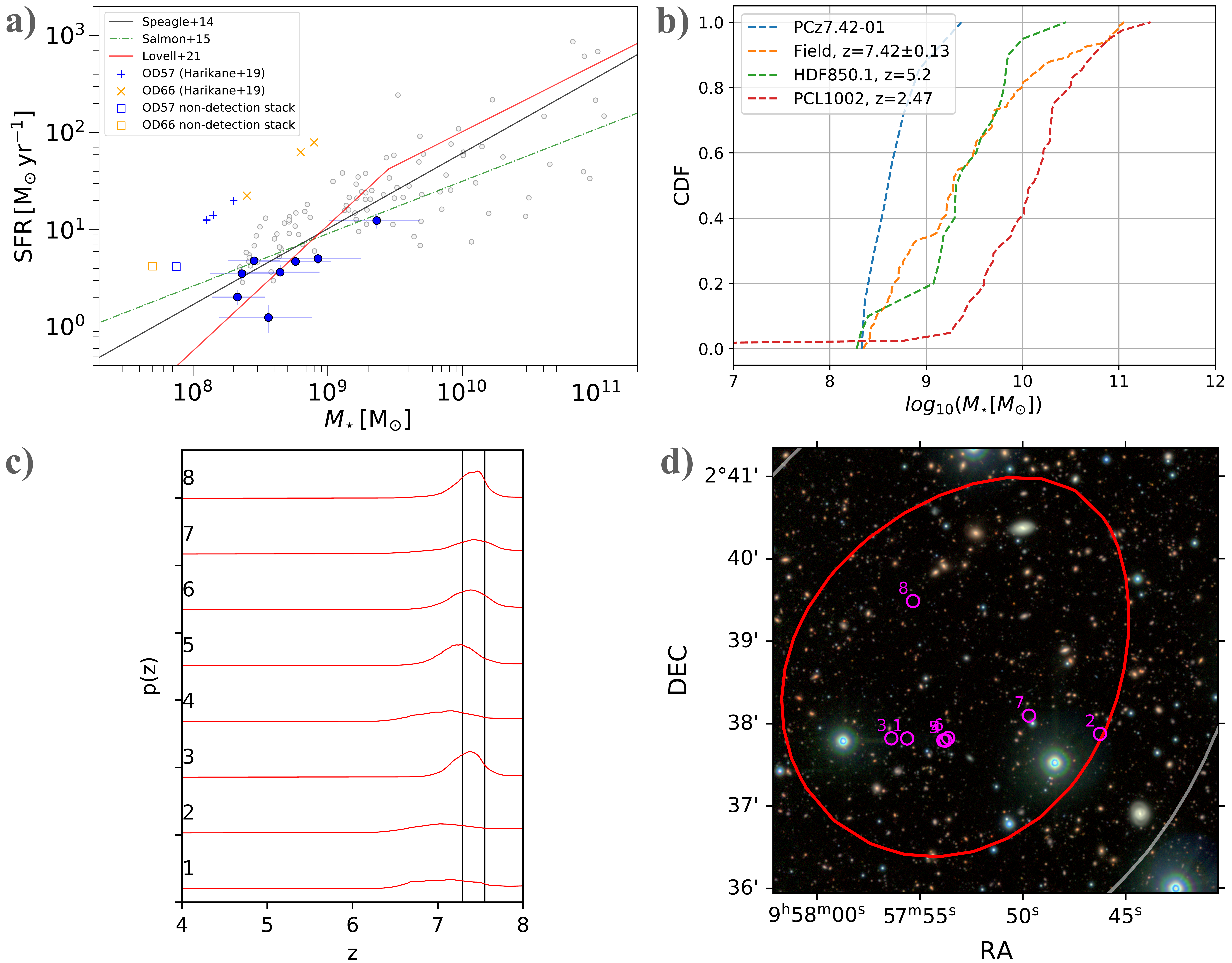}
    \caption{}
        \label{fig:PCz7.42-01}
\end{figure}

\noindent{\bf COSMOS2020-PCz7.42-01:}
We identify 8 sources inside the $4\sigma$ contour of this overdensity, of which 0 are $z$-band dropouts (Table \ref{tab:PCz7.42-01data}). The galaxies span a volume of $6.5\times4.3\times54.1\,{\rm cMpc}$. (Fig.~\ref{fig:PCz7.42-01}c). We infer a dark matter halo mass of 
$M_{\rm DM}\approx 1-4\times 10^{11}\,M_{\odot}$ using the methods described in \ref{subsection:dark-matter-halo}. The stellar mass CDF for the 8 galaxies is shown in Fig.~\ref{fig:PCz7.42-01}b along with the CDF of the field population. A two-sample K-S test of the two CDFs yields a statistic of $0.53$ and a $p$-value of $0.02$.
\begin{table*}[h]
\caption{Same as table \ref{tab:show19data}, but for PCz7.42-01}
\label{tab:PCz7.42-01data}
\centering
\begin{tabular}{l l l l l l l l l}
\hline
\hline
\# & ID & R.A.\, (J2000) & Decl.\, (J2000) & $z_{\rm Lephare}$ & $z_{\rm EAZY}$ & $\log(M_{\rm \star, LePhare})$ & $SFR_{\rm UV}$ & $K_{\rm S}$ \\ \hline
   &    & {\it hh:mm:ss.ss} & {\it dd:mm:ss.ss} &                   &                &                                                ${\rm dex}$        &                        ${\rm [M_{\odot}\,yr^{-1}]}$                          &     ${\rm [AB\,mag]}$                         \\ \hline
1 & 39068 & 09:57:55.62 & 02:37:49.20 & 7.40$^{+1.20}_{-0.56}$ & 7.45$^{+0.19}_{-0.17}$ & 8.4$^{+0.3}_{-0.2}$ & 4$^{+1}_{-1}$ & 26.99\\
2 & 40168 & 09:57:46.25 & 02:37:52.44 & 7.39$^{+1.01}_{-0.57}$ & 7.27$^{+1.16}_{-0.46}$ & 8.6$^{+0.3}_{-0.4}$ & 1$^{+1}_{-1}$ & nan\\
3 & 41007 & 09:57:56.39 & 02:37:49.26 & 7.39$^{+0.41}_{-0.21}$ & 7.30$^{+0.13}_{-7.18}$ & 8.5$^{+0.2}_{-0.2}$ & 5$^{+1}_{-1}$ & 28.48\\
4 & 42562 & 09:57:53.76 & 02:37:47.65 & 7.34$^{+1.13}_{-0.50}$ & 6.98$^{+0.23}_{-0.07}$ & 8.3$^{+0.2}_{-0.2}$ & 2$^{+1}_{-1}$ & 27.94\\
5 & 43841 & 09:57:53.86 & 02:37:47.76 & 7.29$^{+0.48}_{-0.25}$ & 7.23$^{+0.11}_{-0.11}$ & 8.6$^{+0.3}_{-0.3}$ & 4$^{+1}_{-1}$ & 27.6\\
6 & 44452 & 09:57:53.62 & 02:37:49.56 & 7.41$^{+0.44}_{-0.27}$ & 7.39$^{+0.11}_{-0.11}$ & 8.8$^{+0.3}_{-0.3}$ & 5$^{+1}_{-1}$ & 26.17\\
7 & 85572 & 09:57:49.69 & 02:38:05.73 & 7.46$^{+0.71}_{-0.47}$ & 7.47$^{+0.11}_{-0.14}$ & 8.9$^{+0.3}_{-0.3}$ & 5$^{+1}_{-1}$ & 26.42\\
8 & 384723 & 09:57:55.33 & 02:39:29.21 & 7.41$^{+0.23}_{-0.22}$ & 7.46$^{+0.07}_{-0.07}$ & 9.4$^{+0.3}_{-0.4}$ & 12$^{+2}_{-2}$ & nan\\
\hline
\end{tabular}
\end{table*}

\clearpage

\bigskip

\begin{figure}[h]
    \centering
    \includegraphics[width=0.8\textwidth]{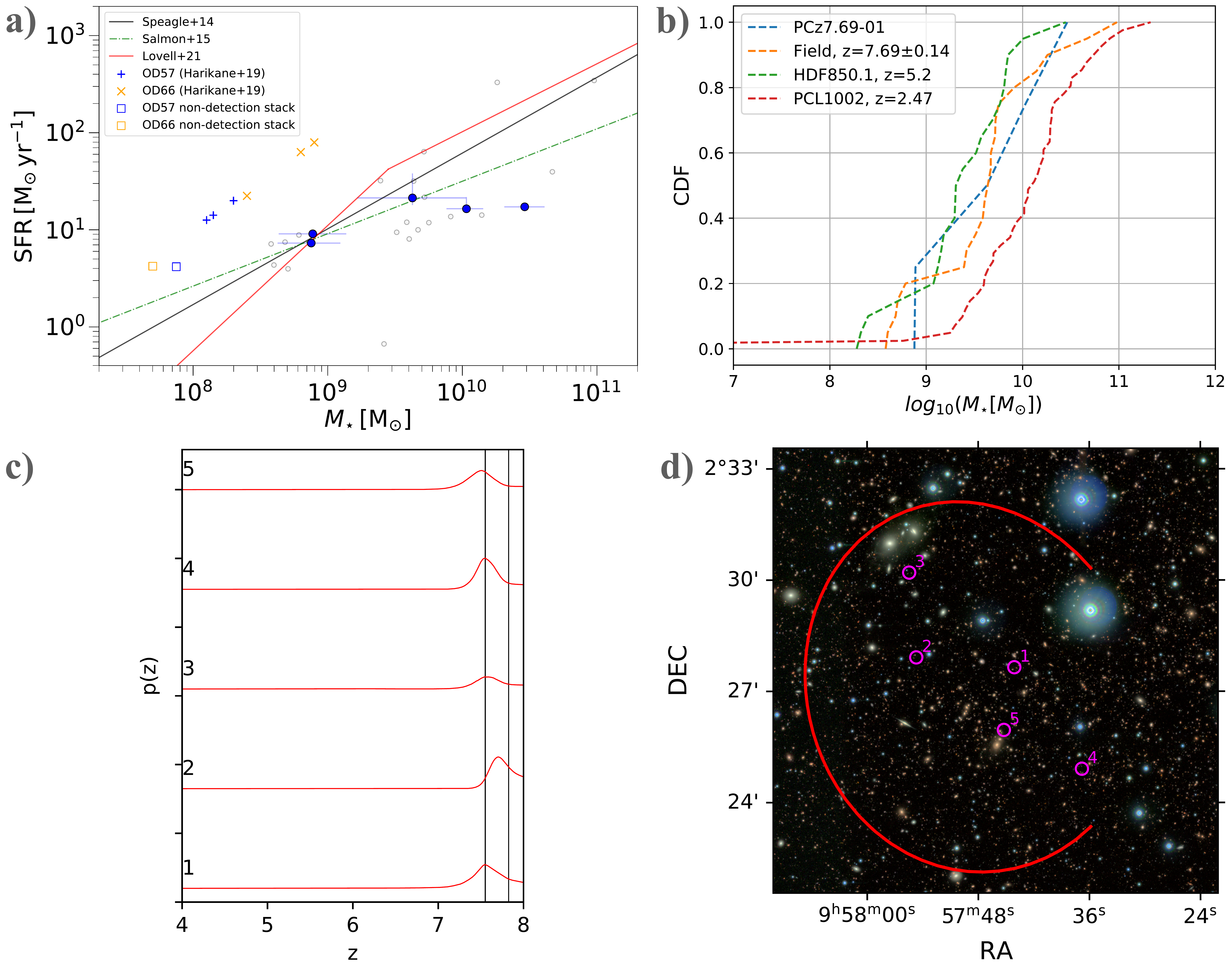}
        \caption{}
        \label{fig:PCz7.69-01}
\end{figure}

\noindent{\bf COSMOS2020-PCz7.69-01:}
We identify 5 sources inside the $4\sigma$ contour of this overdensity, of which 0 are $z$-band dropouts (Table \ref{tab:PCz7.69-01data}). The galaxies span a volume of $12.0\times13.6\times49.4\,{\rm cMpc}$. (Fig.~\ref{fig:PCz7.69-01}c). We infer a dark matter halo mass of 
$M_{\rm DM}\approx 9\times 10^{11}\,M_{\odot}$ using the methods described in \ref{subsection:dark-matter-halo}. The stellar mass CDF for the 5 galaxies is shown in Fig.~\ref{fig:PCz7.69-01}b along with the CDF of the field population. A two-sample K-S test of the two CDFs yields a statistic of $0.24$ and a $p$-value of $0.92$.
\begin{table*}[h]
\caption{Same as table \ref{tab:show19data}, but for PCz7.69-01}
\label{tab:PCz7.69-01data}
\centering
\begin{tabular}{l l l l l l l l l}
\hline
\hline
\# & ID & R.A.\, (J2000) & Decl.\, (J2000) & $z_{\rm Lephare}$ & $z_{\rm EAZY}$ & $\log(M_{\rm \star, LePhare})$ & $SFR_{\rm UV}$ & $K_{\rm S}$ \\ 
   &    & {\it hh:mm:ss.ss} & {\it dd:mm:ss.ss} &                   &                &                                                ${\rm dex}$        &                        ${\rm [M_{\odot}\,yr^{-1}]}$                          &     ${\rm [AB\,mag]}$                         \\ \hline
1 & 20030 & 09:57:44.11 & 02:27:38.97 & 7.64$^{+0.51}_{-0.24}$ & 7.39$^{+0.12}_{-0.12}$ & 10.5$^{+0.1}_{-0.2}$ & 17$^{+1}_{-1}$ & 25.92\\
2 & 71035 & 09:57:54.70 & 02:27:54.93 & 7.80$^{+0.33}_{-0.16}$ & 7.67$^{+0.08}_{-0.07}$ & 10.0$^{+0.1}_{-0.1}$ & 16$^{+5}_{-1}$ & 25.84\\
3 & 550618 & 09:57:55.46 & 02:30:12.19 & 7.66$^{+1.01}_{-6.56}$ & 7.65$^{+1.22}_{-0.08}$ & 8.9$^{+0.2}_{-0.3}$ & 7$^{+1}_{-1}$ & 26.42\\
4 & 605774 & 09:57:36.81 & 02:24:54.68 & 7.64$^{+0.65}_{-0.17}$ & 7.59$^{+0.06}_{-0.06}$ & 8.9$^{+0.2}_{-0.3}$ & 9$^{+1}_{-1}$ & 27.48\\
5 & 801177 & 09:57:45.23 & 02:25:57.54 & 7.69$^{+1.15}_{-0.30}$ & 7.51$^{+0.09}_{-0.10}$ & 9.6$^{+0.4}_{-0.4}$ & 21$^{+17}_{-3}$ & 26.21\\
\hline
\end{tabular}
\end{table*}
\clearpage
\end{appendix}

\end{document}